\documentstyle[lscape,rotate,epsf]{mn}

\def\gaeq{\stackrel{>}{\scriptstyle \sim}}
\def\laeq{\stackrel{<}{\scriptstyle \sim}}
\title[Dust-enshrouded AGB stars in the Solar Neighbourhood]
{Dust-enshrouded Asymptotic Giant Branch Stars in the Solar Neighbourhood
} 
\author[E. A. Olivier et al.]
{Enrico Olivier $^{1,2,3}$, Patricia Whitelock $^{1}$ and Fred Marang $^{1}$\\
$^{1}$ South African Astronomical Observatory, PO Box 9, Observatory,
7935, South Africa email: paw@saao.ac.za.\\
$^{2}$ Department of Physics, University of the Western Cape, 
Private Bag X17, Bellville, 7535, South Africa.\\ 
$^3$ Present address: The Research School of Astronomy and Astrophysics, 
Private Bag, Weston Creek PO, Canberra, ACT, 2611,\\ Australia.
email: rickyo@mso.anu.edu.au.}
\begin{document}

\maketitle

\begin{abstract}
A study is made of a sample of 58 dust-enshrouded Asymptotic Giant Branch (AGB) 
stars (including 2 possible post AGB stars), of which 27 are carbon-rich and
31 are oxygen-rich. These objects were originally identified by Jura \&
Kleinmann as nearby (within about 1\,kpc of the sun) AGB stars with high
mass-loss rates ($\dot{M}>10^{-6}\,M_{\odot}yr^{-1}$). Ground-based near-infrared
photometry, data obtained by the IRAS satellite and kinematic data (radial
and outflow velocities) from the literature are combined to investigate the
properties of these stars. The light amplitude in the near-infrared is found
to be correlated with period, and this amplitude decreases with increasing
wavelength.  Statistical tests show that there is no reason to suspect any
difference in the period distributions of the carbon- and oxygen-rich stars
for periods less than 1000 days. There are no carbon-rich stars with periods
longer than 1000 days in the sample. The colours are consistent with those
of cool stars with evolved circumstellar dust-shells.  Luminosities and
distances are estimated using a period-luminosity relation. Mass-loss rates,
estimated from the $60\,\mu$m fluxes, show a correlation with various
infrared colours and pulsation period. The mass-loss rate is tightly
correlated with the $K-[12]$ colour. The kinematics and scale-height of the
sample shows that the sources with periods less than 1000 days must have low
mass main-sequence progenitors. It is argued that the three oxygen-rich
stars with periods over 1000 days probably had intermediate mass
main-sequence progenitors. For the other stars an average progenitor mass of
about $1.3\,M_{\odot}$ is estimated with a final white dwarf mass of
$0.6\,M_{\odot}$. The average lifetime of stars in this high mass-loss AGB
phase is estimated to be about $4\times10^{4}$ years, which suggests that
these stars will undergo at most one more thermal pulse before leaving the
AGB if current theoretical relations between thermal interpulse-period 
and core-mass are correct. 
\end{abstract}

\begin{keywords}

\end{keywords}
	
\section{Introduction}
 Stars of low to intermediate main-sequence mass, up to about
6-8$\,M_{\odot}$, will eventually evolve up the Asymptotic Giant Branch
(AGB) (Iben \& Renzini 1983).  At the tip of the AGB, where they combine a
low effective temperatures ($T_{eff} \laeq 3000\,K$) with high luminosities
($L \gaeq 3000\,L_{\odot}$), the stars pulsate and are subject to rapid mass
loss. Typical examples of such stars are the Miras and the OH/IR\footnote{A
small fraction of OH/IR stars are supergiants, i.e. their progenitors were
massive stars.} stars (e.g. Feast \& Whitelock 1987; Habing 1996). Their
light output varies on timescales from around one hundred days to well over
a thousand days. They also lose mass rapidly ($10^{-7} \laeq
\dot{M} \laeq 10^{-4}\,M_{\odot}\,yr^{-1}$). They follow various
period-luminosity relations and are at or near the maximum luminosity they
will ever achieve during their evolution. This makes them very useful as
distance indicators and as probes of galactic structure. They are also
important contributors of dust into the interstellar medium, which may be
crucial for the formation of new stars and planets.

Jura \& Kleinmann (1989 henceforth JK89) identified a group of mass losing
AGB stars within about 1\,kpc of the sun. In 1989 a monitoring
programme of stars in the JK89 sample visible from the southern hemisphere
was initiated from SAAO. This entailed obtaining near-infrared photometry at
$JHKL$ over a time interval of several years, from which pulsation periods
could be determined. The aim of this study was to investigate the physical
properties of the sample, in particular variability, colours, mass-loss
rates and kinematics, using SAAO photometry together with infrared
photometry and kinematic data from the literature. 

Finally these data are used to estimate the average progenitor mass and
lifetime of stars in this high mass-loss phase.

\section{The Sample}
 JK89 aimed to develop a more quantitative understanding of mass loss from
local AGB stars. They identified AGB stars undergoing heavy mass loss
($\dot{M}>10^{-6}\,M_{\odot}\,yr^{-1}$), henceforth referred to as
dust-enshrouded, within about 1\,kpc of the sun, and within
$-33^{\circ}<\delta{(1950)}<82^{\circ}$. Their selection was based on
infrared photometry from the Two Micron Sky Survey, IRAS and the AFGL survey
from which they identify 63 sources. They give reasons for thinking this
sample was reasonably complete. 

A few comments should be made at this point. JK89 assume a luminosity of
$10^{4}\,L_{\odot}$ for all of these AGB stars. It will be seen later that,
if the period luminosity relation (PL) is correct, this leads to the
inclusion of more luminous OH/IR stars outside the 1\,kpc sphere and up to
$\sim{2}$\,kpc away. They note that some red supergiants will satisfy their
selection criteria; those they largely excluded, but included the star 3
Puppis, an A supergiant, with the idea that it might be undergoing the
transition from AGB star to planetary nebula. However, Plets, Waelkens \&
Trams (1995) convincingly showed that 3 Puppis is a genuine massive
supergiant, and we therefore exclude it from this study. JK89 also included
the carbon-rich planetary nebulae NGC 7027 (e.g. Justtanont et al. 2000) 
in their sample; this object is excluded from the present study. The
remaining 61 sources are listed in Table 1 (this includes 3 M supergiants, 
$17513-2313$, $18135-1641$ and $18204-1344$ 
(Ukita \& Goldsmith 1984; Winfrey et al. 1994) of which we only became aware 
part the way through this project). Near-infrared data are tabulated for two 
of them, but none are used in the analysis.

\begin{table}
\centering
\caption{Infrared Sources}
\scriptsize{
\begin{tabular}{lcccclcccc} 
\multicolumn{1}{c}{{INFRARED}}&
\multicolumn{1}{c}{{CHEM.}}&
\multicolumn{1}{c}{{SPEC.}}&
\multicolumn{1}{c}{{VARIABLE}}&
\multicolumn{1}{c}{{NO. OF}}\\
\multicolumn{1}{c}{{SOURCE}}&
\multicolumn{1}{c}{{TYPE}}&
\multicolumn{1}{c}{{TYPE}}&
\multicolumn{1}{c}{NAME$^{\dagger}$}&
\multicolumn{1}{c}{REF.$^{*}$}\\ \hline
\\
IRAS $00042+4248$ &  O &  M10    &    KU And      &    47\\          
IRAS $01037+1219$ &  O &  M9     &    WX Psc      &   216\\ 
IRAS $01159+7220$ &  O &  Se     &    S  Cas      &    57\\ 
IRAS $02270-2619$ &  C &         &    R  For      &    91\\ 
IRAS $02316+6455$ &  O &  M9     &    V656 Cas    &    39\\ 
IRAS $02351-2711$ &  O &  M9     &    UU For      &    37\\ 
IRAS $03229+4721$ &  C &  C      &    V384 Per    &   101\\ 
IRAS $03507+1115$ &  O &  M6e    &    IK Tau      &   277\\ 
IRAS $04307+6210$ &  C &         &                &    38\\  
IRAS $04566+5606$ &  O &  M8.5   &    TX Cam      &   147\\  
IRAS $05073+5248$ &  O &  M      &    NV Aur      &   120\\ 
IRAS $05411+6957$ &  O &  M9     &    BX Cam      &    52\\ 
IRAS $05559+7430$ &  O &  M7     &    V  Cam      &    72\\     
IRAS $06176-1036$ &  C &  B8V    & (Red Rectangle)&   272\\ 
IRAS $06300+6058$ &  O &  M9     &    AP Lyn      &    74\\ 
IRAS $06500+0829$ &  O &  M9     &    GX Mon      &    52\\ 
IRAS $08088-3243$ &  C &         &    V346 Pup    &    31\\ 
IRAS $09116-2439$ &  C &  C      &                &    27\\ 
IRAS $09429-2148$ &  O &  M9     &    IW Hya      &    74\\ 
IRAS $09452+1330$ &  C &  C      &    CW Leo      &  1004\\ 
IRAS $10131+3049$ &  C &  Ce     &    RW LMi      &   218\\ 
IRAS $10491-2059$ &  C &  C      &    V  Hya      &   164\\ 
IRAS $12447+0425$ &  C &  Re     &    RU Vir      &    82\\ 
IRAS $17049-2440$ &  C &         &                &    37\\ 
IRAS $17119+0859$ &  O &  M9     &    V2108 Oph   &    51\\ 
IRAS $17297+1747$ &  O &  M2     &    V833 Her    &    50\\ 
IRAS $17360-3012$ &  O &         &    V1019 Sco   &    40\\ 
IRAS $17411-3154$ &  O &         &                &    31\\ 
IRAS $17513-2313$ &  O &  M5(I)     &    V774 Sgr    &    15\\ 
IRAS $18009-2019$ &  O &  M8     &    V4120 Sgr   &    43\\ 
IRAS $18040-0941$ &  C &  C      &    FX Ser      &    43\\ 
IRAS $18135-1641$ &  O &  M5(I)     &                &    15\\
IRAS $18194-2708$ &  C &         &                &    33\\
IRAS $18204-1344$ &  O &  M8(I)     &                &    39\\
IRAS $18240+2326$ &  C &  C      &                &    51\\
IRAS $18333+0533$ &  O &         &    NX Ser      &    44\\
IRAS $18348-0526$ &  O &  M      &    V437 Sct    &   164\\
IRAS $18349+1023$ &  O &  M9     &    V1111 Oph   &    85\\
IRAS $18397+1738$ &  C &  C      &    V821 Her    &    75\\
IRAS $18398-0220$ &  C &  C      &    V1417 Aql   &    46\\
IRAS $18413+1354$ &  O &  M7     &    V837 Her    &    35\\
IRAS $18560-2954$ &  O &  M9     &    V3953 Sgr   &    28\\
IRAS $19008+0726$ &  C &  R      &    V1418 Aql   &    43\\
IRAS $19059-2219$ &  O &  M8     &    V3880 Sgr   &    53\\
IRAS $19093-3256$ &  O &  M9     &    V342 Sgr    &    20\\
IRAS $19126-0708$ &  O &  S      &    W Aql       &    97\\
IRAS $19175-0807$ &  C &  C      &    V1420 Aql   &    52\\
IRAS $19321+2757$ &  C &  C      &    V1965 Cyg   &    48\\
IRAS $20077-0625$ &  O &  M      &    V1300 Aql   &    73\\
IRAS $20396+4757$ &  C &  N      &    V Cyg       &   162\\
IRAS $20440-0105$ &  O &  M9     &    FP Aqr      &    30\\
IRAS $20570+2714$ &  C &  Ce     &                &    35\\
IRAS $21032-0024$ &  C &  Ce     &    RV Aqr      &    54\\
IRAS $21286+1055$ &  O &  M7e    &    UU Peg      &    35\\
IRAS $21320+3850$ &  C &  C      &    V1426 Cyg   &    72\\
IRAS $21456+6422$ &  O &  M6     &    RT Cep      &    18\\
IRAS $23166+1655$ &  C &         &    LL Peg      &   109\\
IRAS $23320+4316$ &  C &  C      &    LP And      &    98\\
IRAS $23496+6131$ &  O &  M9     &    V657 Cas    &    24\\
RAFGL $ 1406    $ &  C &         &    IY Hya      &    40\\
RAFGL $ 2688    $ &  C &  F5Iae  &  (Egg Nebula)  &   347\\
\end{tabular}
}
\begin{flushleft}
\small{
\vspace{0.4cm}
${\dagger}$-Names in brackets are most commonly used.\\
${*}$-From 1960 to 1999.\\
}
\end{flushleft}
\end{table}

 Chemical types (O and C for oxygen- and carbon-rich, respectively) for all
the sources were taken from Loup et al. (1993), and spectral types obtained
from the SIMBAD 
data base for 51 of the sources are also listed in Table 1. Although 
$04307+6210$ is listed as M6 in SIMBAD, Groenewegen (1994) suggests this is
incorrect and the star is actually carbon rich; therefore no spectral type
is assigned to it here. The supergiants are marked (I). 

The final sample thus consists of 27 carbon-rich stars and 31 oxygen-rich
stars (excluding the 3 supergiants) of which 2 are S-type. Almost all of the
stars are of very late spectral
type (M, C, N, R and S), with two exceptions: the Red Rectangle and the Egg
Nebula. Both are well-known carbon-rich bipolar reflection nebulae (see, e.g. 
Sahai et al. 1998; Waters et al. 1998). The Red Rectangle is a wide
binary system with an oxygen-rich dust disk (Waters et al. 1998). These two
objects are generally thought to be post-AGB stars on their way to become
Planetary Nebulae.

The Benson et al. (1990) catalogue lists sources which have been examined
for OH, SiO and $\rm H_{2}O$ maser emission.  Of the 28 oxygen-rich stars in
our sample which have been examined for OH and $\rm H_{2}O$, 20 were
detected in OH and 20 in $\rm H_{2}O$. While 19 of the 21 sources searched
for SiO were detected.  Interestingly, 4 of the 11 carbon-rich stars which
were searched for SiO also had positive detections. We can thus estimate
that 70 percent of the oxygen-rich stars have OH and $\rm H_{2}O$ maser
emission and about 90 percent have SiO emission.

\setcounter{table}{1}
\onecolumn 
\begin{landscape}
\begin{table}
\vspace{1.5 cm}
\caption{Near-infrared data. The  Date is given as JD--2440000.00.}
\footnotesize{
\begin{tabular}{lrrrrclrrrrclrrrrclrrrr}\\
\multicolumn{1}{c}{\bf{DATE}}&
\multicolumn{1}{c}{\bf{J}}&
\multicolumn{1}{c}{\bf{H}}&
\multicolumn{1}{c}{\bf{K}}&
\multicolumn{1}{c}{\bf{L}}&
&
\multicolumn{1}{c}{\bf{DATE}}&
\multicolumn{1}{c}{\bf{J}}&
\multicolumn{1}{c}{\bf{H}}&
\multicolumn{1}{c}{\bf{K}}&
\multicolumn{1}{c}{\bf{L}}&
&
\multicolumn{1}{c}{\bf{DATE}}&
\multicolumn{1}{c}{\bf{J}}&
\multicolumn{1}{c}{\bf{H}}&
\multicolumn{1}{c}{\bf{K}}&
\multicolumn{1}{c}{\bf{L}}&
&
\multicolumn{1}{c}{\bf{DATE}}&
\multicolumn{1}{c}{\bf{J}}&
\multicolumn{1}{c}{\bf{H}}&
\multicolumn{1}{c}{\bf{K}}&
\multicolumn{1}{c}{\bf{L}}\\
\\
\multicolumn{5}{l}{\bf{IRAS $01037+1219$}}                 &\hspace{0.25cm}&\multicolumn{5}{l}{\bf{IRAS $01037+1219$  .....continued}} &\hspace{0.25cm}&\multicolumn{5}{l}{\bf{IRAS $02351-2711$  .....continued}} &\hspace{0.25cm}&\multicolumn{5}{l}{\bf{IRAS $03507+1115$  .....continued}} \\
$06304.57  $ &  $  7.04$ & $  3.90$ & $  1.70$ & $ -0.68$  &\hspace{0.25cm}&$08589.35  $ &  $  8.71$ & $  5.32$ & $  3.04$ & $  0.58$  &\hspace{0.25cm}&$09213.62  $ &  $  3.63$ & $  2.22$ & $  1.38$ & $  0.51$  &\hspace{0.25cm}&$04960.35  $ &  $  2.15$ & $  0.56$ & $ -0.56$ & $ -1.84$  \\
$06334.49  $ &  $  6.98$ & $  3.84$ & $  1.68$ & $ -0.69$  &\hspace{0.25cm}&$08855.58  $ &  $  6.60$ & $  3.75$ & $  1.74$ & $ -0.58$  &\hspace{0.25cm}&$09232.49  $ &  $  3.60$ & $  2.21$ & $  1.35$ & $  0.50$  &\hspace{0.25cm}&$05218.61  $ &  $  1.87$ & $  0.18$ & $ -0.71$ & $ -1.77$  \\
$06356.44  $ &  $  7.00$ & $  3.84$ & $  1.69$ & $ -0.69$  &\hspace{0.25cm}&$08874.55  $ &  $  6.57$ & $  3.70$ & $  1.71$ & $ -0.59$  &\hspace{0.25cm}&$          $ &  $      $ & $      $ & $      $ & $      $  &\hspace{0.25cm}&$05251.56  $ &  $  2.24$ & $  0.49$ & $ -0.52$ & $ -1.60$  \\
$06373.44  $ &  $  6.99$ & $  3.86$ & $  1.72$ & $ -0.67$  &\hspace{0.25cm}&$08930.43  $ &  $  6.51$ & $  3.64$ & $  1.67$ & $ -0.60$  &\hspace{0.25cm}&\multicolumn{5}{l}{\bf{IRAS $03507+1115$}}                 &\hspace{0.25cm}&$05335.33  $ &  $  2.73$ & $  0.87$ & $ -0.30$ & $ -1.56$  \\
$06392.36  $ &  $  7.08$ & $  3.92$ & $  1.76$ & $ -0.64$  &\hspace{0.25cm}&$08961.34  $ &  $  6.52$ & $  3.63$ & $  1.68$ & $ -0.56$  &\hspace{0.25cm}&$03816.49  $ &  $  1.54$ & $  0.05$ & $ -0.78$ & $ -1.84$  &\hspace{0.25cm}&$05602.62  $ &  $  1.25$ & $ -0.30$ & $ -1.14$ & $ -2.25$  \\
$06426.30  $ &  $  7.27$ & $  4.05$ & $  1.92$ & $ -0.46$  &\hspace{0.25cm}&$08988.30  $ &  $  6.57$ & $  3.70$ & $  1.76$ & $ -0.52$  &\hspace{0.25cm}&$03818.45  $ &  $  1.61$ & $  0.08$ & $ -0.76$ & $ -1.78$  &\hspace{0.25cm}&$05612.57  $ &  $  1.33$ & $ -0.26$ & $ -1.11$ & $ -2.27$  \\
$06640.67  $ &  $  9.66$ & $  5.84$ & $  3.28$ & $  0.63$  &\hspace{0.25cm}&$09212.61  $ &  $  8.40$ & $  5.22$ & $  2.99$ & $  0.50$  &\hspace{0.25cm}&$03821.41  $ &  $  1.78$ & $  0.27$ & $ -0.67$ & $ -1.75$  &\hspace{0.25cm}&$05616.51  $ &  $  1.33$ & $ -0.23$ & $ -1.09$ & $ -2.16$  \\
$06657.61  $ &  $  9.81$ & $  5.90$ & $  3.30$ & $  0.63$  &\hspace{0.25cm}&$09232.54  $ &  $  8.52$ & $  5.30$ & $  3.03$ & $  0.55$  &\hspace{0.25cm}&$03848.42  $ &  $  2.00$ & $  0.40$ & $ -0.49$ & $ -1.55$  &\hspace{0.25cm}&$05647.52  $ &  $  1.58$ & $ -0.05$ & $ -0.94$ & $ -2.02$  \\
$06694.55  $ &  $  9.80$ & $  5.79$ & $  3.34$ & $  0.49$  &\hspace{0.25cm}&$09300.40  $ &  $  8.68$ & $  5.43$ & $  3.06$ & $  0.41$  &\hspace{0.25cm}&$03860.30  $ &  $  2.18$ & $  0.59$ & $ -0.34$ & $ -1.45$  &\hspace{0.25cm}&$05681.34  $ &  $  1.92$ & $  0.20$ & $ -0.75$ & $ -1.85$  \\
$06713.45  $ &  $  9.79$ & $  5.93$ & $  3.28$ & $  0.56$  &\hspace{0.25cm}&$09582.57  $ &  $  6.19$ & $  3.34$ & $  1.36$ & $ -0.81$  &\hspace{0.25cm}&$03865.37  $ &  $  2.17$ & $  0.54$ & $ -0.40$ & $ -1.51$  &\hspace{0.25cm}&$05710.30  $ &  $  2.29$ & $  0.50$ & $ -0.51$ & $ -1.58$  \\
$06741.41  $ &  $  9.74$ & $  5.89$ & $  3.25$ & $  0.47$  &\hspace{0.25cm}&$09614.53  $ &  $  6.30$ & $  3.43$ & $  1.45$ & $ -0.73$  &\hspace{0.25cm}&$03868.37  $ &  $  2.25$ & $  0.61$ & $ -0.35$ & $ -1.47$  &\hspace{0.25cm}&$06309.58  $ &  $  2.81$ & $  0.89$ & $ -0.28$ & $ -1.55$  \\
$06750.38  $ &  $  9.76$ & $  5.90$ & $  3.26$ & $  0.52$  &\hspace{0.25cm}&$09637.43  $ &  $  6.41$ & $  3.53$ & $  1.54$ & $ -0.62$  &\hspace{0.25cm}&$03879.35  $ &  $  2.39$ & $  0.72$ & $ -0.25$ & $ -1.40$  &\hspace{0.25cm}&$06335.59  $ &  $  2.68$ & $  0.83$ & $ -0.32$ & $ -1.59$  \\
$06754.33  $ &  $  9.71$ & $  5.90$ & $  3.26$ & $  0.47$  &\hspace{0.25cm}&$09672.36  $ &  $  6.65$ & $  3.73$ & $  1.72$ & $ -0.42$  &\hspace{0.25cm}&$03892.32  $ &  $  2.55$ & $  0.81$ & $ -0.21$ & $ -1.35$  &\hspace{0.25cm}&$06378.51  $ &  $  1.75$ & $  0.24$ & $      $ & $ -1.99$  \\
$06774.33  $ &  $  9.57$ & $  5.78$ & $  3.15$ & $  0.39$  &\hspace{0.25cm}&$10004.45  $ &  $  8.55$ & $  5.11$ & $  2.65$ & $  0.15$  &\hspace{0.25cm}&$03896.33  $ &  $  2.51$ & $  0.82$ & $ -0.19$ & $ -1.39$  &\hspace{0.25cm}&$06439.29  $ &  $  1.27$ & $  0.00$ & $      $ & $ -2.23$  \\
$06775.31  $ &  $  9.81$ & $  5.75$ & $  3.11$ & $  0.38$  &\hspace{0.25cm}&$10019.44  $ &  $  7.87$ & $  4.67$ & $  2.33$ & $ -0.10$  &\hspace{0.25cm}&$04096.58  $ &  $  1.13$ & $ -0.09$ & $ -0.94$ & $ -2.17$  &\hspace{0.25cm}&$06491.25  $ &  $  1.16$ & $ -0.40$ & $ -1.26$ & $ -2.36$  \\
$06784.28  $ &  $  9.23$ & $  5.52$ & $  2.95$ & $  0.23$  &\hspace{0.25cm}&$10063.33  $ &  $  6.74$ & $  3.85$ & $  1.75$ & $ -0.55$  &\hspace{0.25cm}&$04101.62  $ &  $  1.09$ & $ -0.18$ & $ -1.00$ & $ -2.19$  &\hspace{0.25cm}&$06645.64  $ &  $  2.46$ & $  0.54$ & $ -0.52$ & $ -1.62$  \\
$06801.27  $ &  $  8.28$ & $  4.88$ & $  2.49$ & $ -0.09$  &\hspace{0.25cm}&$10089.28  $ &  $  6.66$ & $  3.76$ & $  1.64$ & $ -0.58$  &\hspace{0.25cm}&$04158.58  $ &  $  1.05$ & $ -0.32$ & $ -1.10$ & $ -2.19$  &\hspace{0.25cm}&$06658.65  $ &  $  2.65$ & $  0.71$ & $ -0.40$ & $ -1.55$  \\
$07005.66  $ &  $  6.78$ & $  3.72$ & $  1.65$ & $ -0.69$  &\hspace{0.25cm}&$10298.63  $ &  $  6.37$ & $  3.42$ & $  1.44$ & $ -0.59$  &\hspace{0.25cm}&$04161.52  $ &  $  1.06$ & $ -0.32$ & $ -1.10$ & $ -2.22$  &\hspace{0.25cm}&$06690.59  $ &  $  3.07$ & $  1.02$ & $ -0.17$ & $ -1.41$  \\
$07024.55  $ &  $  6.82$ & $  3.74$ & $  1.69$ & $ -0.60$  &\hspace{0.25cm}&$10360.49  $ &  $  6.88$ & $  3.83$ & $  1.79$ & $ -0.23$  &\hspace{0.25cm}&$04245.32  $ &  $  1.53$ & $ -0.02$ & $ -0.85$ & $ -1.87$  &\hspace{0.25cm}&$06723.57  $ &  $  3.28$ & $  1.13$ & $ -0.13$ & $ -1.40$  \\
$07072.49  $ &  $  7.00$ & $  3.90$ & $  1.84$ & $ -0.47$  &\hspace{0.25cm}&$10417.35  $ &  $  7.63$ & $  4.38$ & $  2.20$ & $  0.14$  &\hspace{0.25cm}&$04252.31  $ &  $  1.59$ & $  0.04$ & $ -0.81$ & $ -1.85$  &\hspace{0.25cm}&$06752.44  $ &  $  3.06$ & $  0.98$ & $ -0.23$ & $ -1.51$  \\
$07114.40  $ &  $  7.32$ & $  4.17$ & $  2.12$ & $ -0.22$  &\hspace{0.25cm}&$10717.50  $ &  $  7.03$ & $  3.89$ & $  1.70$ & $ -0.56$  &\hspace{0.25cm}&$04284.28  $ &  $  1.90$ & $  0.28$ & $ -0.61$ & $ -1.68$  &\hspace{0.25cm}&$06777.37  $ &  $  2.73$ & $  0.79$ & $ -0.38$ & $ -1.70$  \\
$07127.36  $ &  $  7.46$ & $  4.29$ & $  2.25$ & $ -0.07$  &\hspace{0.25cm}&$10736.46  $ &  $  6.83$ & $  3.81$ & $  1.63$ & $ -0.62$  &\hspace{0.25cm}&$04578.42  $ &  $  1.32$ & $ -0.15$ & $ -1.05$ & $ -2.25$  &\hspace{0.25cm}&$06783.37  $ &  $  2.73$ & $  0.76$ & $ -0.41$ & $ -1.67$  \\
$07147.03  $ &  $  7.68$ & $  4.43$ & $  2.35$ & $  0.04$  &\hspace{0.25cm}&$10805.30  $ &  $  6.39$ & $  3.37$ & $  1.30$ & $ -0.81$  &\hspace{0.25cm}&$04582.36  $ &  $  1.34$ & $ -0.09$ & $      $ & $ -2.24$  &\hspace{0.25cm}&$06803.30  $ &  $  2.58$ & $  0.68$ & $ -0.45$ & $ -1.76$  \\
$07367.67  $ &  $  9.94$ & $  6.07$ & $  3.34$ & $  0.50$  &\hspace{0.25cm}&$          $ &  $      $ & $      $ & $      $ & $      $  &\hspace{0.25cm}&$04604.36  $ &  $  1.32$ & $ -0.19$ & $ -1.08$ & $ -2.15$  &\hspace{0.25cm}&$06826.28  $ &  $  2.38$ & $  0.56$ & $ -0.52$ & $ -1.81$  \\
$07461.42  $ &  $  8.38$ & $  5.02$ & $  2.61$ & $ -0.01$  &\hspace{0.25cm}&\multicolumn{5}{l}{\bf{IRAS $02270-2619$}}                 &\hspace{0.25cm}&$04605.34  $ &  $  1.25$ & $ -0.22$ & $ -1.12$ & $ -2.32$  &\hspace{0.25cm}&$07014.69  $ &  $  1.40$ & $ -0.24$ & $ -1.13$ & $ -2.20$  \\
$07735.60  $ &  $  7.03$ & $  3.97$ & $  1.98$ & $ -0.35$  &\hspace{0.25cm}&$10361.45  $ &  $  4.64$ & $  2.88$ & $  1.54$ & $  0.21$  &\hspace{0.25cm}&$04607.32  $ &  $  1.26$ & $ -0.24$ & $ -1.18$ & $ -2.36$  &\hspace{0.25cm}&$07056.60  $ &  $  1.77$ & $  0.04$ & $ -0.89$ & $ -1.97$  \\
$07792.51  $ &  $  7.60$ & $  4.40$ & $  2.35$ & $  0.01$  &\hspace{0.25cm}&$10399.48  $ &  $  4.49$ & $  2.87$ & $  1.45$ & $  0.07$  &\hspace{0.25cm}&$04614.33  $ &  $  1.27$ & $      $ & $      $ & $ -2.32$  &\hspace{0.25cm}&$07074.61  $ &  $  2.00$ & $  0.24$ & $ -0.72$ & $ -1.90$  \\
$07817.42  $ &  $  7.93$ & $  4.63$ & $  2.50$ & $  0.14$  &\hspace{0.25cm}&$10437.32  $ &  $  4.20$ & $  2.54$ & $  1.27$ & $ -0.10$  &\hspace{0.25cm}&$04623.33  $ &  $  1.27$ & $      $ & $      $ & $ -2.29$  &\hspace{0.25cm}&$07114.54  $ &  $  2.60$ & $  0.68$ & $ -0.38$ & $ -1.56$  \\
$07835.38  $ &  $  8.15$ & $  4.82$ & $  2.64$ & $  0.25$  &\hspace{0.25cm}&$10467.28  $ &  $  3.85$ & $  2.22$ & $  1.05$ & $ -0.30$  &\hspace{0.25cm}&$04627.30  $ &  $  1.37$ & $      $ & $      $ & $ -2.23$  &\hspace{0.25cm}&$07144.36  $ &  $  2.96$ & $  1.00$ & $ -0.14$ & $ -1.42$  \\
$08109.61  $ &  $  8.21$ & $  4.80$ & $  2.40$ & $ -0.21$  &\hspace{0.25cm}&$10499.25  $ &  $  3.70$ & $  2.06$ & $  0.92$ & $ -0.38$  &\hspace{0.25cm}&$04634.29  $ &  $  1.31$ & $      $ & $      $ & $ -2.25$  &\hspace{0.25cm}&$07186.32  $ &  $  3.23$ & $  1.17$ & $ -0.05$ & $ -1.40$  \\
$08164.51  $ &  $  6.87$ & $  3.94$ & $  1.87$ & $ -0.53$  &\hspace{0.25cm}&$10673.63  $ &  $  4.72$ & $  2.90$ & $  1.54$ & $  0.12$  &\hspace{0.25cm}&$04911.46  $ &  $  2.81$ & $  0.89$ & $ -0.32$ & $ -1.66$  &\hspace{0.25cm}&$07386.64  $ &  $  1.42$ & $ -0.10$ & $ -1.04$ & $ -2.26$  \\
$08211.34  $ &  $  6.66$ & $  3.74$ & $  1.72$ & $ -0.60$  &\hspace{0.25cm}&$10754.38  $ &  $  4.40$ & $  2.64$ & $  1.34$ & $  0.08$  &\hspace{0.25cm}&$04914.43  $ &  $  2.68$ & $  0.80$ & $ -0.41$ & $ -1.81$  &\hspace{0.25cm}&$07506.38  $ &  $  1.88$ & $  0.14$ & $ -0.83$ & $ -1.93$  \\
$08232.37  $ &  $  6.53$ & $  3.67$ & $  1.67$ & $ -0.61$  &\hspace{0.25cm}&$10796.35  $ &  $  4.18$ & $  2.51$ & $  1.25$ & $ -0.03$  &\hspace{0.25cm}&$04918.45  $ &  $  2.73$ & $  0.85$ & $ -0.34$ & $ -1.73$  &\hspace{0.25cm}&$07530.34  $ &  $  2.00$ & $  0.26$ & $ -0.73$ & $ -1.87$  \\
$08473.66  $ &  $  7.59$ & $  4.45$ & $  2.42$ & $  0.18$  &\hspace{0.25cm}&$10854.29  $ &  $  3.49$ & $  1.94$ & $  0.83$ & $ -0.37$  &\hspace{0.25cm}&$04947.38  $ &  $  2.50$ & $  0.77$ & $ -0.45$ & $ -1.76$  &\hspace{0.25cm}&$07805.58  $ &  $  1.55$ & $ -0.01$ & $ -0.99$ & $ -2.23$  \\
$08495.59  $ &  $  7.84$ & $  4.63$ & $  2.57$ & $  0.30$  &\hspace{0.25cm}&$          $ &  $      $ & $      $ & $      $ & $      $  &\hspace{0.25cm}&$04949.40  $ &  $  2.48$ & $  0.72$ & $ -0.43$ & $ -1.72$  &\hspace{0.25cm}&$07821.50  $ &  $  1.45$ & $ -0.10$ & $ -1.05$ & $ -2.30$  \\
$08520.53  $ &  $  8.09$ & $  4.87$ & $  2.73$ & $  0.41$  &\hspace{0.25cm}&\multicolumn{5}{l}{\bf{IRAS $02351-2711$}}                 &\hspace{0.25cm}&$04954.40  $ &  $  2.32$ & $  0.68$ & $ -0.48$ & $ -1.78$  &\hspace{0.25cm}&$07841.45  $ &  $  1.32$ & $ -0.22$ & $ -1.15$ & $ -2.37$  \\
$08554.43  $ &  $  8.46$ & $  5.06$ & $  2.89$ & $  0.36$  &\hspace{0.25cm}&$09170.68  $ &  $  3.32$ & $  1.99$ & $  1.24$ & $  0.44$  &\hspace{0.25cm}&$04958.34  $ &  $  2.19$ & $  0.55$ & $ -0.57$ & $ -1.84$  &\hspace{0.25cm}&$07868.45  $ &  $  1.21$ & $ -0.36$ & $ -1.27$ & $ -2.46$  \\
\end{tabular}
}
\end{table}
\setcounter{table}{1}
\begin{table}
\vspace{1.5 cm}
\caption{ ...continued. Near-infrared data. The  Date is given as
JD--2440000.00.}
\footnotesize{
\begin{tabular}{lrrrrclrrrrclrrrrclrrrr}\\
\multicolumn{1}{c}{\bf{DATE}}&
\multicolumn{1}{c}{\bf{J}}&
\multicolumn{1}{c}{\bf{H}}&
\multicolumn{1}{c}{\bf{K}}&
\multicolumn{1}{c}{\bf{L}}&
&
\multicolumn{1}{c}{\bf{DATE}}&
\multicolumn{1}{c}{\bf{J}}&
\multicolumn{1}{c}{\bf{H}}&
\multicolumn{1}{c}{\bf{K}}&
\multicolumn{1}{c}{\bf{L}}&
&
\multicolumn{1}{c}{\bf{DATE}}&
\multicolumn{1}{c}{\bf{J}}&
\multicolumn{1}{c}{\bf{H}}&
\multicolumn{1}{c}{\bf{K}}&
\multicolumn{1}{c}{\bf{L}}&
&
\multicolumn{1}{c}{\bf{DATE}}&
\multicolumn{1}{c}{\bf{J}}&
\multicolumn{1}{c}{\bf{H}}&
\multicolumn{1}{c}{\bf{K}}&
\multicolumn{1}{c}{\bf{L}}\\
\\
\multicolumn{5}{l}{\bf{IRAS $03507+1115$  .....continued}} &\hspace{0.25cm}&\multicolumn{5}{l}{\bf{IRAS $06176-1036$}}                 &\hspace{0.25cm}&\multicolumn{5}{l}{\bf{IRAS $09429-2148$}}                 &\hspace{0.25cm}&\multicolumn{5}{l}{\bf{IRAS $09429-2148$  .....continued}} \\
$07905.32  $ &  $  1.21$ & $ -0.42$ & $ -1.33$ & $ -2.46$  &\hspace{0.25cm}&$06867.33  $ &  $  6.61$ & $  4.95$ & $  3.39$ & $  1.30$  &\hspace{0.25cm}&$06356.62  $ &  $  6.16$ & $  4.41$ & $  3.09$ & $  1.46$  &\hspace{0.25cm}&$08024.23  $ &  $  4.79$ & $  3.14$ & $  1.80$ & $  0.09$  \\
$07919.32  $ &  $  1.23$ & $ -0.42$ & $ -1.32$ & $ -2.45$  &\hspace{0.25cm}&$10437.52  $ &  $  6.70$ & $  5.03$ & $  3.44$ & $  1.31$  &\hspace{0.25cm}&$06427.60  $ &  $  6.43$ & $  4.55$ & $  3.16$ & $  1.46$  &\hspace{0.25cm}&$08052.20  $ &  $  4.91$ & $  3.28$ & $  1.93$ & $  0.18$  \\
$08129.62  $ &  $  3.42$ & $  1.24$ & $ -0.06$ & $ -1.43$  &\hspace{0.25cm}&$          $ &  $      $ & $      $ & $      $ & $      $  &\hspace{0.25cm}&$06458.51  $ &  $  6.46$ & $  4.59$ & $  3.18$ & $  1.44$  &\hspace{0.25cm}&$08213.57  $ &  $  5.80$ & $  4.04$ & $  2.66$ & $  1.01$  \\
$08165.58  $ &  $  3.25$ & $  1.11$ & $ -0.17$ & $ -1.51$  &\hspace{0.25cm}&\multicolumn{5}{l}{\bf{IRAS $08088-3243$}}                 &\hspace{0.25cm}&$06465.51  $ &  $  6.47$ & $  4.58$ & $  3.16$ & $  1.43$  &\hspace{0.25cm}&$08256.58  $ &  $  6.07$ & $  4.28$ & $  2.85$ & $  1.17$  \\
$08228.41  $ &  $  3.15$ & $  1.10$ & $ -0.16$ & $ -1.59$  &\hspace{0.25cm}&$09117.32  $ &  $  9.67$ & $  6.64$ & $  4.24$ & $  1.48$  &\hspace{0.25cm}&$06487.43  $ &  $  6.54$ & $  4.64$ & $  3.22$ & $  1.48$  &\hspace{0.25cm}&$08281.50  $ &  $  6.22$ & $  4.38$ & $  2.94$ & $  1.26$  \\
$08252.41  $ &  $  2.52$ & $  0.70$ & $ -0.47$ & $ -1.84$  &\hspace{0.25cm}&$09734.55  $ &  $  8.95$ & $  6.35$ & $  4.20$ & $  1.53$  &\hspace{0.25cm}&$06502.33  $ &  $  6.39$ & $  4.56$ & $  3.15$ & $  1.38$  &\hspace{0.25cm}&$08299.45  $ &  $  6.26$ & $  4.44$ & $  2.98$ & $  1.27$  \\
$08280.33  $ &  $  1.63$ & $  0.09$ & $ -0.87$ & $ -2.14$  &\hspace{0.25cm}&$09832.29  $ &  $  8.77$ & $  6.19$ & $  4.09$ & $  1.56$  &\hspace{0.25cm}&$06506.37  $ &  $  6.35$ & $  4.53$ & $  3.12$ & $  1.34$  &\hspace{0.25cm}&$08327.38  $ &  $  6.39$ & $  4.52$ & $  3.05$ & $  1.31$  \\
$08519.60  $ &  $  2.76$ & $  0.78$ & $ -0.33$ & $ -1.48$  &\hspace{0.25cm}&$10034.53  $ &  $  7.43$ & $  5.01$ & $  3.02$ & $  0.58$  &\hspace{0.25cm}&$06541.36  $ &  $  5.34$ & $  3.80$ & $  2.52$ & $  0.74$  &\hspace{0.25cm}&$08377.26  $ &  $  6.44$ & $  4.57$ & $  3.06$ & $  1.31$  \\
$08630.34  $ &  $  4.09$ & $  1.80$ & $  0.39$ & $ -1.12$  &\hspace{0.25cm}&$10090.52  $ &  $  7.40$ & $  4.98$ & $  3.01$ & $  0.57$  &\hspace{0.25cm}&$06568.21  $ &  $  5.05$ & $  3.59$ & $  2.33$ & $  0.57$  &\hspace{0.25cm}&$08616.60  $ &  $  4.79$ & $  3.10$ & $  1.75$ & $  0.08$  \\
$08873.64  $ &  $  1.87$ & $  0.16$ & $ -0.87$ & $ -2.13$  &\hspace{0.25cm}&$10112.44  $ &  $  7.46$ & $  5.04$ & $  3.05$ & $  0.64$  &\hspace{0.25cm}&$06747.63  $ &  $  4.78$ & $  3.19$ & $  1.97$ & $  0.34$  &\hspace{0.25cm}&$08617.59  $ &  $  4.78$ & $  3.09$ & $  1.75$ & $  0.11$  \\
$08933.48  $ &  $  2.37$ & $  0.54$ & $ -0.53$ & $ -1.79$  &\hspace{0.25cm}&$10478.46  $ &  $  8.62$ & $  6.10$ & $  4.05$ & $  1.40$  &\hspace{0.25cm}&$06778.52  $ &  $  4.85$ & $  3.27$ & $  2.06$ & $  0.41$  &\hspace{0.25cm}&$08639.52  $ &  $  4.75$ & $  3.05$ & $  1.72$ & $  0.08$  \\
$08961.40  $ &  $  2.75$ & $  0.84$ & $ -0.29$ & $ -1.52$  &\hspace{0.25cm}&$10500.38  $ &  $  8.42$ & $  5.92$ & $  3.89$ & $  1.30$  &\hspace{0.25cm}&$06808.54  $ &  $  4.95$ & $  3.37$ & $  2.15$ & $  0.49$  &\hspace{0.25cm}&$08665.47  $ &  $  4.81$ & $  3.08$ & $  1.76$ & $  0.17$  \\
$08990.38  $ &  $  3.19$ & $  1.21$ & $ -0.01$ & $ -1.34$  &\hspace{0.25cm}&$10591.24  $ &  $  7.45$ & $  5.07$ & $  3.12$ & $  0.66$  &\hspace{0.25cm}&$06826.47  $ &  $  5.06$ & $  3.47$ & $  2.24$ & $  0.58$  &\hspace{0.25cm}&$08701.38  $ &  $  4.89$ & $  3.19$ & $  1.87$ & $  0.27$  \\
$09212.68  $ &  $  2.59$ & $  0.82$ & $ -0.41$ & $ -1.92$  &\hspace{0.25cm}&$10796.45  $ &  $  8.85$ & $  6.18$ & $  4.05$ & $  1.40$  &\hspace{0.25cm}&$06834.43  $ &  $  5.10$ & $  3.50$ & $  2.28$ & $  0.62$  &\hspace{0.25cm}&$08721.29  $ &  $  5.01$ & $  3.30$ & $  1.96$ & $  0.34$  \\
$09236.62  $ &  $  2.25$ & $  0.56$ & $ -0.61$ & $ -2.03$  &\hspace{0.25cm}&$10916.39  $ &  $  9.73$ & $  6.97$ & $  4.71$ & $  1.91$  &\hspace{0.25cm}&$06847.47  $ &  $  5.22$ & $  3.60$ & $  2.36$ & $  0.75$  &\hspace{0.25cm}&$08761.21  $ &  $  5.20$ & $  3.46$ & $  2.14$ & $  0.56$  \\
$09581.67  $ &  $  3.88$ & $  1.64$ & $  0.21$ & $ -1.31$  &\hspace{0.25cm}&$10978.21  $ &  $  9.20$ & $  6.51$ & $  4.31$ & $  1.62$  &\hspace{0.25cm}&$06873.39  $ &  $  5.38$ & $  3.76$ & $  2.52$ & $  0.88$  &\hspace{0.25cm}&$08992.53  $ &  $  6.65$ & $  4.54$ & $  2.92$ & $  1.17$  \\
$09614.60  $ &  $  3.82$ & $  1.64$ & $  0.21$ & $ -1.33$  &\hspace{0.25cm}&$          $ &  $      $ & $      $ & $      $ & $      $  &\hspace{0.25cm}&$06894.33  $ &  $  5.52$ & $  3.91$ & $  2.64$ & $  0.98$  &\hspace{0.25cm}&$09106.27  $ &  $  5.68$ & $  3.85$ & $  2.36$ & $  0.60$  \\
$09637.56  $ &  $  3.89$ & $  1.63$ & $  0.21$ & $ -1.34$  &\hspace{0.25cm}&\multicolumn{5}{l}{\bf{IRAS $09116-2439$}}                 &\hspace{0.25cm}&$06897.36  $ &  $  5.49$ & $  3.90$ & $  2.63$ & $  0.99$  &\hspace{0.25cm}&$09382.48  $ &  $  5.17$ & $  3.35$ & $  2.01$ & $  0.43$  \\
$09672.42  $ &  $  2.80$ & $  1.03$ & $ -0.24$ & $ -1.76$  &\hspace{0.25cm}&$07175.50  $ &  $      $ & $ 10.93$ & $  7.36$ & $  3.03$  &\hspace{0.25cm}&$06897.37  $ &  $  5.57$ & $  3.93$ & $  2.68$ & $  1.00$  &\hspace{0.25cm}&$09405.46  $ &  $  5.31$ & $  3.46$ & $  2.10$ & $  0.55$  \\
$09709.34  $ &  $  2.41$ & $  0.66$ & $ -0.55$ & $ -2.03$  &\hspace{0.25cm}&$09117.34  $ &  $ 13.93$ & $ 10.31$ & $  6.77$ & $  2.68$  &\hspace{0.25cm}&$06934.28  $ &  $  5.82$ & $  4.14$ & $  2.83$ & $  1.16$  &\hspace{0.25cm}&$09467.39  $ &  $  5.78$ & $  3.83$ & $  2.41$ & $  0.90$  \\
$09728.33  $ &  $  2.34$ & $  0.56$ & $ -0.63$ & $ -2.10$  &\hspace{0.25cm}&$09734.59  $ &  $      $ & $  9.75$ & $  6.39$ & $  2.40$  &\hspace{0.25cm}&$06978.23  $ &  $  6.17$ & $  4.37$ & $  3.02$ & $  1.36$  &\hspace{0.25cm}&$09501.25  $ &  $  6.01$ & $  4.01$ & $  2.56$ & $  0.98$  \\
$09986.61  $ &  $  3.79$ & $  1.49$ & $  0.12$ & $ -1.20$  &\hspace{0.25cm}&$09829.27  $ &  $      $ & $ 10.21$ & $  6.89$ & $  2.91$  &\hspace{0.25cm}&$07121.55  $ &  $  6.37$ & $  4.50$ & $  3.07$ & $  1.29$  &\hspace{0.25cm}&$09677.55  $ &  $  6.73$ & $  4.45$ & $  2.79$ & $  1.04$  \\
$10019.52  $ &  $  3.89$ & $  1.56$ & $  0.14$ & $ -1.27$  &\hspace{0.25cm}&$09887.23  $ &  $      $ & $ 10.39$ & $  7.07$ & $  3.11$  &\hspace{0.25cm}&$07171.52  $ &  $  5.50$ & $  3.90$ & $  2.55$ & $  0.71$  &\hspace{0.25cm}&$09824.34  $ &  $  4.81$ & $  3.00$ & $  1.65$ & $  0.08$  \\
$10052.47  $ &  $  3.68$ & $  1.41$ & $  0.01$ & $ -1.39$  &\hspace{0.25cm}&$10034.56  $ &  $      $ & $  9.15$ & $  5.83$ & $  1.91$  &\hspace{0.25cm}&$07179.50  $ &  $  5.37$ & $  3.78$ & $  2.44$ & $  0.62$  &\hspace{0.25cm}&$09887.27  $ &  $  4.67$ & $  2.82$ & $  1.51$ & $ -0.02$  \\
$10086.35  $ &  $  3.52$ & $  1.35$ & $ -0.04$ & $ -1.51$  &\hspace{0.25cm}&$10086.56  $ &  $      $ & $  8.64$ & $  5.34$ & $  1.51$  &\hspace{0.25cm}&$07192.49  $ &  $  5.21$ & $  3.66$ & $  2.33$ & $  0.51$  &\hspace{0.25cm}&$10034.58  $ &  $  5.06$ & $  3.17$ & $  1.87$ & $  0.43$  \\
$10109.29  $ &  $  3.48$ & $  1.33$ & $ -0.05$ & $ -1.54$  &\hspace{0.25cm}&$10111.46  $ &  $      $ & $  8.47$ & $  5.18$ & $  1.36$  &\hspace{0.25cm}&$07213.49  $ &  $  5.03$ & $  3.51$ & $  2.21$ & $  0.41$  &\hspace{0.25cm}&$10088.59  $ &  $  5.41$ & $  3.47$ & $  2.13$ & $  0.71$  \\
$10125.27  $ &  $  3.09$ & $  1.12$ & $ -0.21$ & $ -1.69$  &\hspace{0.25cm}&$10123.50  $ &  $      $ & $  8.43$ & $  5.15$ & $  1.33$  &\hspace{0.25cm}&$07236.38  $ &  $  4.91$ & $  3.38$ & $  2.08$ & $  0.29$  &\hspace{0.25cm}&$10463.56  $ &  $  4.61$ & $  2.83$ & $  1.57$ & $  0.06$  \\
$10362.54  $ &  $  2.51$ & $  0.50$ & $ -0.61$ & $ -1.71$  &\hspace{0.25cm}&$10222.31  $ &  $      $ & $  8.58$ & $  5.27$ & $  1.42$  &\hspace{0.25cm}&$07268.32  $ &  $  4.80$ & $  3.25$ & $  1.95$ & $  0.20$  &\hspace{0.25cm}&$10500.44  $ &  $  4.56$ & $  2.75$ & $  1.50$ & $  0.00$  \\
$10417.42  $ &  $  3.12$ & $  0.98$ & $ -0.24$ & $ -1.38$  &\hspace{0.25cm}&$10256.20  $ &  $      $ & $  8.74$ & $  5.46$ & $  1.65$  &\hspace{0.25cm}&$07533.54  $ &  $  5.41$ & $  3.76$ & $  2.46$ & $  0.84$  &\hspace{0.25cm}&$10800.47  $ &  $  6.17$ & $  4.06$ & $  2.60$ & $  1.05$  \\
$10437.38  $ &  $  3.33$ & $  1.16$ & $ -0.10$ & $ -1.33$  &\hspace{0.25cm}&$10438.58  $ &  $      $ & $  9.96$ & $  6.68$ & $  2.65$  &\hspace{0.25cm}&$07542.41  $ &  $  5.51$ & $  3.82$ & $  2.51$ & $  0.93$  &\hspace{0.25cm}&$          $ &  $      $ & $      $ & $      $ & $      $  \\
$10469.32  $ &  $  3.52$ & $  1.32$ & $  0   $ & $ -1.31$  &\hspace{0.25cm}&$10483.45  $ &  $      $ & $ 10.26$ & $  6.88$ & $  2.80$  &\hspace{0.25cm}&$07581.42  $ &  $  5.78$ & $  4.04$ & $  2.70$ & $  1.07$  &\hspace{0.25cm}&\multicolumn{5}{l}{\bf{IRAS $09452+1330$}}                 \\
$10503.00  $ &  $  3.40$ & $  1.23$ & $ -0.11$ & $ -1.44$  &\hspace{0.25cm}&$10590.25  $ &  $      $ & $ 10.16$ & $  6.86$ & $  2.78$  &\hspace{0.25cm}&$07617.33  $ &  $  6.03$ & $  4.24$ & $  2.84$ & $  1.20$  &\hspace{0.25cm}&$07215.45  $ &  $  8.38$ & $  5.01$ & $  2.10$ & $ -1.71$  \\
$10721.61  $ &  $  1.80$ & $  0.00$ & $ -1.04$ & $ -2.17$  &\hspace{0.25cm}&$10795.51  $ &  $      $ & $  8.62$ & $  5.31$ & $  1.46$  &\hspace{0.25cm}&$07690.20  $ &  $  6.56$ & $  4.57$ & $  3.05$ & $      $  &\hspace{0.25cm}&$07242.36  $ &  $  8.24$ & $  4.85$ & $  1.95$ & $ -1.72$  \\
$10753.45  $ &  $  1.90$ & $  0.04$ & $ -1.00$ & $ -2.12$  &\hspace{0.25cm}&$10913.38  $ &  $ 12.46$ & $  8.65$ & $  4.71$ & $  1.92$  &\hspace{0.25cm}&$07890.54  $ &  $  5.06$ & $  3.46$ & $  2.10$ & $  0.28$  &\hspace{0.25cm}&$07535.56  $ &  $  6.39$ & $  3.05$ & $  0.28$ & $ -3.23$  \\
$10805.38  $ &  $  2.16$ & $  0.26$ & $ -0.82$ & $ -1.91$  &\hspace{0.25cm}&$10978.23  $ &  $      $ & $  9.03$ & $  5.77$ & $  1.97$  &\hspace{0.25cm}&$07895.59  $ &  $  4.98$ & $  3.41$ & $  2.05$ & $  0.20$  &\hspace{0.25cm}&$07583.42  $ &  $  6.74$ & $  3.40$ & $  0.61$ & $ -3.01$  \\
$          $ &  $      $ & $      $ & $      $ & $      $  &\hspace{0.25cm}&$          $ &  $      $ & $      $ & $      $ & $      $  &\hspace{0.25cm}&$07904.56  $ &  $  4.95$ & $  3.36$ & $  2.01$ & $  0.20$  &\hspace{0.25cm}&$07617.35  $ &  $  7.05$ & $  3.69$ & $  0.90$ & $ -2.67$  \\
$          $ &  $      $ & $      $ & $      $ & $      $  &\hspace{0.25cm}&$          $ &  $      $ & $      $ & $      $ & $      $  &\hspace{0.25cm}&$07921.48  $ &  $  4.85$ & $  3.27$ & $  1.92$ & $  0.11$  &\hspace{0.25cm}&$07669.24  $ &  $  7.45$ & $  4.12$ & $  1.33$ & $ -2.29$  \\
\end{tabular}
}
\end{table}
\setcounter{table}{1}
\begin{table}
\vspace{1.5 cm}
\caption{ ...continued. Near-infrared data. The  Date is given as
JD--2440000.00.}
\footnotesize{
\begin{tabular}{lrrrrclrrrrclrrrrclrrrr}\\
\multicolumn{1}{c}{\bf{DATE}}&
\multicolumn{1}{c}{\bf{J}}&
\multicolumn{1}{c}{\bf{H}}&
\multicolumn{1}{c}{\bf{K}}&
\multicolumn{1}{c}{\bf{L}}&
&
\multicolumn{1}{c}{\bf{DATE}}&
\multicolumn{1}{c}{\bf{J}}&
\multicolumn{1}{c}{\bf{H}}&
\multicolumn{1}{c}{\bf{K}}&
\multicolumn{1}{c}{\bf{L}}&
&
\multicolumn{1}{c}{\bf{DATE}}&
\multicolumn{1}{c}{\bf{J}}&
\multicolumn{1}{c}{\bf{H}}&
\multicolumn{1}{c}{\bf{K}}&
\multicolumn{1}{c}{\bf{L}}&
&
\multicolumn{1}{c}{\bf{DATE}}&
\multicolumn{1}{c}{\bf{J}}&
\multicolumn{1}{c}{\bf{H}}&
\multicolumn{1}{c}{\bf{K}}&
\multicolumn{1}{c}{\bf{L}}\\
\\
\multicolumn{5}{l}{\bf{IRAS $09452+1330$  .....continued}} &\hspace{0.25cm}&\multicolumn{5}{l}{\bf{IRAS $10491-2059$  .....continued}} &\hspace{0.25cm}&\multicolumn{5}{l}{\bf{IRAS $10491-2059$  .....continued}} &\hspace{0.25cm}&\multicolumn{5}{l}{\bf{IRAS $17049-2440$}}                 \\
$07903.55  $ &  $  7.70$ & $  4.48$ & $  1.70$ & $ -1.94$  &\hspace{0.25cm}&$06486.41  $ &  $  1.86$ & $  0.32$ & $ -0.72$ & $ -1.88$  &\hspace{0.25cm}&$10802.56  $ &  $      $ & $  0.74$ & $ -0.35$ & $      $  &\hspace{0.25cm}&$09824.50  $ &  $      $ & $  7.66$ & $  4.73$ & $  1.24$  \\
$08254.57  $ &  $  6.83$ & $  3.52$ & $  0.73$ & $ -2.87$  &\hspace{0.25cm}&$06502.42  $ &  $  1.98$ & $  0.42$ & $ -0.64$ & $ -1.84$  &\hspace{0.25cm}&$10802.56  $ &  $  2.41$ & $  0.73$ & $      $ & $ -1.67$  &\hspace{0.25cm}&$09886.47  $ &  $      $ & $  8.15$ & $  5.15$ & $  1.56$  \\
$08296.49  $ &  $  7.22$ & $  3.96$ & $  1.15$ & $ -2.51$  &\hspace{0.25cm}&$06506.39  $ &  $  2.01$ & $  0.43$ & $ -0.64$ & $ -1.83$  &\hspace{0.25cm}&$10809.58  $ &  $  2.41$ & $  0.73$ & $ -0.36$ & $ -1.66$  &\hspace{0.25cm}&$09911.47  $ &  $      $ & $  8.35$ & $  5.34$ & $  1.76$  \\
$08388.27  $ &  $  7.86$ & $  4.61$ & $  1.81$ & $ -1.91$  &\hspace{0.25cm}&$06540.42  $ &  $  2.20$ & $  0.60$ & $ -0.53$ & $ -1.76$  &\hspace{0.25cm}&$          $ &  $      $ & $      $ & $      $ & $      $  &\hspace{0.25cm}&$09945.31  $ &  $      $ & $  8.65$ & $  5.61$ & $  1.89$  \\
$08634.54  $ &  $  6.58$ & $  3.26$ & $  0.52$ & $ -3.02$  &\hspace{0.25cm}&$06778.55  $ &  $  1.94$ & $  0.44$ & $ -0.59$ & $ -1.75$  &\hspace{0.25cm}&\multicolumn{5}{l}{\bf{IRAS $12447+0425$}}                 &\hspace{0.25cm}&$10180.63  $ &  $      $ & $  9.56$ & $  6.45$ & $  2.65$  \\
$08706.37  $ &  $  6.26$ & $  2.95$ & $  0.20$ & $ -3.31$  &\hspace{0.25cm}&$06808.57  $ &  $  1.80$ & $  0.32$ & $ -0.66$ & $ -1.85$  &\hspace{0.25cm}&$06597.24  $ &  $  4.63$ & $  2.94$ & $  1.71$ & $  0.23$  &\hspace{0.25cm}&$10208.52  $ &  $      $ & $  9.60$ & $  6.38$ & $  2.62$  \\
$08734.32  $ &  $  6.18$ & $  2.85$ & $  0.09$ & $ -3.42$  &\hspace{0.25cm}&$06828.53  $ &  $  1.69$ & $  0.24$ & $ -0.74$ & $ -1.86$  &\hspace{0.25cm}&$06805.58  $ &  $  5.08$ & $  3.42$ & $  2.07$ & $  0.46$  &\hspace{0.25cm}&$10233.46  $ &  $      $ & $  9.46$ & $  6.25$ & $  2.51$  \\
$08992.56  $ &  $  7.77$ & $  4.43$ & $  1.54$ & $ -2.24$  &\hspace{0.25cm}&$06834.53  $ &  $  1.65$ & $  0.20$ & $ -0.76$ & $ -1.92$  &\hspace{0.25cm}&$06828.62  $ &  $  4.90$ & $  3.28$ & $  1.97$ & $  0.41$  &\hspace{0.25cm}&$10255.42  $ &  $      $ & $  9.18$ & $  6.08$ & $  2.39$  \\
$09000.47  $ &  $  7.85$ & $  4.49$ & $  1.60$ & $ -2.17$  &\hspace{0.25cm}&$06846.50  $ &  $  1.58$ & $  0.14$ & $ -0.80$ & $ -1.94$  &\hspace{0.25cm}&$06844.56  $ &  $  4.63$ & $  3.04$ & $  1.77$ & $  0.20$  &\hspace{0.25cm}&$10295.39  $ &  $      $ & $  8.63$ & $  5.53$ & $  1.91$  \\
$09146.20  $ &  $  8.30$ & $  5.03$ & $  2.13$ & $ -1.68$  &\hspace{0.25cm}&$06877.45  $ &  $  1.45$ & $  0.05$ & $ -0.88$ & $ -2.03$  &\hspace{0.25cm}&$06873.48  $ &  $  4.30$ & $  2.74$ & $  1.52$ & $ -0.01$  &\hspace{0.25cm}&$10317.29  $ &  $      $ & $  8.28$ & $  5.28$ & $  1.76$  \\
$09501.19  $ &  $  6.60$ & $  3.29$ & $  0.43$ & $ -3.21$  &\hspace{0.25cm}&$06898.38  $ &  $  1.39$ & $ -0.02$ & $ -0.92$ & $ -2.10$  &\hspace{0.25cm}&$06894.45  $ &  $  4.22$ & $  2.64$ & $  1.45$ & $ -0.10$  &\hspace{0.25cm}&$10362.23  $ &  $      $ & $  7.94$ & $  4.95$ & $  1.44$  \\
$09674.61  $ &  $  7.88$ & $  4.67$ & $  1.78$ & $ -2.11$  &\hspace{0.25cm}&$06936.27  $ &  $  1.39$ & $ -0.06$ & $ -0.96$ & $ -2.05$  &\hspace{0.25cm}&$06899.39  $ &  $  4.12$ & $  2.62$ & $  1.43$ & $ -0.08$  &\hspace{0.25cm}&$10531.60  $ &  $      $ & $  7.53$ & $  4.57$ & $  1.13$  \\
$09820.30  $ &  $  8.22$ & $  5.07$ & $  2.19$ & $ -1.68$  &\hspace{0.25cm}&$06967.28  $ &  $  1.51$ & $ -0.01$ & $ -0.94$ & $ -2.01$  &\hspace{0.25cm}&$06943.30  $ &  $  4.15$ & $  2.52$ & $  1.37$ & $ -0.09$  &\hspace{0.25cm}&$10692.28  $ &  $      $ & $  8.32$ & $  5.34$ & $  1.80$  \\
$09888.20  $ &  $  7.49$ & $  4.31$ & $  1.46$ & $ -2.34$  &\hspace{0.25cm}&$07121.57  $ &  $  2.21$ & $  0.63$ & $ -0.48$ & $ -1.65$  &\hspace{0.25cm}&$06967.35  $ &  $  4.22$ & $  2.58$ & $  1.41$ & $ -0.08$  &\hspace{0.25cm}&$10912.57  $ &  $ 12.56$ & $  9.32$ & $  6.36$ & $  2.73$  \\
$10111.55  $ &  $  6.32$ & $  3.11$ & $  0.28$ & $ -3.36$  &\hspace{0.25cm}&$07149.55  $ &  $  2.12$ & $  0.53$ & $ -0.47$ & $ -1.75$  &\hspace{0.25cm}&$07172.60  $ &  $  5.27$ & $  3.54$ & $  2.22$ & $  0.66$  &\hspace{0.25cm}&$10980.47  $ &  $      $ & $  9.29$ & $  6.37$ & $  2.71$  \\
$10126.55  $ &  $  6.34$ & $  3.12$ & $  0.29$ & $ -3.32$  &\hspace{0.25cm}&$07176.55  $ &  $  1.88$ & $  0.38$ & $ -0.62$ & $ -1.72$  &\hspace{0.25cm}&$07191.60  $ &  $  5.00$ & $  3.34$ & $  2.07$ & $  0.47$  &\hspace{0.25cm}&$          $ &  $      $ & $      $ & $      $ & $      $  \\
$10468.50  $ &  $  8.20$ & $  5.13$ & $  2.25$ & $ -1.73$  &\hspace{0.25cm}&$07191.54  $ &  $  1.83$ & $  0.34$ & $ -0.65$ & $ -1.78$  &\hspace{0.25cm}&$07236.45  $ &  $  4.95$ & $  3.26$ & $  2.01$ & $  0.43$  &\hspace{0.25cm}&\multicolumn{5}{l}{\bf{IRAS $17119+0859$}}                 \\
$10504.41  $ &  $  8.07$ & $  4.97$ & $  2.09$ & $ -1.85$  &\hspace{0.25cm}&$07219.49  $ &  $  1.80$ & $  0.30$ & $ -0.65$ & $ -1.78$  &\hspace{0.25cm}&$07263.41  $ &  $  4.90$ & $  3.22$ & $  2.00$ & $  0.43$  &\hspace{0.25cm}&$09912.42  $ &  $  4.51$ & $  3.16$ & $  2.25$ & $  0.81$  \\
$10805.60  $ &  $  6.76$ & $  3.47$ & $  0.57$ & $ -3.18$  &\hspace{0.25cm}&$07240.41  $ &  $  1.72$ & $  0.26$ & $ -0.67$ & $ -1.74$  &\hspace{0.25cm}&$07368.19  $ &  $  3.99$ & $  2.35$ & $  1.27$ & $ -0.22$  &\hspace{0.25cm}&$10214.54  $ &  $  5.38$ & $  3.89$ & $  2.65$ & $  0.95$  \\
$          $ &  $      $ & $      $ & $      $ & $      $  &\hspace{0.25cm}&$07288.38  $ &  $  1.73$ & $  0.27$ & $ -0.63$ & $ -1.74$  &\hspace{0.25cm}&$07539.56  $ &  $  5.60$ & $  3.74$ & $  2.29$ & $  0.71$  &\hspace{0.25cm}&$10237.45  $ &  $  4.79$ & $  3.44$ & $  2.27$ & $  0.62$  \\
\multicolumn{5}{l}{\bf{IRAS $10131+3049$}}                 &\hspace{0.25cm}&$07507.56  $ &  $  1.62$ & $  0.14$ & $ -0.81$ & $ -1.91$  &\hspace{0.25cm}&$07605.56  $ &  $  5.56$ & $  3.75$ & $  2.30$ & $  0.68$  &\hspace{0.25cm}&$          $ &  $      $ & $      $ & $      $ & $      $  \\
$10111.54  $ &  $  6.73$ & $  3.98$ & $  1.74$ & $ -0.99$  &\hspace{0.25cm}&$07534.59  $ &  $  1.78$ & $  0.26$ & $ -0.73$ & $ -1.88$  &\hspace{0.25cm}&$07629.43  $ &  $  5.45$ & $  3.63$ & $  2.20$ & $  0.59$  &\hspace{0.25cm}&\multicolumn{5}{l}{\bf{IRAS $17297+1747$}}                 \\
$10468.53  $ &  $  5.48$ & $  2.85$ & $  0.76$ & $ -1.78$  &\hspace{0.25cm}&$07555.53  $ &  $  1.92$ & $  0.39$ & $ -0.65$ & $ -1.82$  &\hspace{0.25cm}&$07669.27  $ &  $  5.28$ & $  3.48$ & $  2.10$ & $  0.56$  &\hspace{0.25cm}&$09912.41  $ &  $  9.87$ & $  6.84$ & $  4.50$ & $  1.45$  \\
$10504.43  $ &  $  5.35$ & $  2.71$ & $  0.65$ & $ -1.84$  &\hspace{0.25cm}&$07618.38  $ &  $  2.22$ & $  0.63$ & $ -0.47$ & $ -1.63$  &\hspace{0.25cm}&$07690.29  $ &  $  5.31$ & $  3.51$ & $  2.11$ & $  0.53$  &\hspace{0.25cm}&$10214.55  $ &  $  8.30$ & $  5.27$ & $  2.92$ & $  0.08$  \\
$10980.20  $ &  $  6.90$ & $  4.13$ & $  1.89$ & $ -0.84$  &\hspace{0.25cm}&$07685.29  $ &  $  2.19$ & $  0.59$ & $ -0.49$ & $ -1.62$  &\hspace{0.25cm}&$07903.59  $ &  $  4.78$ & $  2.97$ & $  1.69$ & $  0.17$  &\hspace{0.25cm}&$10237.47  $ &  $  8.26$ & $  5.21$ & $  2.88$ & $  0.08$  \\
$          $ &  $      $ & $      $ & $      $ & $      $  &\hspace{0.25cm}&$07896.56  $ &  $  1.69$ & $  0.26$ & $ -0.68$ & $ -1.85$  &\hspace{0.25cm}&$07989.43  $ &  $  5.62$ & $  3.73$ & $  2.29$ & $  0.73$  &\hspace{0.25cm}&$          $ &  $      $ & $      $ & $      $ & $      $  \\
\multicolumn{5}{l}{\bf{IRAS $10491-2059$}}                 &\hspace{0.25cm}&$08029.29  $ &  $  1.81$ & $  0.31$ & $ -0.68$ & $ -1.84$  &\hspace{0.25cm}&$08299.56  $ &  $  4.41$ & $  2.61$ & $  1.40$ & $ -0.15$  &\hspace{0.25cm}&\multicolumn{5}{l}{\bf{IRAS $17360-3012$}}                 \\
$05008.49  $ &  $  2.48$ & $  0.85$ & $ -0.33$ & $ -1.59$  &\hspace{0.25cm}&$08073.21  $ &  $  2.02$ & $  0.46$ & $ -0.56$ & $ -1.71$  &\hspace{0.25cm}&$08323.53  $ &  $  4.63$ & $  2.78$ & $  1.53$ & $ -0.06$  &\hspace{0.25cm}&$09824.52  $ &  $  9.58$ & $  6.36$ & $  4.36$ & $  2.12$  \\
$05033.48  $ &  $  2.29$ & $  0.72$ & $ -0.37$ & $ -1.59$  &\hspace{0.25cm}&$10483.50  $ &  $  2.48$ & $  0.88$ & $ -0.20$ & $ -1.62$  &\hspace{0.25cm}&$08388.36  $ &  $  5.31$ & $  3.35$ & $  1.99$ & $  0.41$  &\hspace{0.25cm}&$10210.57  $ &  $ 10.50$ & $  7.28$ & $  5.03$ & $  2.36$  \\
$06074.56  $ &  $  2.03$ & $  0.55$ & $ -0.53$ & $ -1.71$  &\hspace{0.25cm}&$10500.48  $ &  $  2.42$ & $  0.83$ & $ -0.23$ & $ -1.63$  &\hspace{0.25cm}&$08457.21  $ &  $  5.76$ & $  3.81$ & $  2.35$ & $  0.74$  &\hspace{0.25cm}&$10238.51  $ &  $ 10.24$ & $  7.05$ & $  4.84$ & $  2.20$  \\
$06111.49  $ &  $  1.90$ & $  0.48$ & $ -0.58$ & $ -1.75$  &\hspace{0.25cm}&$10511.43  $ &  $  2.44$ & $  0.82$ & $ -0.23$ & $ -1.61$  &\hspace{0.25cm}&$08735.45  $ &  $  4.51$ & $  2.66$ & $  1.36$ & $ -0.20$  &\hspace{0.25cm}&$10358.25  $ &  $  8.73$ & $  5.78$ & $  3.76$ & $  1.37$  \\
$06132.45  $ &  $  1.81$ & $  0.37$ & $ -0.64$ & $ -1.78$  &\hspace{0.25cm}&$10520.47  $ &  $  2.37$ & $  0.77$ & $ -0.28$ & $ -1.65$  &\hspace{0.25cm}&$09172.26  $ &  $  4.70$ & $  2.69$ & $  1.34$ & $ -0.20$  &\hspace{0.25cm}&$10590.56  $ &  $  7.87$ & $  5.06$ & $  3.16$ & $  0.93$  \\
$06151.43  $ &  $  1.75$ & $  0.30$ & $ -0.68$ & $ -1.83$  &\hspace{0.25cm}&$10590.30  $ &  $  2.20$ & $  0.65$ & $ -0.38$ & $ -1.89$  &\hspace{0.25cm}&$09888.22  $ &  $  4.99$ & $  3.22$ & $  1.97$ & $  0.44$  &\hspace{0.25cm}&$10618.48  $ &  $  7.83$ & $  5.02$ & $  3.14$ & $  0.95$  \\
$06181.33  $ &  $  1.75$ & $  0.33$ & $      $ & $ -1.81$  &\hspace{0.25cm}&$10615.23  $ &  $      $ & $      $ & $ -0.42$ & $      $  &\hspace{0.25cm}&$10112.50  $ &  $  4.59$ & $  2.74$ & $  1.51$ & $  0.09$  &\hspace{0.25cm}&$10719.26  $ &  $  7.97$ & $  5.12$ & $  3.27$ & $  1.07$  \\
$06223.30  $ &  $  1.78$ & $  0.34$ & $      $ & $ -1.72$  &\hspace{0.25cm}&$10615.23  $ &  $  2.27$ & $  0.65$ & $ -0.40$ & $ -1.80$  &\hspace{0.25cm}&$10504.52  $ &  $  4.15$ & $  2.40$ & $  1.27$ & $ -0.19$  &\hspace{0.25cm}&$10976.49  $ &  $  9.38$ & $  6.41$ & $  4.48$ & $  2.13$  \\
$06270.21  $ &  $  1.74$ & $  0.31$ & $ -0.67$ & $ -1.76$  &\hspace{0.25cm}&$10639.21  $ &  $  2.29$ & $  0.66$ & $ -0.39$ & $ -1.81$  &\hspace{0.25cm}&$          $ &  $      $ & $      $ & $      $ & $      $  &\hspace{0.25cm}&$          $ &  $      $ & $      $ & $      $ & $      $  \\
$06455.55  $ &  $  1.59$ & $  0.09$ & $ -0.87$ & $ -2.09$  &\hspace{0.25cm}&$10639.22  $ &  $  2.27$ & $  0.62$ & $ -0.45$ & $      $  &\hspace{0.25cm}&$          $ &  $      $ & $      $ & $      $ & $      $  &\hspace{0.25cm}&$          $ &  $      $ & $      $ & $      $ & $      $  \\
\end{tabular}
}
\end{table}
\setcounter{table}{1}
\begin{table}
\vspace{1.5 cm}
\caption{ ...continued. Near-infrared data. The  Date is given as
JD--2440000.00.}
\footnotesize{
\begin{tabular}{lrrrrclrrrrclrrrrclrrrr}\\
\multicolumn{1}{c}{\bf{DATE}}&
\multicolumn{1}{c}{\bf{J}}&
\multicolumn{1}{c}{\bf{H}}&
\multicolumn{1}{c}{\bf{K}}&
\multicolumn{1}{c}{\bf{L}}&
&
\multicolumn{1}{c}{\bf{DATE}}&
\multicolumn{1}{c}{\bf{J}}&
\multicolumn{1}{c}{\bf{H}}&
\multicolumn{1}{c}{\bf{K}}&
\multicolumn{1}{c}{\bf{L}}&
&
\multicolumn{1}{c}{\bf{DATE}}&
\multicolumn{1}{c}{\bf{J}}&
\multicolumn{1}{c}{\bf{H}}&
\multicolumn{1}{c}{\bf{K}}&
\multicolumn{1}{c}{\bf{L}}&
&
\multicolumn{1}{c}{\bf{DATE}}&
\multicolumn{1}{c}{\bf{J}}&
\multicolumn{1}{c}{\bf{H}}&
\multicolumn{1}{c}{\bf{K}}&
\multicolumn{1}{c}{\bf{L}}\\
\\
\multicolumn{5}{l}{\bf{IRAS $17411-3154$}}                 &\hspace{0.25cm}&\multicolumn{5}{l}{\bf{IRAS $18194-2708$  .....continued}} &\hspace{0.25cm}&\multicolumn{5}{l}{\bf{IRAS $18333+0533$  .....continued}} &\hspace{0.25cm}&\multicolumn{5}{l}{\bf{IRAS $18398-0220$  .....continued}} \\
$08419.48  $ &  $ 13.17$ & $ 11.01$ & $  9.74$ & $  3.43$  &\hspace{0.25cm}&$10236.51  $ &  $  8.47$ & $  5.33$ & $  3.03$ & $  0.32$  &\hspace{0.25cm}&$06351.26  $ &  $ 12.27$ & $  7.41$ & $  4.63$ & $  1.96$  &\hspace{0.25cm}&$10006.27  $ &  $  6.56$ & $  4.05$ & $  2.21$ & $  0.14$  \\
$08446.51  $ &  $      $ & $      $ & $  9.64$ & $  3.44$  &\hspace{0.25cm}&$10255.45  $ &  $  8.47$ & $  5.34$ & $  3.03$ & $  0.33$  &\hspace{0.25cm}&$06640.37  $ &  $  9.35$ & $  5.63$ & $  3.26$ & $  0.82$  &\hspace{0.25cm}&$10204.65  $ &  $  6.04$ & $  3.66$ & $  1.95$ & $ -0.08$  \\
$08500.42  $ &  $      $ & $      $ & $      $ & $  3.61$  &\hspace{0.25cm}&$10295.45  $ &  $  8.50$ & $  5.44$ & $  3.15$ & $  0.40$  &\hspace{0.25cm}&$06661.37  $ &  $  9.17$ & $  5.48$ & $  3.15$ & $  0.78$  &\hspace{0.25cm}&$10237.53  $ &  $  5.60$ & $  3.28$ & $  1.75$ & $ -0.28$  \\
$08547.29  $ &  $      $ & $      $ & $      $ & $  3.64$  &\hspace{0.25cm}&$10316.39  $ &  $  8.75$ & $  5.57$ & $  3.22$ & $  0.54$  &\hspace{0.25cm}&$06697.29  $ &  $  9.00$ & $  5.32$ & $  3.05$ & $  0.70$  &\hspace{0.25cm}&$10259.49  $ &  $  5.45$ & $  3.14$ & $  1.52$ & $ -0.41$  \\
$08703.60  $ &  $      $ & $      $ & $      $ & $  3.90$  &\hspace{0.25cm}&$10363.31  $ &  $  8.99$ & $  5.85$ & $  3.49$ & $      $  &\hspace{0.25cm}&$06698.26  $ &  $  9.00$ & $  5.32$ & $  3.04$ & $  0.68$  &\hspace{0.25cm}&$10296.39  $ &  $  5.30$ & $  3.01$ & $  1.40$ & $ -0.50$  \\
$08768.37  $ &  $      $ & $      $ & $      $ & $  3.99$  &\hspace{0.25cm}&$10531.64  $ &  $  9.87$ & $  6.74$ & $  4.28$ & $  1.51$  &\hspace{0.25cm}&$06723.26  $ &  $  8.90$ & $  5.20$ & $  2.94$ & $  0.68$  &\hspace{0.25cm}&$10320.32  $ &  $  5.26$ & $  2.97$ & $  1.36$ & $ -0.53$  \\
$09059.61  $ &  $      $ & $      $ & $      $ & $  4.47$  &\hspace{0.25cm}&$10620.53  $ &  $  9.37$ & $  6.26$ & $  3.87$ & $  1.14$  &\hspace{0.25cm}&$06936.52  $ &  $  9.52$ & $  5.73$ & $  3.50$ & $  1.25$  &\hspace{0.25cm}&$10532.63  $ &  $  6.20$ & $  3.79$ & $  2.04$ & $  0.06$  \\
$09227.39  $ &  $      $ & $      $ & $      $ & $  4.29$  &\hspace{0.25cm}&$10692.34  $ &  $  9.44$ & $  6.23$ & $  3.85$ & $  1.11$  &\hspace{0.25cm}&$07011.41  $ &  $ 10.31$ & $  6.28$ & $  3.87$ & $  1.52$  &\hspace{0.25cm}&$          $ &  $      $ & $      $ & $      $ & $      $  \\
$09491.49  $ &  $      $ & $      $ & $      $ & $  3.41$  &\hspace{0.25cm}&$10716.30  $ &  $  9.11$ & $  5.99$ & $  3.61$ & $  0.84$  &\hspace{0.25cm}&$07072.29  $ &  $ 11.10$ & $  6.79$ & $  4.18$ & $  1.73$  &\hspace{0.25cm}&\multicolumn{5}{l}{\bf{IRAS $18397+1738$}}                 \\
$          $ &  $      $ & $      $ & $      $ & $      $  &\hspace{0.25cm}&$10974.47  $ &  $  8.31$ & $  5.19$ & $  2.99$ & $  0.37$  &\hspace{0.25cm}&$07367.37  $ &  $ 10.15$ & $  6.18$ & $  3.58$ & $  1.03$  &\hspace{0.25cm}&$09824.63  $ &  $  5.34$ & $  3.11$ & $  1.41$ & $ -0.62$  \\
\multicolumn{5}{l}{\bf{IRAS $18009-2019$}}                 &\hspace{0.25cm}&$          $ &  $      $ & $      $ & $      $ & $      $  &\hspace{0.25cm}&$07613.67  $ &  $  8.68$ & $  5.02$ & $  2.83$ & $  0.62$  &\hspace{0.25cm}&$10204.64  $ &  $  5.03$ & $  2.93$ & $  1.25$ & $ -0.81$  \\
$06935.60  $ &  $  3.86$ & $  2.13$ & $  1.12$ & $  0.11$  &\hspace{0.25cm}&\multicolumn{5}{l}{\bf{IRAS $18204-1344$}}                 &\hspace{0.25cm}&$07686.59  $ &  $  9.14$ & $  5.33$ & $  3.12$ & $  0.98$  &\hspace{0.25cm}&$10236.58  $ &  $  4.97$ & $  2.85$ & $  1.20$ & $ -0.83$  \\
$          $ &  $      $ & $      $ & $      $ & $      $  &\hspace{0.25cm}&$09824.58  $ &  $  2.87$ & $  1.31$ & $  0.61$ & $ -0.18$  &\hspace{0.25cm}&$09820.37  $ &  $  9.07$ & $  5.26$ & $  2.98$ & $  0.71$  &\hspace{0.25cm}&$10320.31  $ &  $  5.01$ & $  2.89$ & $  1.24$ & $ -0.77$  \\
\multicolumn{5}{l}{\bf{IRAS $18040-0941$}}                 &\hspace{0.25cm}&$09948.35  $ &  $  2.90$ & $  1.33$ & $  0.62$ & $ -0.17$  &\hspace{0.25cm}&$10204.42  $ &  $ 10.07$ & $  6.01$ & $  3.68$ & $  1.48$  &\hspace{0.25cm}&$10532.64  $ &  $  6.54$ & $  4.39$ & $  2.55$ & $  0.34$  \\
$09824.55  $ &  $  5.83$ & $  3.45$ & $  1.79$ & $ -0.05$  &\hspace{0.25cm}&$10204.60  $ &  $  2.94$ & $  1.38$ & $  0.67$ & $ -0.14$  &\hspace{0.25cm}&$10237.51  $ &  $ 10.34$ & $  6.22$ & $  3.82$ & $  1.59$  &\hspace{0.25cm}&$          $ &  $      $ & $      $ & $      $ & $      $  \\
$09888.45  $ &  $  5.87$ & $  3.48$ & $  1.84$ & $      $  &\hspace{0.25cm}&$10236.53  $ &  $  2.94$ & $  1.38$ & $  0.70$ & $ -0.09$  &\hspace{0.25cm}&$10320.30  $ &  $ 11.10$ & $  6.79$ & $  4.17$ & $  1.76$  &\hspace{0.25cm}&\multicolumn{5}{l}{\bf{IRAS $18413+1354$}}                 \\
$10204.58  $ &  $  6.77$ & $  4.34$ & $  2.56$ & $  0.58$  &\hspace{0.25cm}&$10299.41  $ &  $  2.97$ & $  1.42$ & $  0.73$ & $ -0.10$  &\hspace{0.25cm}&$          $ &  $      $ & $      $ & $      $ & $      $  &\hspace{0.25cm}&$09825.59  $ &  $  5.01$ & $  3.10$ & $  2.08$ & $  0.99$  \\
$10255.47  $ &  $  6.72$ & $  4.29$ & $  2.55$ & $  0.59$  &\hspace{0.25cm}&$10318.24  $ &  $  2.98$ & $  1.42$ & $  0.74$ & $ -0.09$  &\hspace{0.25cm}&\multicolumn{5}{l}{\bf{IRAS $18348-0526$}}                 &\hspace{0.25cm}&$10236.59  $ &  $  4.64$ & $  2.93$ & $  2.01$ & $  1.14$  \\
$10296.34  $ &  $  6.10$ & $  3.76$ & $  2.11$ & $  0.23$  &\hspace{0.25cm}&$10531.64  $ &  $  2.92$ & $  1.39$ & $  0.72$ & $ -0.11$  &\hspace{0.25cm}&$08419.54  $ &  $      $ & $      $ & $  7.97$ & $  2.02$  &\hspace{0.25cm}&$          $ &  $      $ & $      $ & $      $ & $      $  \\
$10589.55  $ &  $  7.37$ & $  4.83$ & $  3.01$ & $  0.96$  &\hspace{0.25cm}&$10716.31  $ &  $  2.95$ & $  1.40$ & $  0.69$ & $ -0.09$  &\hspace{0.25cm}&$08496.31  $ &  $      $ & $      $ & $  8.32$ & $  2.28$  &\hspace{0.25cm}&\multicolumn{5}{l}{\bf{IRAS $18560-2954$}}                 \\
$          $ &  $      $ & $      $ & $      $ & $      $  &\hspace{0.25cm}&$          $ &  $      $ & $      $ & $      $ & $      $  &\hspace{0.25cm}&$08552.27  $ &  $      $ & $      $ & $  8.59$ & $  2.36$  &\hspace{0.25cm}&$09117.65  $ &  $  3.53$ & $  2.01$ & $  1.15$ & $  0.06$  \\
\multicolumn{5}{l}{\bf{IRAS $18135-1641$}}                 &\hspace{0.25cm}&\multicolumn{5}{l}{\bf{IRAS $18240+2326$}}                 &\hspace{0.25cm}&$08703.62  $ &  $      $ & $      $ & $  9.35$ & $  2.80$  &\hspace{0.25cm}&$09825.55  $ &  $  2.17$ & $  0.94$ & $  0.30$ & $ -0.60$  \\
$08778.59  $ &  $  2.99$ & $  1.57$ & $  0.93$ & $  0.14$  &\hspace{0.25cm}&$09824.59  $ &  $      $ & $  8.66$ & $  5.47$ & $  1.63$  &\hspace{0.25cm}&$08768.56  $ &  $      $ & $      $ & $  9.74$ & $  3.03$  &\hspace{0.25cm}&$09948.39  $ &  $  2.29$ & $  0.95$ & $  0.30$ & $ -0.48$  \\
$09888.95  $ &  $  3.02$ & $  1.65$ & $  1.01$ & $  0.23$  &\hspace{0.25cm}&$10204.61  $ &  $      $ & $  8.67$ & $  5.42$ & $  1.54$  &\hspace{0.25cm}&$09095.65  $ &  $      $ & $      $ & $  9.28$ & $  2.62$  &\hspace{0.25cm}&$10006.30  $ &  $  2.66$ & $  1.29$ & $  0.57$ & $ -0.25$  \\
$10204.59  $ &  $  3.06$ & $  1.65$ & $  0.99$ & $  0.21$  &\hspace{0.25cm}&$10236.55  $ &  $      $ & $  8.44$ & $  5.26$ & $  1.42$  &\hspace{0.25cm}&$09236.28  $ &  $      $ & $      $ & $  7.96$ & $  1.77$  &\hspace{0.25cm}&$10204.67  $ &  $  4.16$ & $  2.47$ & $  1.41$ & $  0.14$  \\
$10236.50  $ &  $  3.05$ & $  1.65$ & $  0.99$ & $  0.22$  &\hspace{0.25cm}&$10318.26  $ &  $ 11.83$ & $  8.18$ & $  4.99$ & $  1.23$  &\hspace{0.25cm}&$09491.52  $ &  $      $ & $      $ & $  6.77$ & $  0.94$  &\hspace{0.25cm}&$10238.58  $ &  $  4.18$ & $  2.51$ & $  1.45$ & $  0.19$  \\
$10259.45  $ &  $  3.04$ & $  1.62$ & $  0.97$ & $  0.19$  &\hspace{0.25cm}&$10565.62  $ &  $      $ & $ 10.02$ & $  6.72$ & $  2.63$  &\hspace{0.25cm}&$          $ &  $      $ & $      $ & $      $ & $      $  &\hspace{0.25cm}&$10299.44  $ &  $  3.14$ & $  1.86$ & $  1.00$ & $ -0.17$  \\
$10296.38  $ &  $  2.96$ & $  1.58$ & $  0.94$ & $  0.16$  &\hspace{0.25cm}&$          $ &  $      $ & $      $ & $      $ & $      $  &\hspace{0.25cm}&\multicolumn{5}{l}{\bf{IRAS $18349+1023$}}                 &\hspace{0.25cm}&$10320.41  $ &  $  2.74$ & $  1.53$ & $  0.78$ & $ -0.27$  \\
$10317.36  $ &  $  2.94$ & $  1.56$ & $  0.93$ & $  0.17$  &\hspace{0.25cm}&\multicolumn{5}{l}{\bf{IRAS $18333+0533$}}                 &\hspace{0.25cm}&$08703.65  $ &  $      $ & $      $ & $      $ & $ -0.96$  &\hspace{0.25cm}&$10589.63  $ &  $  2.94$ & $  1.47$ & $  0.73$ & $ -0.08$  \\
$10531.62  $ &  $  2.95$ & $  1.61$ & $  0.97$ & $  0.18$  &\hspace{0.25cm}&$05897.40  $ &  $  8.79$ & $  5.22$ & $  3.04$ & $  0.77$  &\hspace{0.25cm}&$09824.61  $ &  $  3.00$ & $  1.48$ & $  0.60$ & $ -0.35$  &\hspace{0.25cm}&$10619.53  $ &  $  3.21$ & $  1.69$ & $  0.90$ & $  0.03$  \\
$          $ &  $      $ & $      $ & $      $ & $      $  &\hspace{0.25cm}&$05923.35  $ &  $  8.64$ & $  5.10$ & $  2.95$ & $  0.74$  &\hspace{0.25cm}&$10204.63  $ &  $  2.43$ & $  1.05$ & $  0.26$ & $ -0.79$  &\hspace{0.25cm}&$10677.41  $ &  $  3.84$ & $  2.15$ & $  1.24$ & $  0.29$  \\
\multicolumn{5}{l}{\bf{IRAS $18194-2708$}}                 &\hspace{0.25cm}&$05958.23  $ &  $  8.50$ & $  4.99$ & $  2.89$ & $      $  &\hspace{0.25cm}&$10236.56  $ &  $  2.65$ & $  1.18$ & $  0.41$ & $ -0.59$  &\hspace{0.25cm}&$10755.26  $ &  $  4.13$ & $  2.42$ & $  1.40$ & $  0.28$  \\
$09824.57  $ &  $ 10.35$ & $  7.07$ & $  4.49$ & $  1.52$  &\hspace{0.25cm}&$06191.58  $ &  $  9.42$ & $  5.71$ & $  3.53$ & $  1.36$  &\hspace{0.25cm}&$10259.48  $ &  $  2.87$ & $  1.36$ & $  0.52$ & $ -0.49$  &\hspace{0.25cm}&$10974.49  $ &  $  2.09$ & $  0.82$ & $  0.19$ & $ -0.68$  \\
$09887.50  $ &  $ 10.27$ & $  7.01$ & $  4.46$ & $  1.50$  &\hspace{0.25cm}&$06193.63  $ &  $  9.50$ & $  5.74$ & $  3.55$ & $  1.40$  &\hspace{0.25cm}&$10320.30  $ &  $  3.81$ & $  2.10$ & $  1.08$ & $ -0.04$  &\hspace{0.25cm}&$          $ &  $      $ & $      $ & $      $ & $      $  \\
$09945.37  $ &  $      $ & $  6.72$ & $  4.20$ & $      $  &\hspace{0.25cm}&$06222.50  $ &  $  9.81$ & $  5.96$ & $  3.70$ & $  1.50$  &\hspace{0.25cm}&$          $ &  $      $ & $      $ & $      $ & $      $  &\hspace{0.25cm}&\multicolumn{5}{l}{\bf{IRAS $19008+0726$}}                 \\
$10031.26  $ &  $  9.88$ & $  6.65$ & $  4.15$ & $  1.25$  &\hspace{0.25cm}&$06271.43  $ &  $ 10.52$ & $  6.42$ & $  4.01$ & $  1.70$  &\hspace{0.25cm}&\multicolumn{5}{l}{\bf{IRAS $18398-0220$}}                 &\hspace{0.25cm}&$09117.65  $ &  $  7.51$ & $  4.78$ & $  2.64$ & $  0.19$  \\
$10204.60  $ &  $  8.46$ & $  5.36$ & $  3.03$ & $  0.25$  &\hspace{0.25cm}&$06309.37  $ &  $ 11.03$ & $  6.83$ & $  4.23$ & $  1.80$  &\hspace{0.25cm}&$09824.63  $ &  $  5.88$ & $  3.41$ & $  1.65$ & $ -0.33$  &\hspace{0.25cm}&$09825.59  $ &  $  6.97$ & $  4.38$ & $  2.34$ & $ -0.10$  \\
\end{tabular}
}
\end{table}
\setcounter{table}{1}
\begin{table}
\vspace{1.5 cm}
\caption{ ...continued. Near-infrared data. The  Date is given as JD--2440000.00.}
\footnotesize{
\begin{tabular}{lrrrrclrrrrclrrrrclrrrr}\\
\multicolumn{1}{c}{\bf{DATE}}&
\multicolumn{1}{c}{\bf{J}}&
\multicolumn{1}{c}{\bf{H}}&
\multicolumn{1}{c}{\bf{K}}&
\multicolumn{1}{c}{\bf{L}}&
&
\multicolumn{1}{c}{\bf{DATE}}&
\multicolumn{1}{c}{\bf{J}}&
\multicolumn{1}{c}{\bf{H}}&
\multicolumn{1}{c}{\bf{K}}&
\multicolumn{1}{c}{\bf{L}}&
&
\multicolumn{1}{c}{\bf{DATE}}&
\multicolumn{1}{c}{\bf{J}}&
\multicolumn{1}{c}{\bf{H}}&
\multicolumn{1}{c}{\bf{K}}&
\multicolumn{1}{c}{\bf{L}}&
&
\multicolumn{1}{c}{\bf{DATE}}&
\multicolumn{1}{c}{\bf{J}}&
\multicolumn{1}{c}{\bf{H}}&
\multicolumn{1}{c}{\bf{K}}&
\multicolumn{1}{c}{\bf{L}}\\
\\
\multicolumn{5}{l}{\bf{IRAS $19008+0726$  .....continued}} &\hspace{0.25cm}&\multicolumn{5}{l}{\bf{IRAS $19126-0708$  .....continued}} &\hspace{0.25cm}&\multicolumn{5}{l}{\bf{IRAS $20077-0625$  .....continued}} &\hspace{0.25cm}&\multicolumn{5}{l}{\bf{IRAS $21286+1055$  .....continued}} \\
$10204.66  $ &  $  7.76$ & $  5.09$ & $  3.00$ & $  0.44$  &\hspace{0.25cm}&$05576.32  $ &  $  2.41$ & $  0.83$ & $  0.08$ & $ -1.08$  &\hspace{0.25cm}&$08785.59  $ &  $  5.85$ & $  3.34$ & $  1.66$ & $ -0.26$  &\hspace{0.25cm}&$10976.64  $ &  $  5.05$ & $  3.51$ & $  2.50$ & $  1.35$  \\
$10236.60  $ &  $  7.46$ & $  4.80$ & $  2.76$ & $  0.32$  &\hspace{0.25cm}&$10262.48  $ &  $  1.63$ & $  0.52$ & $ -0.19$ & $ -1.54$  &\hspace{0.25cm}&$08898.29  $ &  $  5.65$ & $  3.12$ & $  1.49$ & $ -0.36$  &\hspace{0.25cm}&$          $ &  $      $ & $      $ & $      $ & $      $  \\
$10320.37  $ &  $  6.77$ & $  4.18$ & $  2.19$ & $ -0.15$  &\hspace{0.25cm}&$10275.42  $ &  $  1.55$ & $  0.41$ & $      $ & $ -1.58$  &\hspace{0.25cm}&$09142.64  $ &  $  7.83$ & $  4.71$ & $  2.73$ & $  0.77$  &\hspace{0.25cm}&\multicolumn{5}{l}{\bf{IRAS $23166+1655$}}                 \\
$          $ &  $      $ & $      $ & $      $ & $      $  &\hspace{0.25cm}&$10275.96  $ &  $  1.55$ & $  0.41$ & $ -0.31$ & $      $  &\hspace{0.25cm}&$09172.52  $ &  $  8.02$ & $  4.89$ & $  2.85$ & $  0.81$  &\hspace{0.25cm}&$10615.69  $ &  $      $ & $      $ & $  9.89$ & $      $  \\
\multicolumn{5}{l}{\bf{IRAS $19059-2219$}}                 &\hspace{0.25cm}&$          $ &  $      $ & $      $ & $      $ & $      $  &\hspace{0.25cm}&$09214.41  $ &  $  8.44$ & $  5.17$ & $  3.02$ & $  0.87$  &\hspace{0.25cm}&                                                           \\
$09117.66  $ &  $  6.06$ & $  4.08$ & $  2.78$ & $  1.44$  &\hspace{0.25cm}&\multicolumn{5}{l}{\bf{IRAS $19175-0807$}}                 &\hspace{0.25cm}&$09501.63  $ &  $  5.60$ & $  3.04$ & $  1.43$ & $ -0.39$  &\hspace{0.25cm}&                                                           \\
$09825.56  $ &  $  3.93$ & $  2.47$ & $  1.66$ & $  0.66$  &\hspace{0.25cm}&$09117.68  $ &  $  7.02$ & $  4.17$ & $  2.06$ & $ -0.32$  &\hspace{0.25cm}&$09581.48  $ &  $  5.72$ & $  3.14$ & $  1.55$ & $ -0.21$  &\hspace{0.25cm}&                                                           \\
$10006.31  $ &  $  5.15$ & $  3.48$ & $  2.47$ & $  1.47$  &\hspace{0.25cm}&$09825.60  $ &  $  6.51$ & $  3.88$ & $  1.99$ & $ -0.30$  &\hspace{0.25cm}&$09614.40  $ &  $  5.87$ & $  3.29$ & $  1.68$ & $ -0.07$  &\hspace{0.25cm}&                                                           \\
$10259.51  $ &  $  3.87$ & $  2.44$ & $  1.64$ & $  0.65$  &\hspace{0.25cm}&$10035.25  $ &  $  8.48$ & $  5.71$ & $  3.52$ & $  0.89$  &\hspace{0.25cm}&$09637.30  $ &  $  5.97$ & $  3.38$ & $  1.77$ & $  0.04$  &\hspace{0.25cm}&                                                           \\
$10320.42  $ &  $  3.82$ & $  2.35$ & $  1.56$ & $  0.66$  &\hspace{0.25cm}&$10212.64  $ &  $  7.98$ & $  5.16$ & $  2.95$ & $  0.42$  &\hspace{0.25cm}&$09904.62  $ &  $  8.30$ & $  5.07$ & $  2.95$ & $  0.85$  &\hspace{0.25cm}&                                                           \\
$10716.36  $ &  $  4.24$ & $  2.85$ & $  1.94$ & $  0.88$  &\hspace{0.25cm}&$10259.55  $ &  $  8.21$ & $  5.35$ & $  3.12$ & $  0.53$  &\hspace{0.25cm}&$          $ &  $      $ & $      $ & $      $ & $      $  &\hspace{0.25cm}&                                                           \\
$10738.33  $ &  $  4.13$ & $  2.74$ & $  1.85$ & $  0.81$  &\hspace{0.25cm}&$10295.49  $ &  $  7.69$ & $  4.86$ & $  2.68$ & $  0.20$  &\hspace{0.25cm}&\multicolumn{5}{l}{\bf{IRAS $20440-0105$}}                 &\hspace{0.25cm}&                                                           \\
$          $ &  $      $ & $      $ & $      $ & $      $  &\hspace{0.25cm}&$10316.44  $ &  $  7.28$ & $  4.48$ & $  2.35$ & $ -0.02$  &\hspace{0.25cm}&$07373.49  $ &  $  2.24$ & $  1.23$ & $  0.81$ & $  0.27$  &\hspace{0.25cm}&                                                           \\
\multicolumn{5}{l}{\bf{IRAS $19093-3256$}}                 &\hspace{0.25cm}&$          $ &  $      $ & $      $ & $      $ & $      $  &\hspace{0.25cm}&$07416.41  $ &  $  2.28$ & $  1.25$ & $  0.82$ & $      $  &\hspace{0.25cm}&                                                           \\
$09825.58  $ &  $  2.61$ & $  1.51$ & $  0.92$ & $  0.03$  &\hspace{0.25cm}&\multicolumn{5}{l}{\bf{IRAS $19321+2757$}}                 &\hspace{0.25cm}&$07695.63  $ &  $  3.17$ & $  2.17$ & $  1.60$ & $  0.80$  &\hspace{0.25cm}&                                                           \\
$09948.40  $ &  $  2.93$ & $  1.69$ & $  1.03$ & $  0.29$  &\hspace{0.25cm}&$09825.60  $ &  $  6.41$ & $  4.22$ & $  2.55$ & $  0.63$  &\hspace{0.25cm}&$10976.60  $ &  $  3.82$ & $  2.48$ & $  1.77$ & $  1.02$  &\hspace{0.25cm}&                                                           \\
$10006.32  $ &  $  3.49$ & $  2.16$ & $  1.41$ & $  0.63$  &\hspace{0.25cm}&$10204.64  $ &  $  6.00$ & $  3.63$ & $  1.88$ & $ -0.20$  &\hspace{0.25cm}&$          $ &  $      $ & $      $ & $      $ & $      $  &\hspace{0.25cm}&                                                           \\
$10035.26  $ &  $  3.77$ & $  2.37$ & $  1.58$ & $  0.74$  &\hspace{0.25cm}&$10238.60  $ &  $  6.14$ & $  3.72$ & $  1.95$ & $ -0.13$  &\hspace{0.25cm}&\multicolumn{5}{l}{\bf{IRAS $20570+2714$}}                 &\hspace{0.25cm}&                                                           \\
$10259.52  $ &  $  2.61$ & $  1.48$ & $  0.87$ & $  0.05$  &\hspace{0.25cm}&$10296.44  $ &  $  6.79$ & $  4.26$ & $  2.39$ & $  0.19$  &\hspace{0.25cm}&$09117.67  $ &  $  8.26$ & $  5.31$ & $  2.97$ & $  0.23$  &\hspace{0.25cm}&                                                           \\
$10299.45  $ &  $  2.73$ & $  1.54$ & $  0.92$ & $  0.16$  &\hspace{0.25cm}&$10320.34  $ &  $  7.07$ & $  4.52$ & $  2.60$ & $  0.33$  &\hspace{0.25cm}&$10296.48  $ &  $  9.56$ & $  6.59$ & $  4.07$ & $  1.05$  &\hspace{0.25cm}&                                                           \\
$10358.27  $ &  $  3.25$ & $  1.94$ & $  1.27$ & $  0.56$  &\hspace{0.25cm}&$10976.54  $ &  $  7.71$ & $  5.24$ & $  3.23$ & $  0.74$  &\hspace{0.25cm}&$          $ &  $      $ & $      $ & $      $ & $      $  &\hspace{0.25cm}&                                                           \\
$10619.55  $ &  $  2.60$ & $  1.49$ & $  0.89$ & $  0.09$  &\hspace{0.25cm}&$          $ &  $      $ & $      $ & $      $ & $      $  &\hspace{0.25cm}&\multicolumn{5}{l}{\bf{IRAS $21032-0024$}}                 &\hspace{0.25cm}&                                                           \\
$          $ &  $      $ & $      $ & $      $ & $      $  &\hspace{0.25cm}&\multicolumn{5}{l}{\bf{IRAS $20077-0625$}}                 &\hspace{0.25cm}&$08786.61  $ &  $  5.34$ & $  3.26$ & $  1.75$ & $  0.06$  &\hspace{0.25cm}&                                                           \\
$10677.42  $ &  $  2.69$ & $  1.46$ & $  0.86$ & $  0.17$  &\hspace{0.25cm}&$07013.51  $ &  $  6.84$ & $  4.08$ & $  2.29$ & $  0.33$  &\hspace{0.25cm}&$08875.46  $ &  $  4.50$ & $  2.57$ & $  1.20$ & $ -0.38$  &\hspace{0.25cm}&                                                           \\
$10756.27  $ &  $  3.41$ & $  2.06$ & $  1.36$ & $  0.67$  &\hspace{0.25cm}&$07034.40  $ &  $  7.00$ & $  4.21$ & $  2.39$ & $  0.43$  &\hspace{0.25cm}&$09146.57  $ &  $  5.42$ & $  3.38$ & $  1.90$ & $  0.32$  &\hspace{0.25cm}&                                                           \\
$10974.50  $ &  $  2.50$ & $  1.42$ & $  0.87$ & $  0.11$  &\hspace{0.25cm}&$07068.32  $ &  $  7.22$ & $  4.37$ & $  2.53$ & $  0.55$  &\hspace{0.25cm}&$09222.46  $ &  $  5.23$ & $  3.25$ & $  1.79$ & $  0.25$  &\hspace{0.25cm}&                                                           \\
$          $ &  $      $ & $      $ & $      $ & $      $  &\hspace{0.25cm}&$07083.34  $ &  $  7.27$ & $  4.46$ & $  2.59$ & $  0.61$  &\hspace{0.25cm}&$09640.28  $ &  $  5.43$ & $  3.32$ & $  1.82$ & $  0.27$  &\hspace{0.25cm}&                                                           \\
\multicolumn{5}{l}{\bf{IRAS $19126-0708$}}                 &\hspace{0.25cm}&$07365.46  $ &  $  6.24$ & $  3.67$ & $  1.88$ & $ -0.11$  &\hspace{0.25cm}&$09673.26  $ &  $  5.20$ & $  3.12$ & $  1.66$ & $  0.09$  &\hspace{0.25cm}&                                                           \\
$02251.00  $ &  $  3.04$ & $  1.51$ & $  0.54$ & $ -1.21$  &\hspace{0.25cm}&$08105.48  $ &  $      $ & $  3.17$ & $  1.47$ & $      $  &\hspace{0.25cm}&$10261.62  $ &  $  3.86$ & $  2.07$ & $   .85$ & $ -0.59$  &\hspace{0.25cm}&                                                           \\
$02629.00  $ &  $  3.44$ & $  1.78$ & $  0.81$ & $ -0.81$  &\hspace{0.25cm}&$08110.50  $ &  $  5.64$ & $  3.13$ & $  1.42$ & $ -0.41$  &\hspace{0.25cm}&$          $ &  $      $ & $      $ & $      $ & $      $  &\hspace{0.25cm}&                                                           \\
$02910.00  $ &  $  3.24$ & $  2.02$ & $  1.12$ & $ -0.70$  &\hspace{0.25cm}&$08139.38  $ &  $  5.50$ & $  2.98$ & $  1.32$ & $ -0.49$  &\hspace{0.25cm}&\multicolumn{5}{l}{\bf{IRAS $21286+1055$}}                 &\hspace{0.25cm}&                                                           \\
$02975.00  $ &  $  2.94$ & $  1.41$ & $  0.52$ & $ -1.18$  &\hspace{0.25cm}&$08178.28  $ &  $  5.42$ & $  2.91$ & $  1.26$ & $ -0.55$  &\hspace{0.25cm}&$10016.31  $ &  $  4.43$ & $  2.97$ & $  1.99$ & $  0.93$  &\hspace{0.25cm}&                                                           \\
$03014.00  $ &  $  2.77$ & $  1.29$ & $  0.36$ & $ -1.22$  &\hspace{0.25cm}&$08201.26  $ &  $  5.45$ & $  2.91$ & $  1.29$ & $ -0.49$  &\hspace{0.25cm}&$10035.29  $ &  $  4.34$ & $  2.94$ & $  1.96$ & $  0.82$  &\hspace{0.25cm}&                                                           \\
$03347.41  $ &  $  3.74$ & $  2.26$ & $  1.40$ & $ -0.04$  &\hspace{0.25cm}&$08457.45  $ &  $  7.10$ & $  4.20$ & $  2.39$ & $  0.60$  &\hspace{0.25cm}&$10237.65  $ &  $  3.08$ & $  1.87$ & $  1.14$ & $  0.26$  &\hspace{0.25cm}&                                                           \\
$05200.34  $ &  $  3.50$ & $  1.74$ & $  0.88$ & $ -0.22$  &\hspace{0.25cm}&$08491.46  $ &  $  7.40$ & $  4.39$ & $  2.49$ & $  0.66$  &\hspace{0.25cm}&$10296.50  $ &  $  3.35$ & $  2.00$ & $  1.25$ & $  0.39$  &\hspace{0.25cm}&                                                           \\
$05545.41  $ &  $  3.22$ & $  1.69$ & $ -0.09$ & $ -1.30$  &\hspace{0.25cm}&$08575.25  $ &  $  8.04$ & $  4.92$ & $  2.83$ & $  0.72$  &\hspace{0.25cm}&$10320.43  $ &  $  3.63$ & $  2.19$ & $  1.42$ & $  0.56$  &\hspace{0.25cm}&                                                           \\
$05569.37  $ &  $  2.32$ & $  0.77$ & $  0.03$ & $ -1.17$  &\hspace{0.25cm}&$08760.62  $ &  $  5.87$ & $  3.38$ & $  1.69$ & $ -0.28$  &\hspace{0.25cm}&$10358.31  $ &  $  4.16$ & $  2.66$ & $  1.79$ & $  0.82$  &\hspace{0.25cm}&                                                           \\
\end{tabular}
}
\end{table}
\twocolumn
\end{landscape}

From the number of references listed in Tables 1, CW Leo stands out as the
best studied source in this sample. About 50 molecular species have been
detected in its envelope (Wallerstein \& Knapp 1998) which is more than for
any other source. The full sample consists mainly of cool variable stars
with a few possible post-AGB objects.

\section{Observations}
\subsection{Near-infrared Photometry}
 Near-infrared observations were obtained in the $JHKL$ bands (1.2, 1.65,
2.2 and 3.45$\mu$m) for 42 of the 45 southern sources, defined henceforth as
stars with $\delta<+30^{\circ}$. Some of these 42 stars were selected
exclusively from JK89, while the others were already part of other
monitoring programmes. The remaining three southern stars 
($06500+0829$,  $17513-2313$ and RAFGL 1406) were not observed because
they were in crowded fields and confusion rendered observations with a
single channel photometer impossible (note that Winfrey et al. (1994)
suggest that  $17513-2313$ is actually a supergiant and it is therefore
not referred to again). 

The near-infrared observations for the 42 southern sources, accurate to 
$\pm0.1$\,mag and better (0.03\,mag in $JHK$ and 0.05\,mag in $L$ in the
majority of cases), are listed in Table 2. The magnitudes from the
0.75\,m reflector are on the SAAO system as defined by Carter (1990). The 
$HKL$ bands for the 1.9\,m reflector are assumed to be identical to those of 
the 0.75\,m reflector, while the $J$-magnitude is converted to the SAAO
system. 

In some instances, no observations are listed because the sources were too
faint to be reliably measured through the relevant filter with a particular
telescope. The times of the observations are given as Julian dates from
which 244\,0000 days has been subtracted.  In addition to the data from
Table~2, published SAAO observations from Whitelock et al. (1994, 1997a)
and Lloyd Evans (1997) have been used in the analysis, together with several
unpublished observations from Lloyd Evans (private communication) with a sum
total of 848 measurements.
 
Mean magnitudes calculated from the average of observed values at maximum and
minimum light are listed in Table 3 and compared with the Fourier mean in
appendix A. These two averages compare fairly well, differences were less
than 0.3\,mag for the $J$-band and less than about 0.1\,mag for the
$HKL$ bands. Table 3 also lists mean $K$-magnitudes
calculated as above from values listed in the literature, for the 20
remaining stars with no near-infrared SAAO photometry.

\subsection{Mid- and Far-infrared Observations}
\subsubsection{Photometry}
 Infrared fluxes at 12, 25 and $60\mu$m were taken from the IRAS Point
Source Catalog (PSC) and colour corrected as specified in the IRAS
Explanatory Supplement (ES). The results are listed in Table 4 together with
the IRAS flux-quality designations and the IRAS variability index (V.I.),
which gives the probability that the source is variable in units of 0.1
(see IRAS ES). Table 4 also contains $11\mu$m magnitudes from the
\emph{Revised Airforce Geophysics Laboratory, RAFGL}, survey.

\subsubsection{Low Resolution Spectra (LRS)}
 The IRAS Low Resolution spectral classifications, from Olnon \& Raimond
(1986) or Loup et al. (1993), are listed in Table 4. Note that 
$18135-1641$ does not have a published classification so we examined 
the spectrum in the IRAS Low Resolution Spectra electronic database at the
University of Calgary. It shows a weak $10\mu$m emission feature, but no
$18\mu$m feature, which is present in the spectra of all the other
oxygen-rich stars. It is actually a supergiant as discussed below (section
4).

\section{Variability}
 Using the data in Table 2, the southern sources were classified as variable
in the near-infrared if the difference between the maximum and minimum
magnitudes at $K$ and/or at $L$ was greater than 0.1 mag; thirty seven of
the stars are variable and 2 non-variable. The only two decisively
non-variable stars, $18135-1641$ and $18204-1344$, are the supergiants (see
section 2). There were insufficient observations to determine if the other 6
stars are variable or not. Table 5 lists the variability information.

Period determinations, using Fourier transforms of the $K$ light curve, were
made for all variables with measurements on at least 10 epochs. Plausible,
although not definitive, periods were also established for some sources with
as few as 8 measurements. For the very red source $17411-3154$, a period could
only be estimated from the $L$ data. Periods, listed in Table 5, were thus
determined for a total of 20 stars, and phased $K$ light curves for these
are shown in Fig.~1. Where possible the Fourier transform analysis was
carried out at the other wavelengths and in all cases the periods found were
within 2 percent of those listed in Table 5. No period determination was
possible for $19126-0708$ (W Aql), although it had been observed on 12
epochs. This is because 6 of the epochs were between 1968 and 1982 when
this carbon star was undergoing a dust obscuration episode (Mattei \& Foster
1997). Such an event would disrupt the apparent periodicity.

The average of periods found in the literature are listed in Table 5, with
references, for the 47 stars concerned. Where there are periods in common
none of them differs significantly from those determined here. In the
following analysis our own determinations, where available, are used in
preference to others, except for $17360-3012$ and $18348-0526$ where our
determination is uncertain due to the small number of measurements
available.

\subsection{Phased Light-curves}           
Using the periods found for these 20 stars, the relative data were phased, 
arbitrarily assuming zero phase at JD 2440000.00, where JD is the
heliocentric Julian date. These phased light-curves are plotted in Fig.~1  
together with least squares fits to the data of the form:


\begin{landscape}
\begin{table}
\setcounter{table}{2}
\caption{Average near-infrared magnitudes}
\vspace{1.5cm}    
\centering
\normalsize{
\begin{tabular}{lcccclcccc}
\multicolumn{1}{c}{\bf{NAME}}&
\multicolumn{1}{c}{$\hspace{0.5cm}{\bf{J}}\hspace{0.8cm}$}&
\multicolumn{1}{c}{$\hspace{0.8cm}{\bf{H}}\hspace{0.8cm}$}&
\multicolumn{1}{c}{$\hspace{0.8cm}{\bf{K}}\hspace{0.8cm}$}&
\multicolumn{1}{c}{$\hspace{0.8cm}{\bf{L}}\hspace{0.8cm}$}&
\multicolumn{1}{c}{\bf{NAME}}&
\multicolumn{1}{c}{$\hspace{0.8cm}{\bf{J}}\hspace{0.8cm}$}&
\multicolumn{1}{c}{$\hspace{0.8cm}{\bf{H}}\hspace{0.8cm}$}&
\multicolumn{1}{c}{$\hspace{0.8cm}{\bf{K}}\hspace{0.8cm}$}&
\multicolumn{1}{c}{$\hspace{0.8cm}{\bf{L}}\hspace{0.8cm}$}\\ 
\\
\hline
IRAS $00042+4248$ & $      $ & $      $ & $  2.52^{\dagger}$ & $      $&IRAS $18194-2708$ & $  9.33$ & $  6.13$ & $  3.74          $ & $  0.88$\\
IRAS $01037+1219$ & $  8.06$ & $  4.71$ & $  2.32          $ & $ -0.09$&IRAS $18240+2326$ & $ 11.83$ & $  9.10$ & $  5.86          $ & $  1.93$\\
IRAS $01159+7220$ & $      $ & $      $ & $  2.30^{\dagger}$ & $      $&IRAS $18333+0533$ & $ 10.38$ & $  6.20$ & $  3.73          $ & $  1.29$\\
IRAS $02270-2619$ & $  4.57$ & $  2.74$ & $  1.42          $ & $ -0.06$&IRAS $18348-0526$ & $      $ & $      $ & $  8.26          $ & $  1.98$\\
IRAS $02316+6455$ & $      $ & $      $ & $  3.10^{*}      $ & $      $&IRAS $18349+1023$ & $  3.12$ & $  1.57$ & $  0.67          $ & $ -0.50$\\
IRAS $02351-2711$ & $  2.92$ & $  1.67$ & $  0.98          $ & $  0.19$&IRAS $18397+1738$ & $  5.76$ & $  3.62$ & $  1.87          $ & $ -0.24$\\
IRAS $03229+4721$ & $      $ & $      $ & $  1.65^{*}      $ & $      $&IRAS $18398-0220$ & $  5.91$ & $  3.51$ & $  1.79          $ & $ -0.19$\\
IRAS $03507+1115$ & $  2.57$ & $  0.69$ & $ -0.47          $ & $ -1.79$&IRAS $18413+1354$ & $  4.83$ & $  3.01$ & $  2.05          $ & $  1.06$\\
IRAS $04307+6210$ & $      $ & $      $ & $  1.80^{\dagger}$ & $      $&IRAS $18560-2954$ & $  3.13$ & $  1.67$ & $  0.82          $ & $ -0.19$\\
IRAS $04566+5606$ & $      $ & $      $ & $ -0.50^{\dagger}$ & $      $&IRAS $19008+0726$ & $  7.27$ & $  4.63$ & $  2.59          $ & $  0.14$\\
IRAS $05073+5248$ & $      $ & $      $ & $  2.95^{*}      $ & $      $&IRAS $19059-2219$ & $  4.94$ & $  3.21$ & $  2.17          $ & $  1.06$\\
IRAS $05411+6957$ & $      $ & $      $ & $  1.05^{*}      $ & $      $&IRAS $19093-3256$ & $  3.14$ & $  1.90$ & $  1.22          $ & $  0.39$\\
IRAS $05559+7430$ & $      $ & $      $ & $  1.49^{\dagger}$ & $      $&IRAS $19126-0708$ & $  2.65$ & $  1.33$ & $  0.54          $ & $ -0.81$\\
IRAS $06176-1036$ & $  6.65$ & $  4.99$ & $  3.41          $ & $  1.30$&IRAS $19175-0807$ & $  7.49$ & $  4.79$ & $  2.76          $ & $  0.28$\\
IRAS $06300+6058$ & $      $ & $      $ & $  1.05^{\dagger}$ & $      $&IRAS $19321+2757$ & $  6.86$ & $  4.44$ & $  2.55          $ & $  0.27$\\
IRAS $06500+0829$ & $      $ & $      $ & $  1.07^{\dagger}$ & $      $&IRAS $20077-0625$ & $  6.93$ & $  4.04$ & $  2.14          $ & $  0.16$\\
IRAS $08088-3243$ & $  8.57$ & $  5.97$ & $  3.86          $ & $  1.24$&IRAS $20396+4757$ & $      $ & $      $ & $  0.41^{\dagger}$ & $      $\\
IRAS $09116-2439$ & $ 13.20$ & $  9.68$ & $  6.03          $ & $  2.22$&IRAS $20440-0105$ & $  3.03$ & $  1.85$ & $  1.29          $ & $  0.64$\\
IRAS $09429-2148$ & $  5.65$ & $  3.69$ & $  2.36          $ & $  0.73$&IRAS $20570+2714$ & $  8.91$ & $  5.95$ & $  3.52          $ & $  0.64$\\
IRAS $09452+1330$ & $  7.28$ & $  3.99$ & $  1.17          $ & $ -2.55$&IRAS $21032-0024$ & $  4.65$ & $  2.72$ & $  1.37          $ & $ -0.13$\\
IRAS $10131+3049$ & $  6.12$ & $  3.42$ & $  1.27          $ & $ -1.34$&IRAS $21286+1055$ & $  4.06$ & $  2.69$ & $  1.82          $ & $  0.81$\\
IRAS $10491-2059$ & $  2.97$ & $  1.18$ & $ -0.09          $ & $ -1.66$&IRAS $21320+3850$ & $      $ & $      $ & $  1.57^{\dagger}$ & $      $\\
IRAS $12447+0425$ & $  4.88$ & $  3.08$ & $  1.81          $ & $  0.26$&IRAS $21456+6422$ & $      $ & $      $ & $  1.49^{\dagger}$ & $      $\\
IRAS $17049-2440$ & $ 12.56$ & $  8.57$ & $  5.51          $ & $  1.93$&IRAS $23166+1655$ & $      $ & $ 14.58$ & $ 10.50          $ & $  4.27$\\
IRAS $17119+0859$ & $  4.94$ & $  3.52$ & $  2.45          $ & $  0.78$&IRAS $23320+4316$ & $      $ & $      $ & $  3.55^{*}      $ & $      $\\
IRAS $17297+1747$ & $  9.07$ & $  6.02$ & $  3.69          $ & $  0.76$&IRAS $23496+6131$ & $      $ & $      $ & $  2.40^{*}      $ & $      $\\
IRAS $17360-3012$ & $  9.16$ & $  6.15$ & $  4.09          $ & $  1.64$&RAFGL $1406     $ & $      $ & $      $ & $  1.98^{\dagger}$ & $      $\\
IRAS $17411-3154$ & $ 13.17$ & $ 11.01$ & $  9.69          $ & $  3.94$&RAFGL $2688     $ & $      $ & $      $ & $  8.40^{\dagger}$ & $      $\\
IRAS $18009-2019$ & $  3.86$ & $  2.13$ & $  1.12          $ & $  0.11$&&&&& \\
IRAS $18040-0941$ & $  6.60$ & $  4.14$ & $  2.40          $ & $  0.45$&&&&& \\  
\end{tabular}
}
\vspace{0.5cm}
\begin{flushleft}                 
\small{${*}$ from Jones et al. (1990)\\
${\dagger}$ from Gezari et al. (1993)\\}
\end{flushleft}
\end{table}
\end{landscape}  
\onecolumn
\setcounter{table}{3}
\begin{table}
\caption{Mid-infrared and IRAS Data.}
\centering
\small{
\begin{tabular}{lrrrrccc} 
\multicolumn{1}{c}{\bf{NAME}}&
\multicolumn{1}{c}{${\bf{[11\mu{m}]}}$}&
\multicolumn{1}{c}{$\bf{F_{12\mu{m}}}$}&
\multicolumn{1}{c}{$\bf{F_{25\mu{m}}}$}&
\multicolumn{1}{c}{$\bf{F_{60\mu{m}}}$}&
\multicolumn{1}{c}{$\bf{F.D.U.^{\dagger}}$}&
\multicolumn{1}{c}{\bf{LRS}}&
\multicolumn{1}{c}{\bf{V.I.}}\\
      &
      &
\multicolumn{1}{c}{$\bf{(Jy)}$}&
\multicolumn{1}{c}{$\bf{(Jy)}$}&
\multicolumn{1}{c}{$\bf{(Jy)}$}&
                &
\multicolumn{1}{c}{\bf{CLASS}}&
             \\ \hline
IRAS $00042+4248$  & $ -2.5$ & $    391.47$  & $   247.39$ & $    45.29$& BBD  & $  26 $  & $ 0$\\ 
IRAS $01037+1219$  & $ -3.4$ & $   1060.69$  & $   770.10$ & $   171.52$& BBC  &$\,\,\,\,\,\,  4no^{*}$  & $ 1$\\ 
IRAS $01159+7220$  & $ -2.9$ & $    245.57$  & $   140.93$ & $    20.35$& BBC  & $  22 $  & $ 1$\\
IRAS $02270-2619$  & $ -2.6$ & $    198.84$  & $    56.09$ & $    12.39$& BBC  & $  43 $  & $ 4$\\ 
IRAS $02316+6455$  & $ -2.7$ & $    361.10$  & $   231.00$ & $    35.13$& BBD  & $\,\,  2n^{*} $  & $ 1$\\  
IRAS $02351-2711$  & $ -2.7$ & $    334.04$  & $   174.63$ & $    23.04$& BBD  & $  29 $  & $ 5$\\ 
IRAS $03229+4721$  & $ -3.2$ & $    427.14$  & $   149.05$ & $    31.05$& CCC  & $  44 $  & $ 9$\\ 
IRAS $03507+1115$  & $ -4.2$ & $   3654.87$  & $  1626.62$ & $   223.78$& BBB  & $  26 $  & $ 0$\\ 
IRAS $04307+6210$  & $ -1.9$ & $    195.65$  & $    68.34$ & $    13.44$& EEC  & $  45 $  & $ 9$\\  
IRAS $04566+5606$  & $ -4.1$ & $   1341.25$  & $   480.48$ & $   104.89$& CFD  & $  27 $  & $ 9$\\ 
IRAS $05073+5248$  & $ -2.4$ & $    227.98$  & $   228.27$ & $    58.72$& BBD  & $  24 $  & $ 9$\\ 
IRAS $05411+6957$  & $ -3.0$ & $    563.75$  & $   289.84$ & $    39.92$& EEC  & $  29 $  & $ 9$\\ 
IRAS $05559+7430$  & $ -1.6$ & $    158.85$  & $    78.47$ & $    13.53$& CBC  & $  24 $  & $ 9$\\ 
IRAS $06176-1036$  & $ -2.7$ & $    455.93$  & $   394.29$ & $   142.91$& BBC  & $  80 $  & $ 1$\\ 
IRAS $06300+6058$  & $ -3.0$ & $    264.31$  & $   167.48$ & $    36.23$& BBE  & $  28 $  & $ 9$\\ 
IRAS $06500+0829$  & $ -2.6$ & $    576.42$  & $   291.21$ & $    84.73$& CCD  & $  28 $  & $ 9$\\ 
IRAS $08088-3243$  & $     $ & $    285.54$  & $   117.61$ & $    24.64$& BBC  & $\,\,  4n^{*} $  & $ 8$\\ 
IRAS $09116-2439$  & $     $ & $    628.22$  & $   307.23$ & $    65.49$& BCD  & $  42 $  & $ 9$\\ 
IRAS $09429-2148$  & $ -1.9$ & $    461.67$  & $   362.10$ & $    55.34$& BBE  & $  28 $  & $ 0$\\ 
IRAS $09452+1330$  & $ -7.7$ & $  41932.79$  & $ 18000.74$ & $  4460.34$& CCB  & $  43 $  & $ 9$\\ 
IRAS $10131+3049$  & $ -5.1$ & $   2740.79$  & $   926.33$ & $   213.87$& CBB  & $  04 $  & $ 2$\\ 
IRAS $10491-2059$  & $ -3.6$ & $    919.06$  & $   349.97$ & $    77.38$& BBD  & $\,\,  4n^{*} $  & $ 0$\\ 
IRAS $12447+0425$  & $ -1.7$ & $    175.96$  & $    51.15$ & $    10.54$& BBD  & $  44 $  & $ 0$\\ 
IRAS $17049-2440$  & $ -3.3$ & $    719.06$  & $   392.97$ & $    93.88$& CBC  & $  42 $  & $ 9$\\    
IRAS $17119+0859$  & $ -2.4$ & $    324.24$  & $   228.96$ & $    31.16$& DCC  & $  28 $  & $ 9$\\ 
IRAS $17297+1747$  & $ -2.9$ & $    465.21$  & $   313.32$ & $    57.71$& BBF  & $  14 $  & $ 9$\\ 
IRAS $17360-3012$  & $ -1.3$ & $    213.51$  & $   264.36$ & $    59.27$& CFD  & $  42 $  & $ 9$\\ 
IRAS $17411-3154$  & $ -3.4$ & $   1480.09$  & $  2524.20$ & $  1160.89$& BBC  & $\,\,  3n^{*} $  & $ 9$\\ 
IRAS $18009-2019$  & $ -3.0$ & $    327.95$  & $   217.51$ & $    36.04$& BDD  & $  29 $  & $ 9$\\ 
IRAS $18040-0941$  & $ -2.1$ & $    182.34$  & $    65.62$ & $    16.22$& BCC  & $  44 $  & $ 9$\\ 
IRAS $18194-2708$  & $ -2.5$ & $    599.07$  & $   206.06$ & $    55.30$& FBD  & $  43 $  & $ 9$\\ 
IRAS $18240+2326$  & $ -2.7$ & $    620.16$  & $   345.47$ & $    69.67$& BBE  & $  42 $  & $ 9$\\ 
IRAS $18333+0533$  & $ -2.9$ & $    288.52$  & $   259.03$ & $    61.53$& BDD  & $ 42,4no^{*} $  & $ 9$\\ 
IRAS $18348-0526$  & $ -2.6$ & $    421.55$  & $   599.56$ & $   398.50$& BCE  & $\,\,  3n^{*} $  & $ 9$\\ 
IRAS $18349+1023$  & $ -3.5$ & $    594.84$  & $   242.10$ & $    51.50$& BCD  & $  26 $  & $ 9$\\ 
IRAS $18397+1738$  & $ -3.5$ & $    470.03$  & $   186.34$ & $    47.44$& BBD  & $  43 $  & $ 9$\\ 
IRAS $18398-0220$  & $ -3.3$ & $    466.38$  & $   186.65$ & $    40.78$& FDF  & $  42 $  & $ 9$\\ 
IRAS $18413+1354$  & $ -2.4$ & $    166.23$  & $   111.92$ & $    16.59$& CBD  & $  29 $  & $ 9$\\ 
IRAS $18560-2954$  & $ -3.2$ & $    525.14$  & $   251.45$ & $    49.69$& FBD  & $  27 $  & $ 9$\\ 
IRAS $19008+0726$  & $ -2.3$ & $    369.27$  & $   137.45$ & $    29.12$& BBD  & $  43 $  & $ 0$\\ 
IRAS $19059-2219$  & $ -2.4$ & $    231.57$  & $   159.48$ & $    26.54$& CCD  & $  28 $  & $ 9$\\ 
IRAS $19093-3256$  & $ -2.8$ & $    256.29$  & $   156.32$ & $    25.69$& BCD  & $  28 $  & $ 9$\\ 
IRAS $19126-0708$  & $ -3.7$ & $   1195.75$  & $   493.44$ & $    86.06$& BBD  & $  22 $  & $ 9$\\ 
IRAS $19175-0807$  & $ -2.3$ & $    342.20$  & $   151.16$ & $    37.83$& BBD  & $  43 $  & $ 9$\\ 
IRAS $19321+2757$  & $ -2.8$ & $    283.93$  & $   132.23$ & $    30.76$& CBC  & $  43 $  & $ 9$\\ 
IRAS $20077-0625$  & $ -3.7$ & $   1116.52$  & $   834.80$ & $   170.94$& BBB  & $  23 $  & $ 6$\\ 
IRAS $20396+4757$  & $ -3.5$ & $    534.40$  & $   176.14$ & $    38.25$& BBD  & $  44 $  & $ 1$\\ 
IRAS $20440-0105$  & $ -1.9$ & $    159.03$  & $    86.10$ & $    13.38$& BBE  & $  27 $  & $ 9$\\ 
IRAS $20570+2714$  & $ -2.5$ & $    239.33$  & $   114.64$ & $    24.89$& BCC  & $  42 $  & $ 9$\\ 
IRAS $21032-0024$  & $ -2.4$ & $    242.68$  & $    86.58$ & $    17.40$& BBC  & $  45 $  & $ 9$\\ 
IRAS $21286+1055$  & $ -2.3$ & $    130.17$  & $    80.86$ & $    14.31$& ABC  & $  26 $  & $ 2$\\ 
IRAS $21320+3850$  & $ -2.0$ & $    210.50$  & $    71.32$ & $    16.03$& CBC  & $  44 $  & $ 0$\\ 
IRAS $21456+6422$  & $ -1.9$ & $    133.45$  & $    78.83$ & $    12.76$& CBB  & $  28 $  & $ 9$\\ 
IRAS $23166+1655$  & $ -3.3$ & $    741.06$  & $   659.45$ & $   203.51$& BBC  & $  02 $  & $ 0$\\ 
IRAS $23320+4316$  & $ -3.5$ & $    839.03$  & $   364.60$ & $    87.96$& BBD  & $  42 $  & $ 0$\\ 
IRAS $23496+6131$  & $ -2.2$ & $    303.54$  & $   194.45$ & $    35.38$& AAC  & $  27 $  & $ 2$\\ 
RAFGL $1406     $  & $ -3.0$ & $          $  & $         $ & $         $&      & $     $  & $  $\\ 
RAFGL $2688     $  & $ -2.6$ & $          $  & $         $ & $         $&      & $     $  & $  $\\ 
\end{tabular}         
}                     
\begin{flushleft}                 
\small{All stars have LRS classifications taken from \emph{IRAS Atlas of Low Resolution
Spectra}\\ 
(Olnon \& Raimond, 1986), except in the cases marked ``${*}$'', which
were taken from Loup et al. (1993)\\
\vspace{0.3cm}
${\dagger}$-Flux Density Uncertainty, as explained in the IRAS Explanatory
Supplement.} 
\end{flushleft}
\end{table}
\twocolumn
\clearpage
\onecolumn
\begin{table}
\caption{Variability data}
\centering
\footnotesize{
\begin{tabular}{lclccccc}
\multicolumn{1}{c}{$\bf {NAME}$} &
\multicolumn{1}{c}{$\bf {P\,(days)}$}&
\multicolumn{1}{c}{$\bf {P_{lit}\,(days)}$}&
\multicolumn{1}{c}{$\bf {VAR}$}&
\multicolumn{1}{|c|}{$\bf {\Delta{J}}$}&
\multicolumn{1}{|c|}{$\bf {\Delta{H}}$}&
\multicolumn{1}{|c|}{$\bf {\Delta{K}}$}&
\multicolumn{1}{|c|}{$\bf {\Delta{L}}$}\\ \hline           
IRAS $00042+1219$ & $    $ &  $\hspace{0.7cm} 750^{5}      $  &   & $     $ &  $     $ &  $     $ &  $     $\\ 
IRAS $01037+1219$ & $ 645$ &  $\hspace{0.7cm} 652^{1,3,5,6}$  & Y & $ 2.87$ &  $ 2.15$ &  $ 1.69$ &  $ 1.34$\\ 
IRAS $01159+7220$ & $    $ &  $\hspace{0.7cm} 612^{5}      $  &   & $     $ &  $     $ &  $     $ &  $     $\\ 
IRAS $02270-2619$ & $ 389$ &  $\hspace{0.7cm} 387^{2,4,5}  $  & Y & $ 1.06$ &  $ 0.88$ &  $ 0.68$ &  $ 0.55$\\ 
IRAS $02316+6455$ & $    $ &  $\hspace{0.7cm} 534^{6}      $  &   & $     $ &  $     $ &  $     $ &  $     $\\ 
IRAS $02351-2711$ & $ 480$ &  $\hspace{0.7cm} 475^{1,3}    $  & Y & $ 1.56$ &  $ 1.24$ &  $ 0.92$ &  $ 0.70$\\ 
IRAS $03229+4721$ & $    $ &  $\hspace{0.7cm} 537^{5,6}    $  &   & $     $ &  $     $ &  $     $ &  $     $\\ 
IRAS $03507+1115$ & $ 469$ &  $\hspace{0.7cm} 464^{1,3,5,6}$  & Y & $ 1.96$ &  $ 1.45$ &  $ 1.08$ &  $ 0.94$\\ 
IRAS $04307+6210$ & $    $ &  $\hspace{0.7cm}              $  &   & $     $ &  $     $ &  $     $ &  $     $\\ 
IRAS $04566+5606$ & $    $ &  $\hspace{0.7cm} 557^{5}      $  &   & $     $ &  $     $ &  $     $ &  $     $\\ 
IRAS $05073+5248$ & $    $ &  $\hspace{0.7cm} 629^{5,6}    $  &   & $     $ &  $     $ &  $     $ &  $     $\\ 
IRAS $05411+6957$ & $    $ &  $\hspace{0.7cm}              $  &   & $     $ &  $     $ &  $     $ &  $     $\\ 
IRAS $05559+7430$ & $    $ &  $\hspace{0.7cm} 523^{5}      $  &   & $     $ &  $     $ &  $     $ &  $     $\\ 
IRAS $06176-1036$ & $    $ &  $\hspace{0.7cm}              $  &   & $     $ &  $     $ &  $     $ &  $     $\\ 
IRAS $06300+6058$ & $    $ &  $\hspace{0.7cm}              $  &   & $     $ &  $     $ &  $     $ &  $     $\\ 
IRAS $06500+0829$ & $    $ &  $\hspace{0.7cm} 527^{5}      $  &   & $     $ &  $     $ &  $     $ &  $     $\\ 
IRAS $08088-3243$ & $ 570$ &  $\hspace{0.7cm} 571^{2}      $  & Y & $ 2.09$ &  $ 1.74$ &  $ 1.47$ &  $ 1.19$\\ 
IRAS $09116-2439$ & $ 670$ &  $\hspace{0.7cm}              $  & Y & $     $ &  $ 2.02$ &  $ 2.03$ &  $ 1.54$\\ 
IRAS $09429-2148$ & $ 640$ &  $\hspace{0.7cm} 643^{3,5}    $  & Y & $ 1.82$ &  $ 1.56$ &  $ 1.39$ &  $ 1.35$\\ 
IRAS $09452+1330$ & $ 652$ &  $\hspace{0.7cm} 635^{2,5,6}  $  & Y & $ 2.14$ &  $ 2.20$ &  $ 2.11$ &  $ 1.81$\\ 
IRAS $10131+3049$ & $    $ &  $\hspace{0.7cm} 640^{5}      $  & Y & $     $ &  $     $ &  $     $ &  $     $\\ 
IRAS $10491-2059$ & $ 530$ &  $\hspace{0.7cm} 530^{2,5}    $  & Y & $     $ &  $     $ &  $     $ &  $     $\\ 
IRAS $12447+0425$ & $ 441$ &  $\hspace{0.7cm} 433^{5}      $  & Y & $ 1.40$ &  $ 1.22$ &  $ 0.99$ &  $ 0.91$\\ 
IRAS $17049-2440$ & $ 775$ &  $\hspace{0.7cm}              $  & Y & $     $ &  $ 2.25$ &  $ 2.13$ &  $ 1.81$\\ 
IRAS $17119+0859$ & $    $ &  $\hspace{0.7cm}              $  & Y & $     $ &  $     $ &  $     $ &  $     $\\ 
IRAS $17297+1747$ & $    $ &  $\hspace{0.7cm} 520^{6}      $  & Y & $     $ &  $     $ &  $     $ &  $     $\\ 
IRAS $17360-3012$ & $(1120)$ &  $\hspace{0.6cm}1150^{3}    $  & Y & $ 1.90$ &  $ 1.99$ &  $ 1.77$ &  $ 1.55$\\ 
IRAS $17411-3154$ & $1440$ &  $\hspace{0.7cm}              $  & Y & $     $ &  $     $ &  $     $ &  $ 1.11$\\ 
IRAS $18009-2019$ & $    $ &  $\hspace{0.7cm}              $  &   & $     $ &  $     $ &  $     $ &  $     $\\ 
IRAS $18040-0941$ & $    $ &  $\hspace{0.7cm}              $  & Y & $     $ &  $     $ &  $     $ &  $     $\\ 
IRAS $18194-2708$ & $ 690$ &  $\hspace{0.7cm}              $  & Y & $ 1.66$ &  $ 1.57$ &  $ 1.35$ &  $ 1.20$\\ 
IRAS $18240+2326$ & $    $ &  $\hspace{0.7cm}              $  & Y & $     $ &  $     $ &  $     $ &  $     $\\ 
IRAS $18333+0533$ & $ 795$ &  $\hspace{0.7cm}              $  & Y & $ 3.00$ &  $ 2.08$ &  $ 1.48$ &  $ 1.17$\\ 
IRAS $18348-0526$ & $(1500)$ &  $\hspace{0.6cm}1570^{3,5,6}$  & Y & $     $ &  $     $ &  $ 3.19$ &  $ 2.11$\\ 
IRAS $18349+1023$ & $    $ &  $\hspace{0.7cm} 500^{6}      $  & Y & $     $ &  $     $ &  $     $ &  $     $\\ 
IRAS $18397+1738$ & $    $ &  $\hspace{0.7cm} 511^{6}      $  & Y & $     $ &  $     $ &  $     $ &  $     $\\ 
IRAS $18398-0220$ & $ 600$ &  $\hspace{0.7cm}              $  & Y & $ 1.25$ &  $ 1.09$ &  $  .93$ &  $ 0.75$\\ 
IRAS $18413+1354$ & $    $ &  $\hspace{0.7cm} 590^{6}      $  &   & $     $ &  $     $ &  $     $ &  $     $\\ 
IRAS $18560-2954$ & $ 575$ &  $\hspace{0.7cm}              $  & Y & $ 2.32$ &  $ 1.81$ &  $ 1.35$ &  $ 1.00$\\ 
IRAS $19008+0726$ & $    $ &  $\hspace{0.7cm} 577^{2}      $  & Y & $     $ &  $     $ &  $     $ &  $     $\\ 
IRAS $19059-2219$ & $    $ &  $\hspace{0.7cm} 511^{3,5}    $  & Y & $     $ &  $     $ &  $     $ &  $     $\\ 
IRAS $19093-3256$ & $ 380$ &  $\hspace{0.7cm} 372^{5}      $  & Y & $ 1.21$ &  $ 0.98$ &  $  .78$ &  $ 0.73$\\ 
IRAS $19126-0708$ & $    $ &  $\hspace{0.7cm} 490^{5}      $  & Y & $     $ &  $     $ &  $     $ &  $     $\\ 
IRAS $19175-0807$ & $    $ &  $\hspace{0.7cm} 676^{2}      $  & Y & $     $ &  $     $ &  $     $ &  $     $\\ 
IRAS $19321+2757$ & $    $ &  $\hspace{0.7cm} 625^{6}      $  & Y & $     $ &  $     $ &  $     $ &  $     $\\ 
IRAS $20077-0625$ & $ 675$ &  $\hspace{0.7cm} 660^{3,5}    $  & Y & $ 2.50$ &  $ 1.88$ &  $ 1.45$ &  $ 1.22$\\ 
IRAS $20396+4757$ & $    $ &  $\hspace{0.7cm} 421^{5}      $  &   & $     $ &  $     $ &  $     $ &  $     $\\ 
IRAS $20440-0105$ & $    $ &  $\hspace{0.7cm}              $  & Y & $     $ &  $     $ &  $     $ &  $     $\\ 
IRAS $20570+2714$ & $    $ &  $\hspace{0.7cm} 750^{6}      $  & Y & $     $ &  $     $ &  $     $ &  $     $\\ 
IRAS $21032-0024$ & $    $ &  $\hspace{0.7cm} 454^{5}      $  & Y & $     $ &  $     $ &  $     $ &  $     $\\ 
IRAS $21286+1055$ & $    $ &  $\hspace{0.7cm} 457^{5}      $  & Y & $     $ &  $     $ &  $     $ &  $     $\\ 
IRAS $21320+3850$ & $    $ &  $\hspace{0.7cm} 470^{5}      $  &   & $     $ &  $     $ &  $     $ &  $     $\\ 
IRAS $21456+6422$ & $    $ &  $\hspace{0.7cm} 622^{5}      $  &   & $     $ &  $     $ &  $     $ &  $     $\\ 
IRAS $23166+1655$ & $    $ &  $\hspace{0.7cm} 700^{1,2}    $  & Y & $     $ &  $     $ &  $     $ &  $     $\\ 
IRAS $23320+4316$ & $    $ &  $\hspace{0.7cm} 620^{6}      $  &   & $     $ &  $     $ &  $     $ &  $     $\\ 
IRAS $23496+6131$ & $    $ &  $\hspace{0.7cm} 419^{6}      $  &   & $     $ &  $     $ &  $     $ &  $     $\\ 
RAFGL $1406     $ & $    $ &  $\hspace{0.7cm}              $  &   & $     $ &  $     $ &  $     $ &  $     $\\ 
RAFGL $2688     $ & $    $ &  $\hspace{0.7cm}              $  &   & $     $ &  $     $ &  $     $ &  $     $\\ 
\end{tabular} 
}                      
\begin{flushleft}                   
\small{                             
REFERENCES-(1) Whitelock et al. (1994); (2) Le Bertre (1993a); (3) Le Bertre(1993b) 
; (4) Whitelock et al. (1997a); (5) Kholopov, P.N. (1985)(GCVS); 
(6) Jones et al. (1990).\\
}                                   
\end{flushleft}                     
\end{table}   
\twocolumn
\clearpage
\onecolumn
\begin{figure}
\caption{Phased Light-curves}
\hspace{2cm}
\epsfxsize=14.5cm
\epsffile{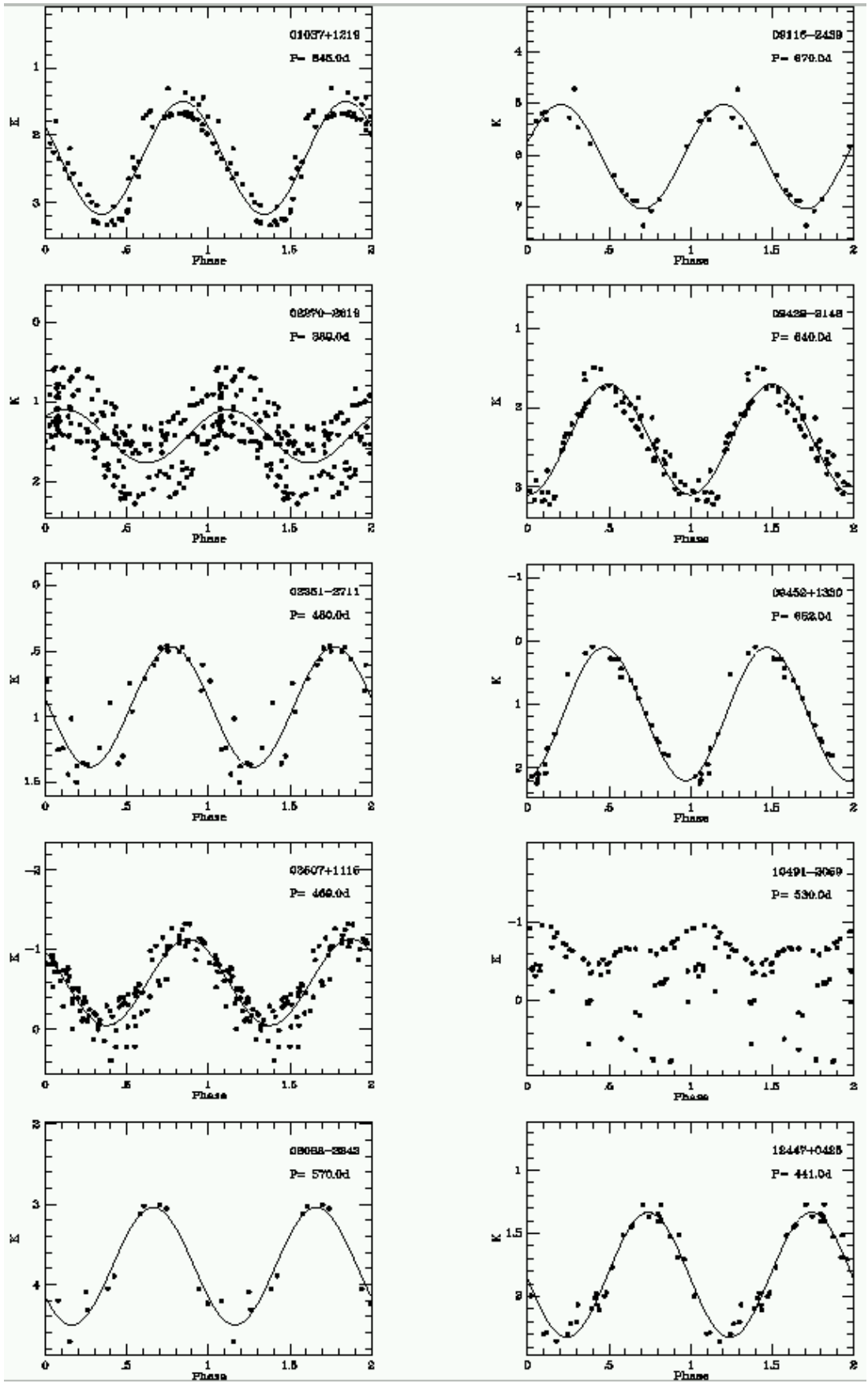}
\end{figure}
\clearpage

\setcounter{figure}{0}
\begin{figure}
\caption{Phased Lightcurves.....continued.}
\hspace{2cm}
\epsfxsize=14.5cm
\epsffile{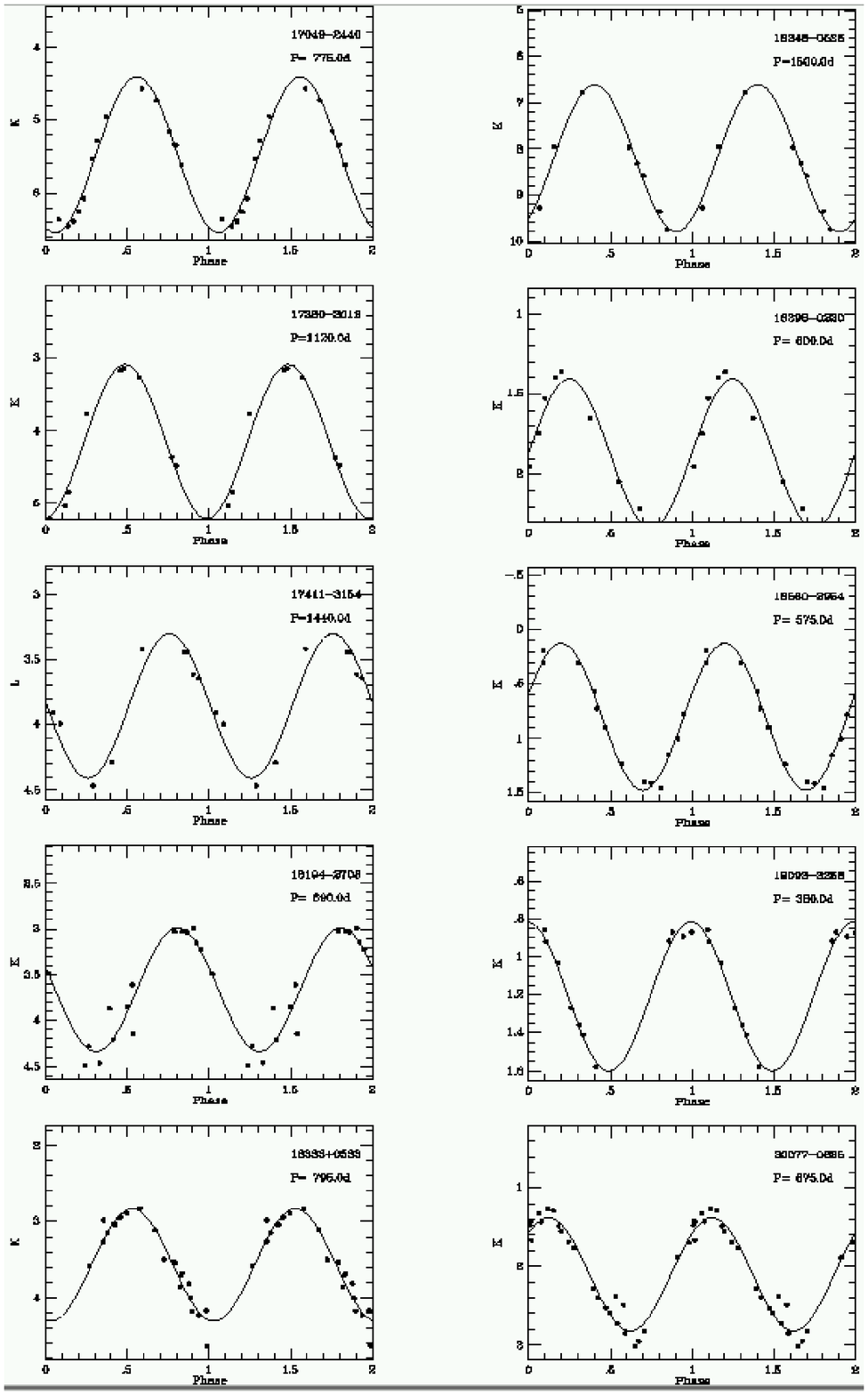}
\end{figure}
\twocolumn
\normalsize

\begin{equation}
M = \frac{{\Delta}M}{2}{\sin ({\phi}-{\phi}_{0})} + M_{av}
\end{equation}
where M is the magnitude in a particular band, ${\Delta}M$ the peak to peak
amplitude, ${\phi}$ related to the phase, $p$, by ${\phi}=2{\pi}p$ and 
${\phi}_{0}$ the phase angle at zero phase. $M_{av}$ is the mean magnitude.

One immediate observation from most of the phased light-curves is that at a
given phase there is a range of magnitudes, rather than one sharply defined
magnitude. This is not, in general, due to observational error, but is
contributed to by various effects.  First, Miras and OH/IR stars are erratic
pulsators in the sense that their light curves do not exactly repeat from
one cycle to the next at visual or infrared wavelengths (Koen \& Lombard
1995; Whitelock, Marang \& Feast 2000) Secondly, some stars, particularly
carbon stars go through faint phases from time to time due to variable dust
obscuration, e.g. $02270-2619$ (R For, see Whitelock et al 1997a). Thirdly, a
very small number of stars show temporal or long term changes in luminosity,
amplitude and/or period, possibly due to evolutionary effects (Whitelock
1999 and references therein).

Many of the light curves in Fig.~1 show clear departures from simple
sinusoids and would be better fitted by the inclusion of one or two
harmonics, in particular, $09429-2148$ and $12447+0425$. The most strikingly
odd curve is that of $10491-2059$ (V Hya) which is discussed below.

\subsubsection{V Hya}
 V Hya is a well studied carbon star and semi-regular variable. It has a
bipolar circumstellar outflow, and models invoke rotation of the central
star or the presence of a binary companion to explain bipolar outflows
(Barnbaum, Morris \& Kahane 1995).

 Fig.~2a illustrates the light curve for V Hya using $K$ magnitudes
measured from SAAO, including material from Lloyd Evans (1997 and private
communication) together with the data from Table 2. In addition to the
regular pulsation V Hya experienced two faint phases around JD\,244\,3500
and 244\,9500. A Fourier analysis of the $K$ data when the star is bright,
between JD\,244\,6000 and 244\,8250 provide a period of 530 days, identical
to the value given in the GCVS. Fig.~2b shows the data phased at this
period and illustrates the double peak caused by harmonics in the light
curve.
\begin{figure}
\centering
\epsfxsize=7.0cm
\epsffile{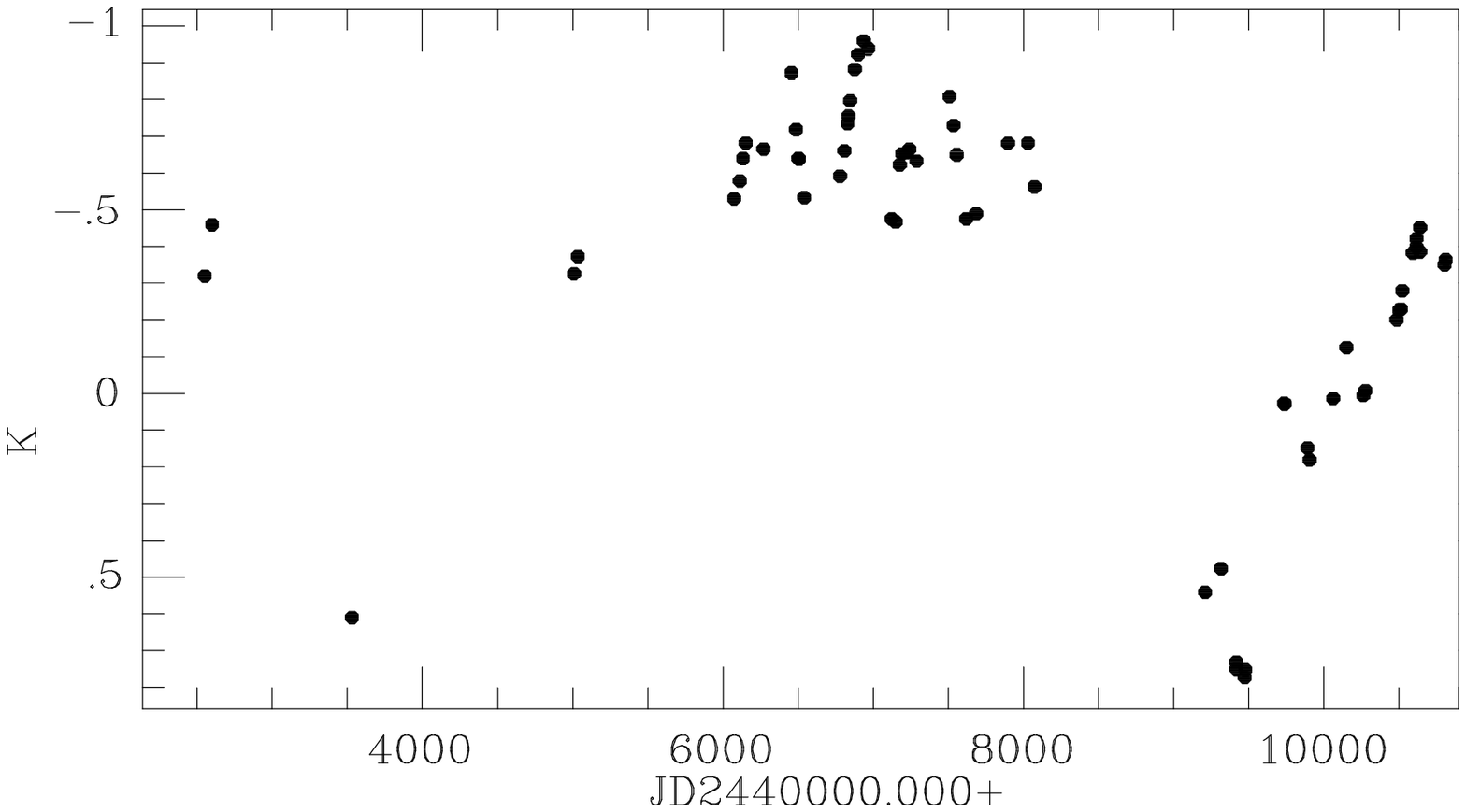}  
\epsfxsize=7.0cm
\epsffile{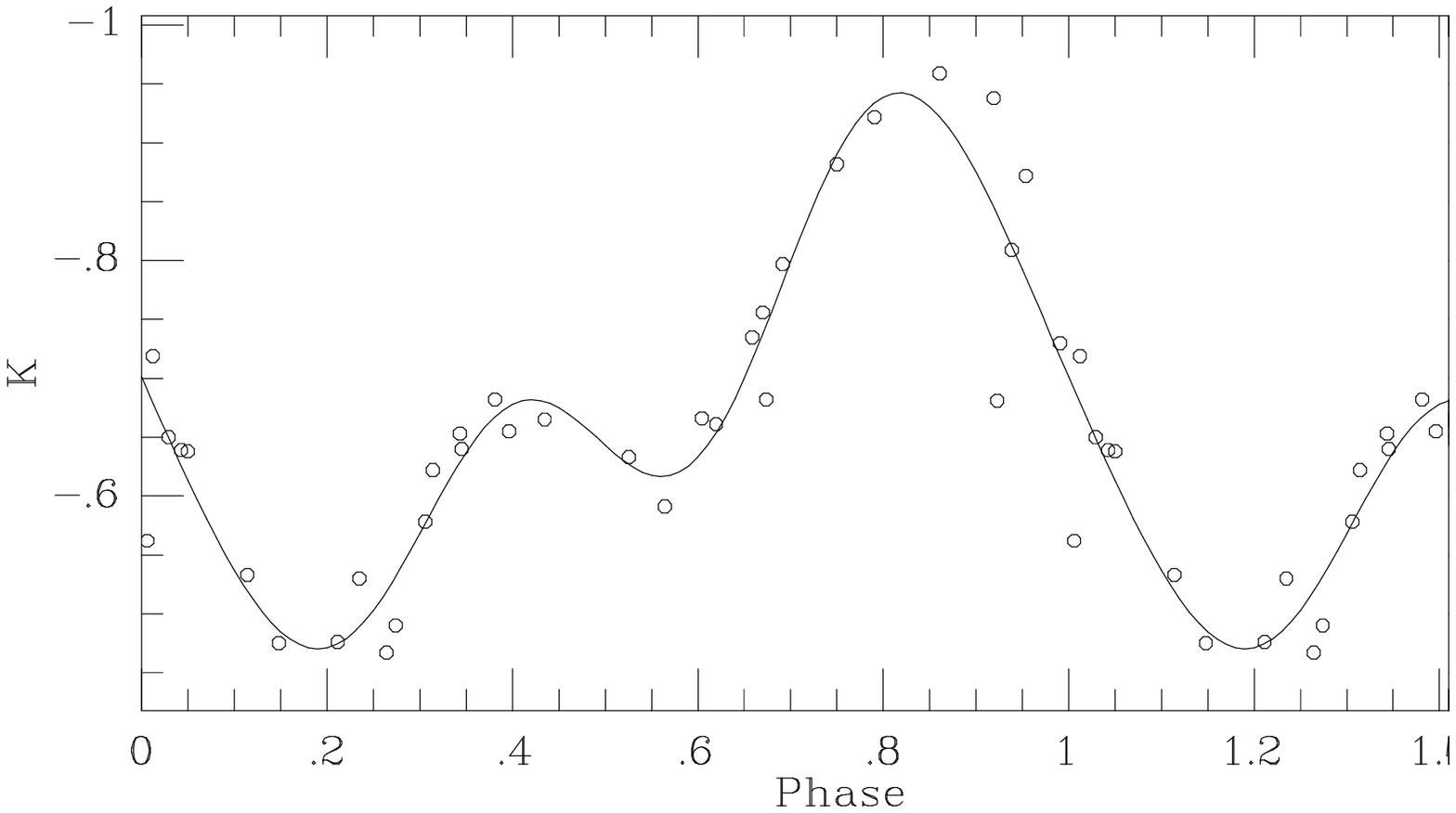}
\caption{(a) $K$ light curve of V Hya. (b) data between JD\,244\,6000 and 
244\,8250 phased at 530 days.}
\end{figure}

 Knapp et al. (1999) analysed 3260 AAVSO visual magnitude estimates covering
almost 35 years. They confirm and refine the very long period found by
Mayall (1965) at $6160\pm400$ days. The two faint phases seen in Fig.~2a are
consistent with this long period. The amplitudes of these 6160 day
variations are at least 2.4, 2.1, 1.7 and 0.7 mag, at $JHKL$, respectively,
while $\Delta V \sim 3.5$ mag (Knapp et al. 1999).

Barnbaum et al. (1995) found evidence for rapid rotation in V Hya, using
spectral line broadening analysis. They suggest that the long term variation
can be attributed to a coupling between the radial pulsation, with period of
530 days, and the rotation of the star with a very similar period. It is not
clear how this interaction would give rise to the very long period without
strongly modulating the shorter period, or why it should result in the
observed reddening of the star during the faint phases.

Lloyd Evans (1997) suggested that the long period may be the result of
obscuration events like those observed in R For. Although the colours would
be consistent with this, the regularity of the long period (Knapp et al.
1999) is quite unlike the effect observed in R For and this explanation
therefore seems unlikely.

Knapp et al. (1999) on the other hand, suggest that the long-period dimming
of this star is due to a thick dust cloud orbiting the star and attached to
a binary companion. Indeed, there is evidence for large numbers of late-type
binaries in the LMC (Wood et al. 1999), so this may not be rare occurrence. 
The colour variations of V~Hya are less extreme than those expected from
optically thin interstellar dust, but they could be explained by large
particles.

\subsection{Period Distributions}
 The period distributions for the oxygen- and carbon-rich stars are plotted
in Fig.~3. Most of the stars in the sample have periods between 500 and 700
days and almost all have periods between 300 and 900 days. There are only
three extreme OH/IR stars with periods in excess of 1000 days. Fig.~3 can be
compared with Whitelock et al. (1994) fig.~8 which shows the period
distribution of IRAS selected Miras in the South Galactic Cap. Most of the
Cap sample have $150<P<450$ days, i.e. shorter than the sample under
discussion, but longer than optically selected Miras.  Longer period Miras
have more massive progenitors and are therefore lower scale heights. It is
not surprising to find more of them in the solar neighbourhood than in the
Galactic Cap.
 
A comparison of the period distributions of the oxygen- and carbon-rich
stars is instructive. There are no carbon-rich stars with periods longer
than 1000 days. From Fig.~3 it can be seen that the maximum of the
carbon-star period distribution seems to be slightly shifted to longer
periods than that of the oxygen-rich stars. A Kolmogorov-Smirnov (KS) test,
when applied to these two distributions for periods less than 1000 days,
yields a probability that the two distributions come from the same parent
population of $0.40$. This probability ($>0.05$) implies that there is no
significant difference in the period distributions for the carbon- and
oxygen-rich stars.

The KS test cannot be used if the very long period stars are included since
it is not very sensitive to differences in the tails of the distributions,
as is well known. In this case the F-test and then the appropriate version
of the student t-test is applied. The F-test gives a probability that the
variances of the two distributions are the same of $2.5 \times 10^{-5}$,
i.e. very unlikely, which is obvious from Fig.~3.  The appropriate version
of the t-test, which takes differences in variances into account, gives a
probability that the two distributions have the same mean of $0.24$.  This
probability ($>0.05$) implies that there is no statistical evidence for any
difference in the mean of the two distributions.

\begin{figure}
\centering
\epsfxsize=7.0cm
\epsffile{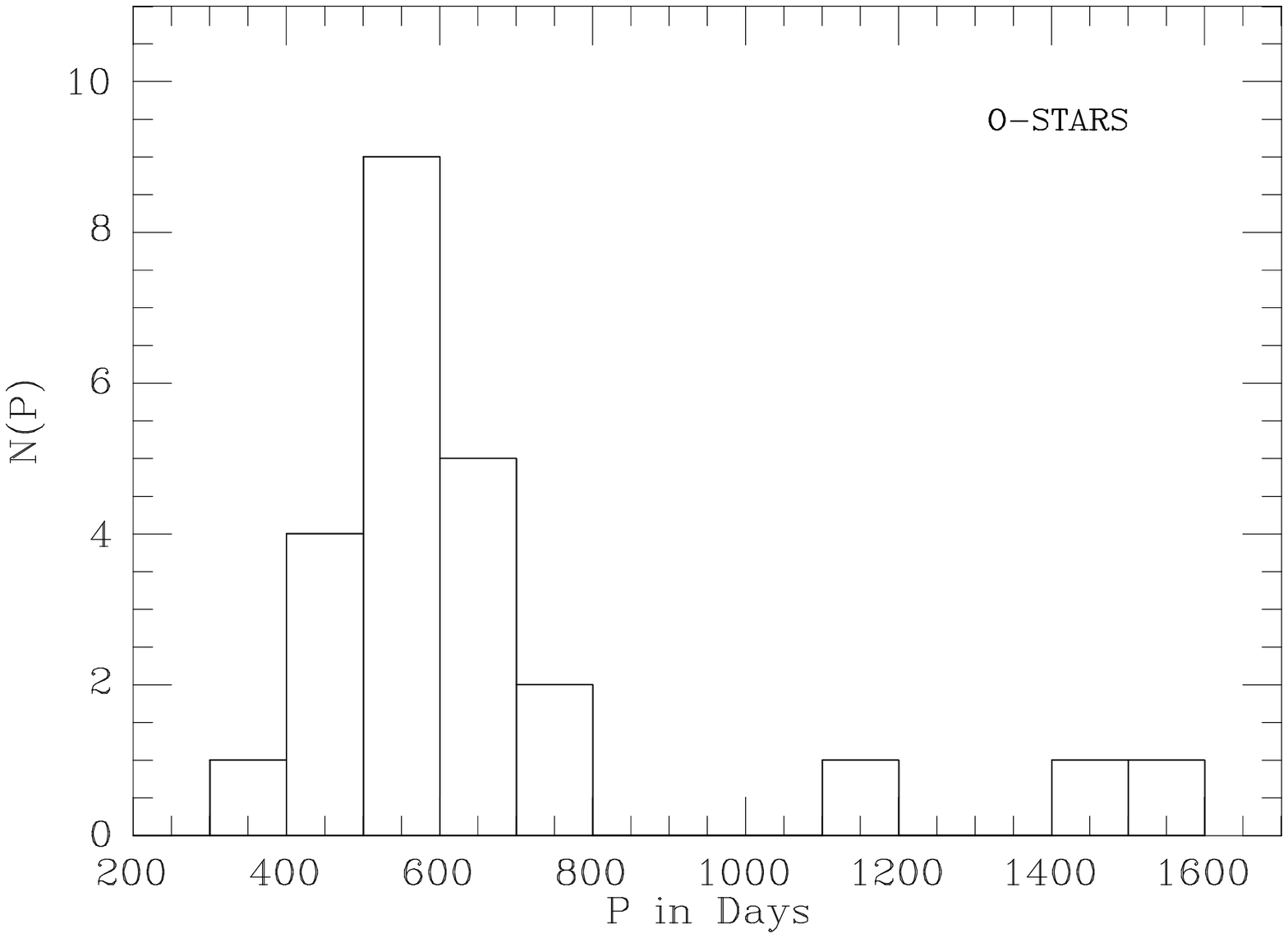}
\epsfxsize=7.0cm
\epsffile{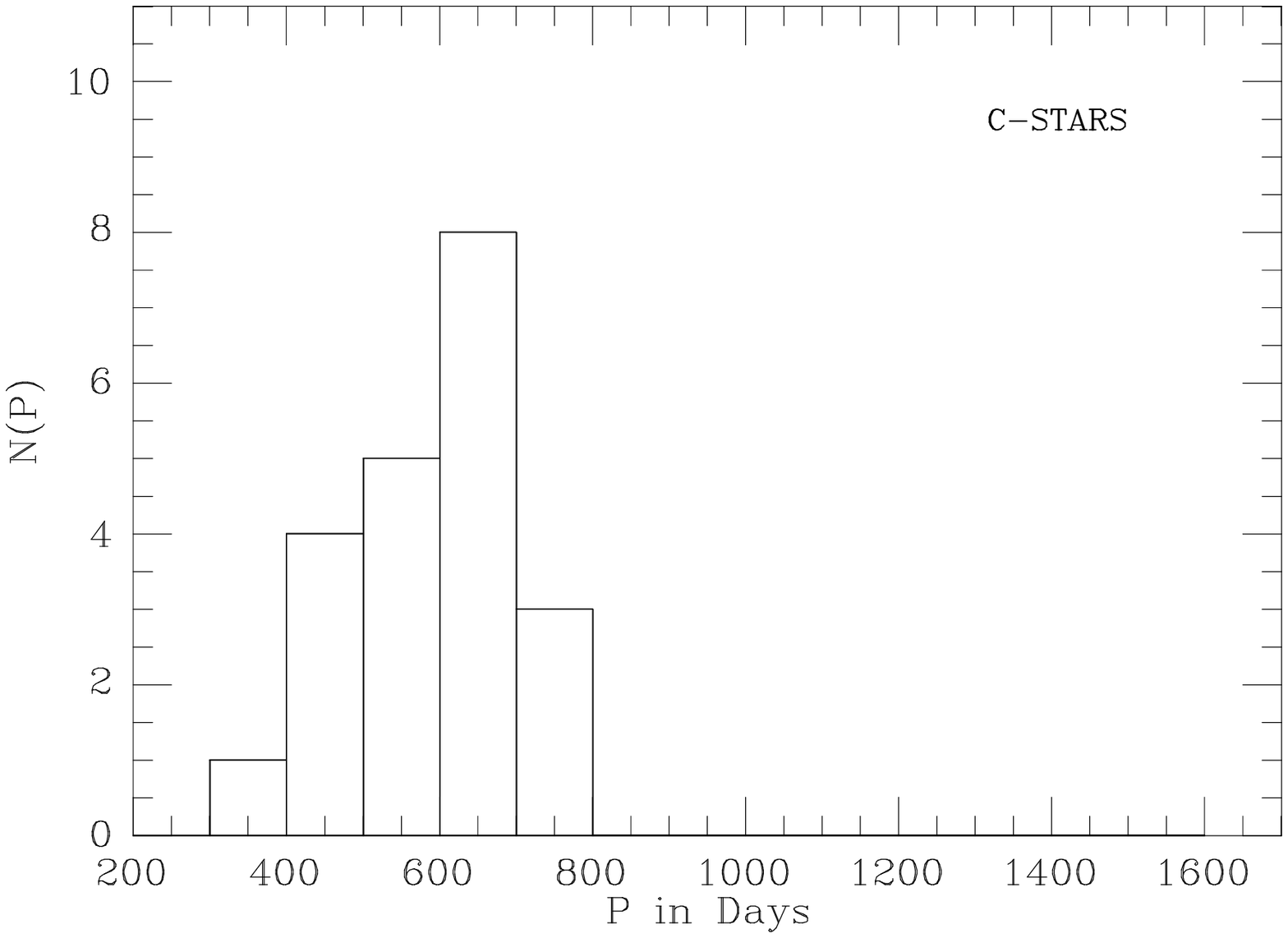}
\caption{Period distributions}
\end{figure}
\subsection{Amplitudes}

Table~5 lists the stars for which periods were determined together with
their amplitudes, from equation~1, at $JHKL$. These amplitudes are plotted
against period in Fig.~4, where a clear correlation can be seen at all
wavelengths.

\begin{figure}
\centering
\epsfxsize=8.0cm
\epsffile{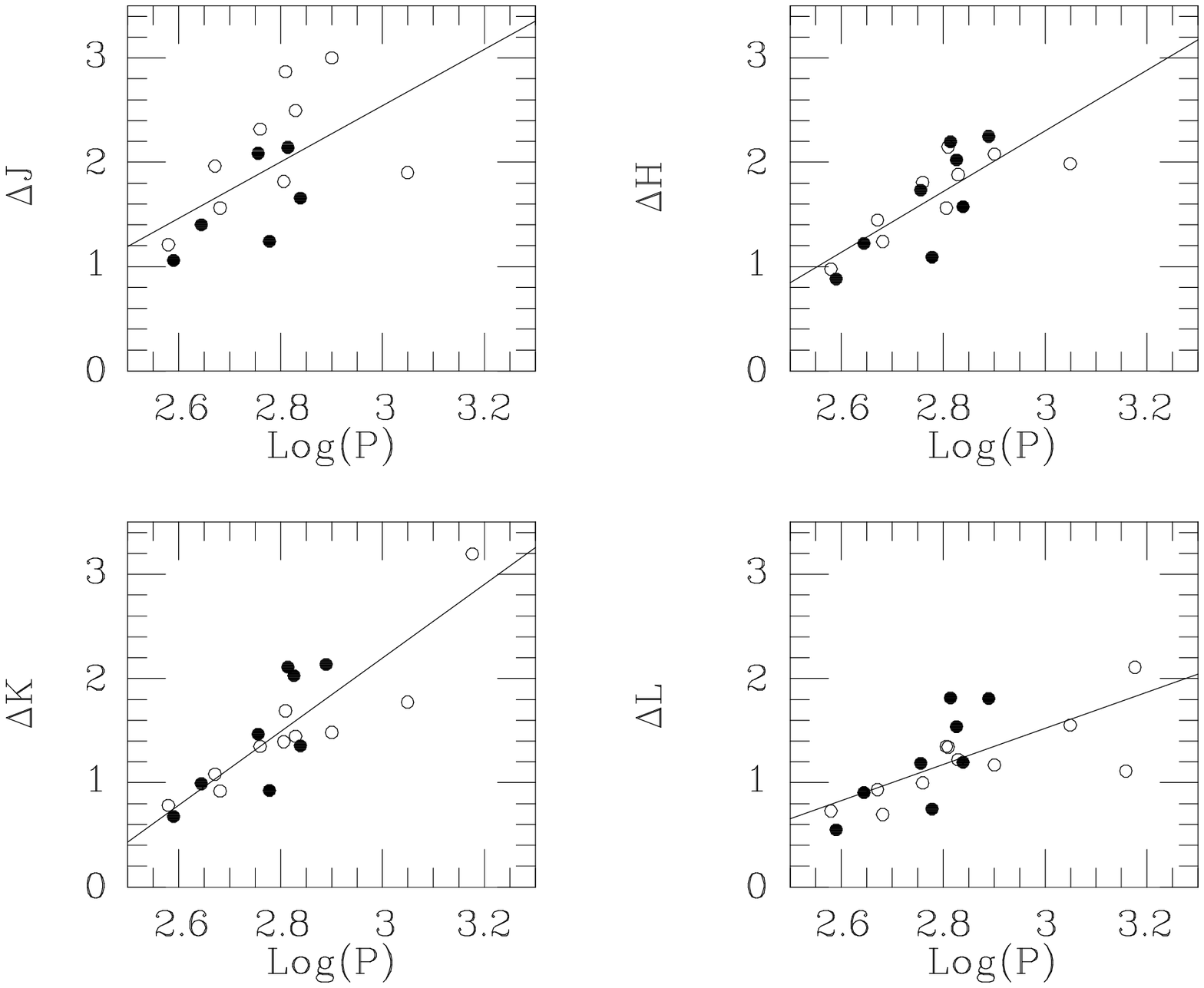}
\caption{Light-Amplitude vs. pulsation period, the lines show least squares 
fits; filled and open circles represent carbon- and oxygen-rich stars, 
respectively.}
\end{figure}

The amplitude clearly decreases with wavelength (see also Feast et al.
1982). Quantitatively, the average amplitudes, relative to that at $J$ are,
for $H$, $K$ and $L$ respectively, $0.84\pm0.1$, $0.69\pm0.14$ and
$0.59\pm0.14$. The linear correlation coefficient between the average
relative amplitude and wavelength is $-0.93$ with a correlation probability
of $Prob=0.07$ (using the relative amplitude at $J$ of $1$). This
probability is considered as marginally significant. The KS test gives the
probability for the null hypothesis that the light amplitude distributions
for the carbon- and oxygen-rich stars (with $P<1000$\,days) came from the
same parent population at $J$, $H$, $K$ and $L$ of 0.25, 0.92, 0.51 and 0.52
respectively. Hence there is no statistical reason to suspect that these
carbon- and oxygen-rich stars have significantly different light-amplitudes
in the near-infrared.

\section{Colours}
\subsection{Average Colours}

  Average near-infrared colours, i.e. consisting of $JHKL$ magnitudes only,
were calculated for each star using the magnitudes at the dates of minimum 
and maximum $K$-light. This average is compared to a Fourier-fit colour in 
Appendix A. The two values are quite comparable with mean differences for 
all the near-infrared colours not exceeding 0.1 mag. Note that due to the
very small phase coverage of  $17411-3154$ in the $K$-band, the derived
$K-L$ colour of this star may not be a good representative of the average
$K-L$ colour.  

The IRAS colours were obtained using the data in Table~4, where the
colour-corrected IRAS fluxes were converted to magnitudes using the zero
points listed in the IRAS-ES. Colours consisting of $JHKL$ and IRAS
photometry, were calculated by combining average magnitudes in the $JHKL$
bands (Table~3) with the IRAS magnitudes.

For the northern sources and the three non-IRAS sources the average $K$
magnitudes were taken from Table~3, to calculate the $K-[12]$ colour. For
the non-IRAS (carbon-rich) sources the average $12\mu$m magnitudes were
estimated from the average $[11]-[12]=0.1\pm0.3$ colour for the carbon-rich
sources and the 11$\,\mu$m magnitudes listed in Table~4. All the derived
infrared colours are given in Table~6.

\onecolumn
\begin{table}
\centering
\caption{Infrared Colours}
\footnotesize{
\begin{tabular}{lrrrrrrr}
\multicolumn{1}{c}{$\bf NAME$}&
\multicolumn{1}{c}{$\bf J-H$}&
\multicolumn{1}{c}{$\bf H-K$} &
\multicolumn{1}{c}{$\bf J-K$} &
\multicolumn{1}{c}{$\bf K-L$}&
\multicolumn{1}{c}{$\bf K-[12]$}&
\multicolumn{1}{c}{$\bf [12]-[25]$}&
\multicolumn{1}{c}{$\bf [25]-[60]$}\\ \hline           
IRAS $00042+4248$&$      $ &  $      $ &  $      $ &  $      $&$ 5.37 $&$ 1.06 $ & $ 0.04  $\\   
IRAS $01037+1219$&$ 3.51 $ &  $ 2.26 $ &  $ 5.77 $ &  $ 2.48 $&$ 6.26 $&$ 1.21 $ & $ 0.25  $\\   
IRAS $01159+7220$&$      $ &  $      $ &  $      $ &  $      $&$ 4.65 $&$ 0.96 $ & $-0.22  $\\   
IRAS $02270-2619$&$ 1.71 $ &  $ 1.33 $ &  $ 3.04 $ &  $ 1.47 $&$ 3.54 $&$ 0.19 $ & $ 0.24  $\\   
IRAS $02316+6455$&$      $ &  $      $ &  $      $ &  $      $&$ 5.87 $&$ 1.07 $ & $-0.16  $\\   
IRAS $02351-2711$&$ 1.25 $ &  $ 0.69 $ &  $ 1.94 $ &  $ 0.79 $&$ 3.66 $&$ 0.86 $ & $-0.32  $\\   
IRAS $03229+4721$&$      $ &  $      $ &  $      $ &  $      $&$ 4.60 $&$ 0.42 $ & $ 0.18  $\\   
IRAS $03507+1115$&$ 1.96 $ &  $ 1.16 $ &  $ 3.12 $ &  $ 1.32 $&$ 4.81 $&$ 0.68 $ & $-0.27  $\\   
IRAS $04307+6210$&$      $ &  $      $ &  $      $ &  $      $&$ 3.90 $&$ 0.42 $ & $ 0.12  $\\   
IRAS $04566+5606$&$      $ &  $      $ &  $      $ &  $      $&$ 3.69 $&$ 0.45 $ & $ 0.23  $\\   
IRAS $05073+5248$&$      $ &  $      $ &  $      $ &  $      $&$ 5.22 $&$ 1.56 $ & $ 0.41  $\\   
IRAS $05411+6957$&$      $ &  $      $ &  $      $ &  $      $&$ 4.30 $&$ 0.84 $ & $-0.27  $\\   
IRAS $05559+7430$&$      $ &  $      $ &  $      $ &  $      $&$ 3.36 $&$ 0.79 $ & $-0.03  $\\   
IRAS $06176-1036$&$ 1.66 $ &  $ 1.58 $ &  $ 3.24 $ &  $ 2.11 $&$ 6.43 $&$ 1.40 $ & $ 0.78  $\\   
IRAS $06300+6058$&$      $ &  $      $ &  $      $ &  $      $&$ 3.48 $&$ 1.06 $ & $ 0.22  $\\   
IRAS $06500+0829$&$      $ &  $      $ &  $      $ &  $      $&$ 4.34 $&$ 0.82 $ & $ 0.54  $\\   
IRAS $08088-3243$&$ 2.59 $ &  $ 2.12 $ &  $ 4.71 $ &  $ 2.62 $&$ 6.37 $&$ 0.60 $ & $ 0.18  $\\   
IRAS $09116-2439$&$      $ &  $ 3.75 $ &  $      $ &  $ 3.56 $&$ 9.40 $&$ 0.78 $ & $ 0.20  $\\   
IRAS $09429-2148$&$ 1.86 $ &  $ 1.34 $ &  $ 3.19 $ &  $ 1.62 $&$ 5.39 $&$ 1.30 $ & $-0.16  $\\   
IRAS $09452+1330$&$ 3.20 $ &  $ 2.82 $ &  $ 6.02 $ &  $ 3.74 $&$ 9.09 $&$ 0.64 $ & $ 0.37  $\\   
IRAS $10131+3049$&$ 2.70 $ &  $ 2.15 $ &  $ 4.85 $ &  $ 2.61 $&$ 6.24 $&$ 0.38 $ & $ 0.29  $\\   
IRAS $10491-2059$&$ 1.79 $ &  $ 1.27 $ &  $ 3.06 $ &  $ 1.55 $&$ 3.69 $&$ 0.51 $ & $ 0.24  $\\   
IRAS $12447+0425$&$ 1.85 $ &  $ 1.29 $ &  $ 3.15 $ &  $ 1.54 $&$ 3.80 $&$ 0.22 $ & $ 0.17  $\\   
IRAS $17049-2440$&$      $ &  $ 3.04 $ &  $      $ &  $ 3.62 $&$ 9.02 $&$ 0.90 $ & $ 0.33  $\\   
IRAS $17119+0859$&$ 1.42 $ &  $ 1.08 $ &  $ 2.50 $ &  $ 1.57 $&$ 5.10 $&$ 1.18 $ & $-0.28  $\\   
IRAS $17297+1747$&$ 3.04 $ &  $ 2.33 $ &  $ 5.37 $ &  $ 2.93 $&$ 6.73 $&$ 1.13 $ & $ 0.04  $\\   
IRAS $17360-3012$&$ 3.01 $ &  $ 2.07 $ &  $ 5.08 $ &  $ 2.43 $&$ 6.28 $&$ 1.79 $ & $ 0.26  $\\   
IRAS $17411-3154$&$      $ &  $      $ &  $      $ &  $ 6.25 $&$13.99 $&$ 2.14 $ & $ 1.04  $\\   
IRAS $18009-2019$&$ 1.73 $ &  $ 1.02 $ &  $ 2.74 $ &  $ 1.01 $&$ 3.78 $&$ 1.11 $ & $-0.07  $\\   
IRAS $18040-0941$&$ 2.46 $ &  $ 1.74 $ &  $ 4.20 $ &  $ 1.95 $&$ 4.43 $&$ 0.45 $ & $ 0.36  $\\   
IRAS $18194-2708$&$ 3.20 $ &  $ 2.39 $ &  $ 5.59 $ &  $ 2.80 $&$ 7.06 $&$ 0.40 $ & $ 0.45  $\\   
IRAS $18240+2326$&$      $ &  $ 3.24 $ &  $      $ &  $ 3.93 $&$ 9.21 $&$ 0.92 $ & $ 0.14  $\\   
IRAS $18333+0533$&$ 4.26 $ &  $ 2.48 $ &  $ 6.74 $ &  $ 2.44 $&$ 6.25 $&$ 1.44 $ & $ 0.32  $\\   
IRAS $18348-0526$&$      $ &  $      $ &  $      $ &  $ 6.27 $&$11.19 $&$ 1.94 $ & $ 1.44  $\\   
IRAS $18349+1023$&$ 1.55 $ &  $ 0.90 $ &  $ 2.45 $ &  $ 1.09 $&$ 3.98 $&$ 0.58 $ & $ 0.20  $\\   
IRAS $18397+1738$&$ 2.14 $ &  $ 1.74 $ &  $ 3.88 $ &  $ 2.12 $&$ 4.92 $&$ 0.56 $ & $ 0.40  $\\   
IRAS $18398-0220$&$ 2.40 $ &  $ 1.72 $ &  $ 4.12 $ &  $ 1.98 $&$ 4.83 $&$ 0.57 $ & $ 0.23  $\\   
IRAS $18413+1354$&$ 1.81 $ &  $ 0.97 $ &  $ 2.78 $ &  $ 0.98 $&$ 3.97 $&$ 1.13 $ & $-0.19  $\\   
IRAS $18560-2954$&$ 1.47 $ &  $ 0.84 $ &  $ 2.31 $ &  $ 1.07 $&$ 3.99 $&$ 0.76 $ & $ 0.12  $\\   
IRAS $19008+0726$&$ 2.63 $ &  $ 2.04 $ &  $ 4.67 $ &  $ 2.45 $&$ 5.38 $&$ 0.49 $ & $ 0.20  $\\   
IRAS $19059-2219$&$ 1.72 $ &  $ 1.04 $ &  $ 2.77 $ &  $ 1.12 $&$ 4.45 $&$ 1.15 $ & $-0.07  $\\   
IRAS $19093-3256$&$ 1.31 $ &  $ 0.70 $ &  $ 2.01 $ &  $ 0.76 $&$ 3.61 $&$ 1.02 $ & $-0.08  $\\   
IRAS $19126-0708$&$ 1.31 $ &  $ 0.79 $ &  $ 2.10 $ &  $      $&$ 4.61 $&$ 0.60 $ & $-0.02  $\\   
IRAS $19175-0807$&$ 2.70 $ &  $ 2.04 $ &  $ 4.74 $ &  $ 2.46 $&$ 5.46 $&$ 0.67 $ & $ 0.38  $\\   
IRAS $19321+2757$&$ 2.42 $ &  $ 1.89 $ &  $ 4.31 $ &  $ 2.29 $&$ 5.06 $&$ 0.73 $ & $ 0.30  $\\   
IRAS $20077-0625$&$ 2.89 $ &  $ 1.90 $ &  $ 4.79 $ &  $ 1.98 $&$ 6.13 $&$ 1.24 $ & $ 0.16  $\\   
IRAS $20396+4757$&$      $ &  $      $ &  $      $ &  $      $&$ 3.60 $&$ 0.35 $ & $ 0.22  $\\   
IRAS $20440-0105$&$ 1.18 $ &  $ 0.56 $ &  $ 1.74 $ &  $ 0.65 $&$ 3.16 $&$ 0.89 $ & $-0.14  $\\   
IRAS $20570+2714$&$ 2.96 $ &  $ 2.43 $ &  $ 5.39 $ &  $ 2.88 $&$ 5.83 $&$ 0.76 $ & $ 0.22  $\\   
IRAS $21032-0024$&$ 1.92 $ &  $ 1.35 $ &  $ 3.27 $ &  $ 1.51 $&$ 3.71 $&$ 0.44 $ & $ 0.14  $\\   
IRAS $21286+1055$&$ 1.38 $ &  $ 0.86 $ &  $ 2.24 $ &  $ 1.01 $&$ 3.48 $&$ 1.04 $ & $ 0.00  $\\   
IRAS $21320+3850$&$      $ &  $      $ &  $      $ &  $      $&$ 3.75 $&$ 0.38 $ & $ 0.26  $\\   
IRAS $21456+6422$&$      $ &  $      $ &  $      $ &  $      $&$ 3.17 $&$ 0.99 $ & $-0.10  $\\   
IRAS $23166+1655$&$      $ &  $      $ &  $      $ &  $ 6.23 $&$14.05 $&$ 1.43 $ & $ 0.61  $\\      
IRAS $23320+4316$&$      $ &  $      $ &  $      $ &  $      $&$ 7.23 $&$ 0.66 $ & $ 0.34  $\\   
IRAS $23496+6131$&$      $ &  $      $ &  $      $ &  $      $&$ 4.98 $&$ 1.08 $ & $ 0.03  $\\   
RAFGL $ 1406    $&$      $ &  $      $ &  $      $ &  $      $&$ 5.1  $&$      $ & $       $\\   
RAFGL $ 2688    $&$      $ &  $      $ &  $      $ &  $      $&$11.1  $&$      $ & $       $\\   
\end{tabular}                                                                               
}                                                                                           
\end{table}
\twocolumn
\normalsize
\subsection{Colour-Colour Diagrams}

 Infrared two colour diagrams are illustrated in Figs.~5 to 8. There is a
clear separation of the oxygen- and carbon-rich stars in these diagrams with
the carbon stars falling to the lower right of the oxygen-rich stars in the
near-infrared two colour diagrams (Figs.~5 and 6) and to the left of them in
the combined diagram, Fig.~7. The $JHKL$ colours of these AGB stars are more
extreme than those of the well studied Miras (e.g. Feast et al. 1982),
or even of IRAS selected Miras in the Galactic Cap (Whitelock et al. 1994).

\begin{figure}
\centering
\epsfxsize=7.5cm
\epsffile{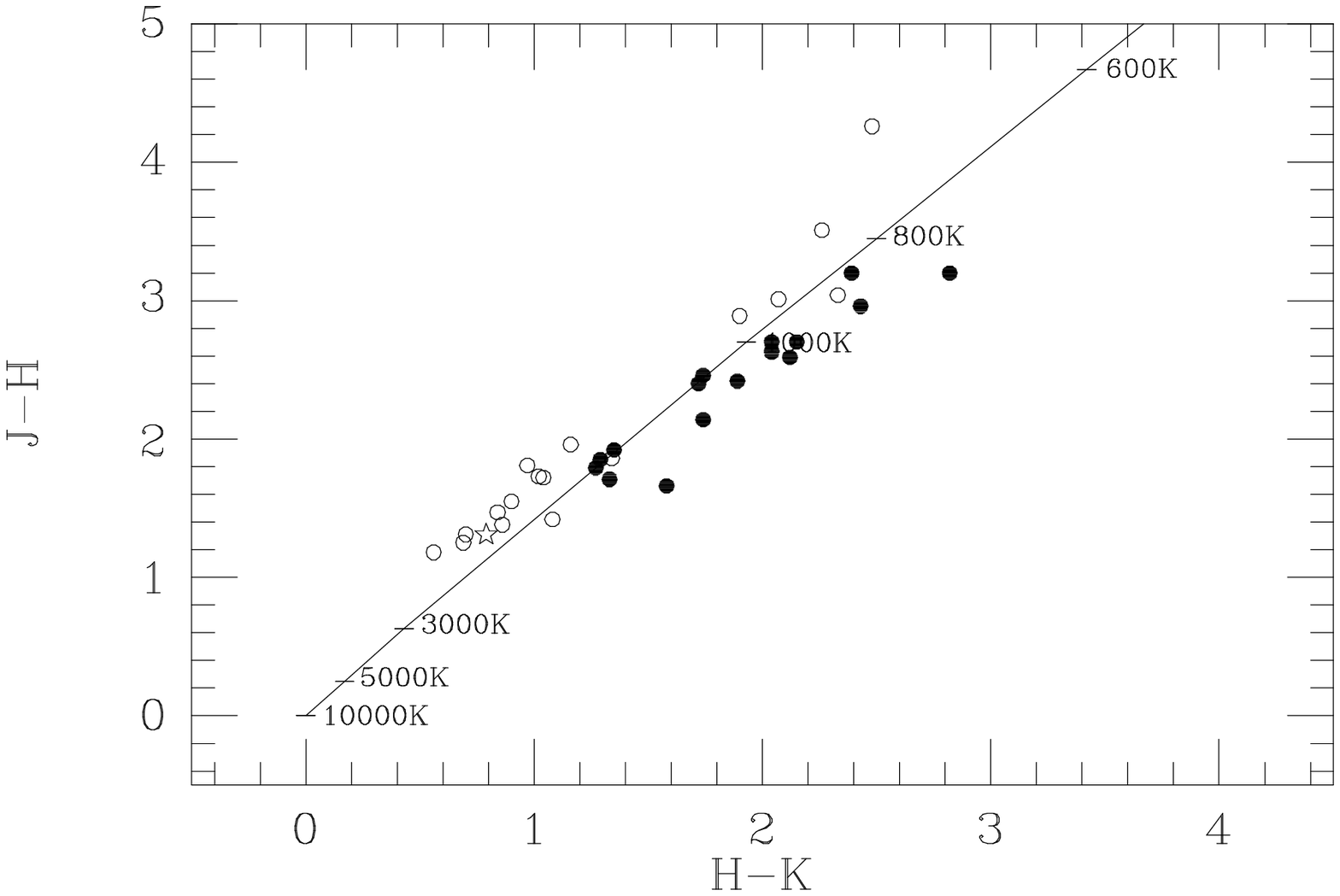}
\caption{$J-H$ versus $H-K$; filled circles represent carbon-rich stars, open
circles oxygen-rich stars and five-point star S-type stars. The solid line
represent loci of blackbodies of different temperatures.}
\end{figure}

\begin{figure}
\centering
\epsfxsize=7.5cm
\epsffile{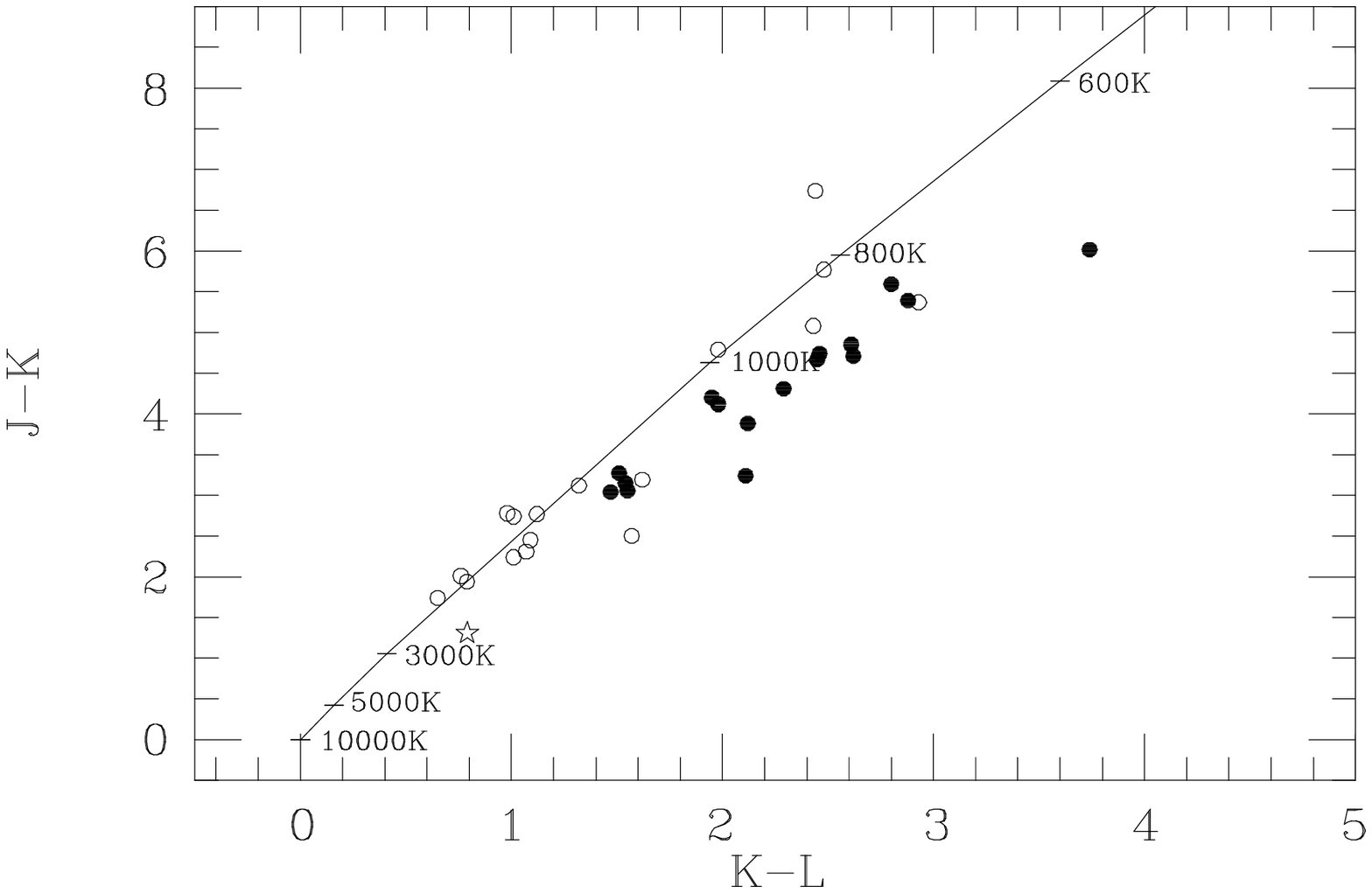}
\caption{J-K versus K-L. Symbols as in Fig.~5.}
\end{figure}

\subsubsection{The $K-[12]$ versus $[12]-[25]$ Diagram}
  Of particular interest is the $K-[12]$ vs. $[12]-[25]$ diagram (Fig.~7).
The colours of the oxygen- and carbon-rich stars separated rather clearly
in this diagram.  This has been noted previously (Le Bertre et al. 1994) and
shown to be the result of the difference in the ratios of the near-infrared
to mid-infrared opacities for the carbon- and oxygen-rich dust.
 
The single carbon star which falls among the oxygen-rich stars is
$06176-1026$ (the Red Rectangle), a well studied, but peculiar star.
It has an IRAS LRS classification of $80$ indicating the presence of the
$11.3\mu$m line in emission in contrast to the other carbon
stars which show $11.3\mu$m absorption, hence its position in Fig.~7. It is
thought to be a pre-planetary nebula with an oxygen-rich circumstellar disk
(see section 2.1).

\begin{figure}
\centering  
\epsfxsize=7.5cm
\epsffile{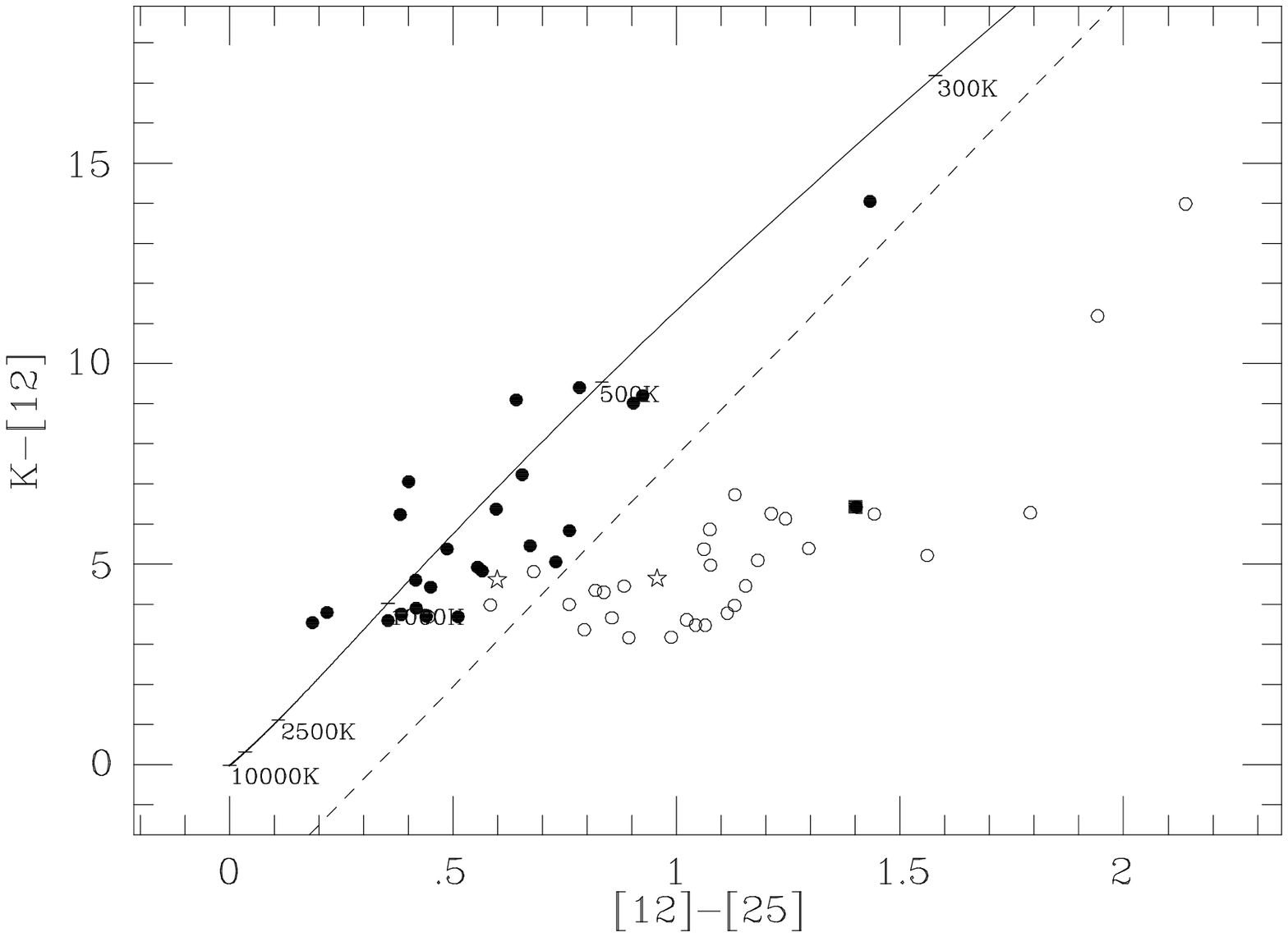}
\caption{K-[12] versus [12]-[25]. Symbols as in Fig.~5. Dotted line
separates the chemical types and is given by equation~3.}
\end{figure}


The dashed line in Fig.~7 has been positioned so as to separate the
carbon- and oxygen-rich stars as far as possible, it is described by:
\begin{equation}
(K-[12]) = 11.5([12]-[25]) - 3.8
\end{equation}
Thus if we define ${\delta}_{C/O}$ as follows:
\begin{equation}
{\delta}_{C/O} = 11.5([12]-[25]) - (K-[12]) - 3.8,
\end{equation}

The two chemical types can be separated by a single
parameter, ${\delta}_{C/O}$, such that for carbon stars ${\delta}_{C/O}<0$
and for most oxygen-rich stars ${\delta}_{C/O}>0$.
This parameter might be useful for separating these objects over the colour
range discussed. On this sample, obviously the one used to define the
parameter, only 6 percent of stars were misclassified.

The two S-type stars were omitted from the above discussion, we note that
one of them falls with the carbon- and the other with the oxygen-rich stars.

\subsubsection{IRAS Two Colour Diagram}
 Van der Veen \& Habing (1988) demonstrated that the position of an AGB star
in an IRAS two-colour diagram, of the type shown in Fig.~8, was indicative
of its chemical type and the thickness of its dust shell. Combined with
information on variability and IRAS spectral-type the diagram provides a
useful diagnostic. The colour indices for this plot ($C_{12/25}$,
$C_{25/60}$) are defined by:
$$C_{12/25} = 2.5\log ({\frac{F_{25\mu{m}}}{F_{12\mu{m}}}}),$$
$$C_{25/60} = 2.5\log ({\frac{F_{60\mu{m}}}{F_{25\mu{m}}}}),$$
where the IRAS fluxes are \emph{not colour corrected}. 

\begin{figure}
\centering  
\epsfxsize=7.5cm
\epsffile{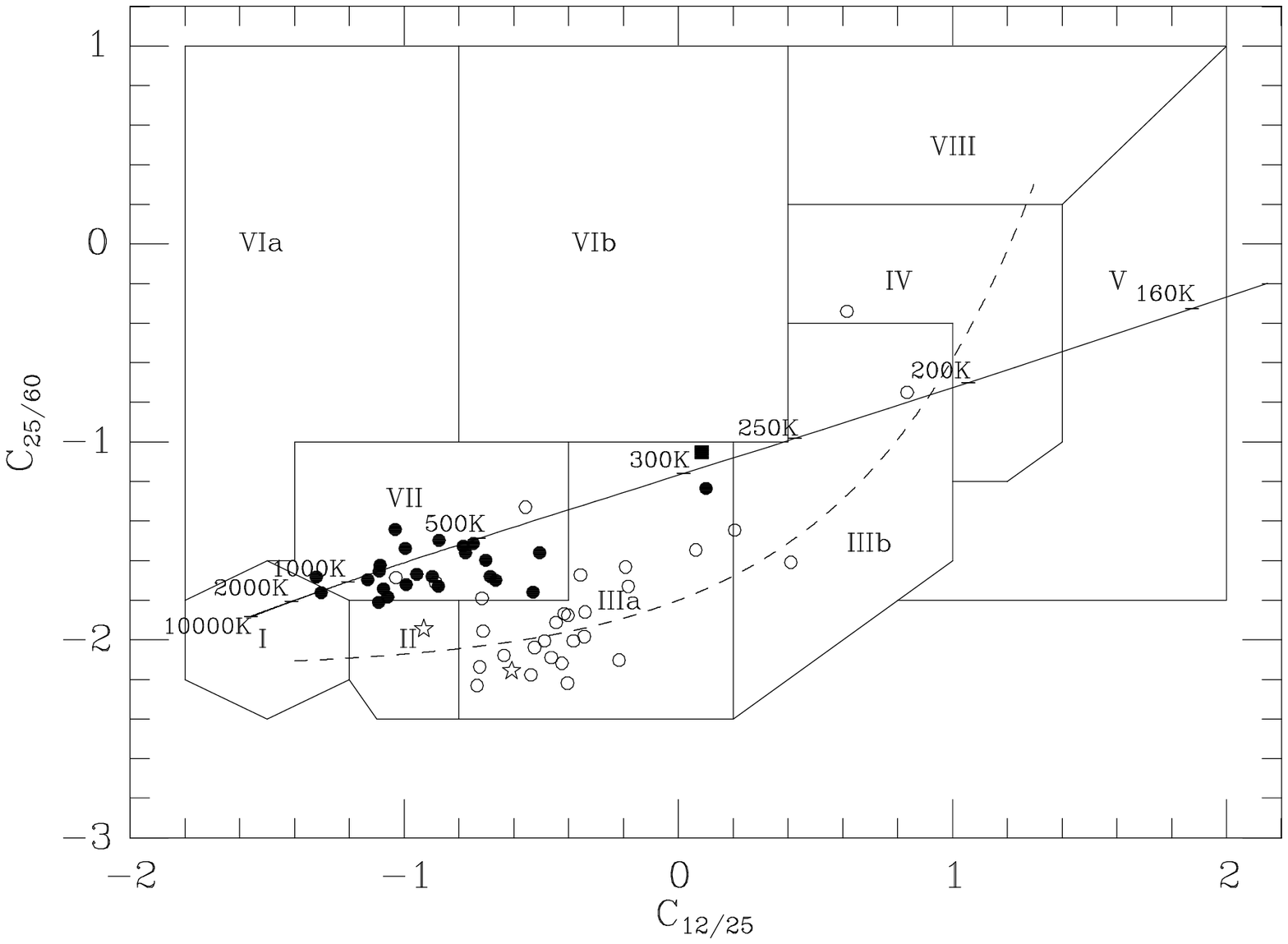}
\caption{$C_{25/60}$ versus $C_{12/25}$. Symbols as in Fig.~5. 
 $18135-1641$ has only an upper limit to its $C_{25/60}$ colour as 
indicated by the arrow.}  
\end{figure}
 
 Following van der Veen \& Habing the diagram is divided into regions, each
of which is dominated by a certain type of object. Stars with low mass-loss
rates will fall in the lower left of the diagram and increasingly higher
mass-loss rates are found as one moves to the upper right. 

Most of the oxygen-rich stars under discussion occupy region IIIa which van
der Veen \& Habing found to contain variable stars with moderate oxygen-rich
dust-shells. Four stars are located in regions IIIb and IV, indicating thick
to very thick oxygen-rich dust-shells and high mass-loss rates. Four
oxygen-rich stars are located in region VII, an area dominated by carbon
stars, but where van der Veen \& Habing also found oxygen-rich stars.

Most of the LRS spectra for oxygen-rich stars in this sample are classified
as 2n, indicating that the $10\mu$m silicate feature is in emission. The two
oxygen-rich variables with the reddest $[12]-[25]$ colour,  $17411-3154$
and  $18348-0526$, have LRS classifications of 3n, i.e. their spectra
show the $10\mu$m silicate feature in absorption. In these stars the
dust-shell has become so cool and optically thick that the silicate feature
has gone into absorption.

Most of the  carbon-stars fall in region VII, which is characterised by 
variable stars with well developed carbon-rich dust-shells. All of these 
stars have an LRS classification of 4n, indicating that the SiC 
$11\mu$m feature is in emission. The Red Rectangle, $06176-1036$, (discussed
above) is located in region IIIa, as is $23166+1655$. The latter object has
an LRS classification of O2, probably due to self absorption of the SiC
feature in the very think shell (Loup et al. 1993).

\section{Bolometric Magnitudes and Distances}
\subsection{Apparent Bolometric Magnitudes}

 Given that these sources were selected on the basis of their high fluxes
and therefore small distances, together with the fact that the bulk of their
energy is emitted at infrared wavelengths it is reasonable to assume that
the effects of interstellar extinction are negligible. The apparent
bolometric magnitude, $m_{bol}$, were calculated by integrating under a
spline curve (Hill 1982), fitted to the $J$, $H$, $K$, $L$, 12- and
25-$\mu{m}$ fluxes as a function of frequency. The end-points were dealt
with by extrapolating a line, which joins the $J$ flux and the point lying
halfway between the $H$ and $K$ fluxes, to zero flux. Zero flux at zero
frequency was assumed. The derived values of $m_{bol}$ are listed in Table~7
for the sources with sufficient photometry.

For the very red sources  $23166+1655$ and  $18348-0526$, which have
no measurements in the $J$ and/or $H$ bands, zero flux at $J$ and $H$ was
assumed when making the spline fit. 

To estimate an upper-limit to the amount
of flux not accounted for by ignoring the tail in the flux distribution, a
Wien approximation, ${\sim}A{\nu}^{3}e^{-b{\nu}}$ (where A and b are
constants), was assumed for flux at frequencies higher than $J$ with the
same slope as the spline at $J$. In more than 75 percent of the cases the
difference between the fluxes calculated in this way and the simple spline
fit was less than 1 percent.  $20440-0105$, which has the smallest
$K-[12]=3.16$, shows the maximum difference of 5 percent. Neglecting $\rm
H_2O$ absorption between the $J$,$H$,$K$ and $L$ bands (Robertson \& Feast
1981) will have a negligible effect on the bolometric magnitude estimated
for stars whose flux originates largely from the circumstellar dust-shell.


\begin{figure}
\centering  
\epsfxsize=7.5cm
\epsffile{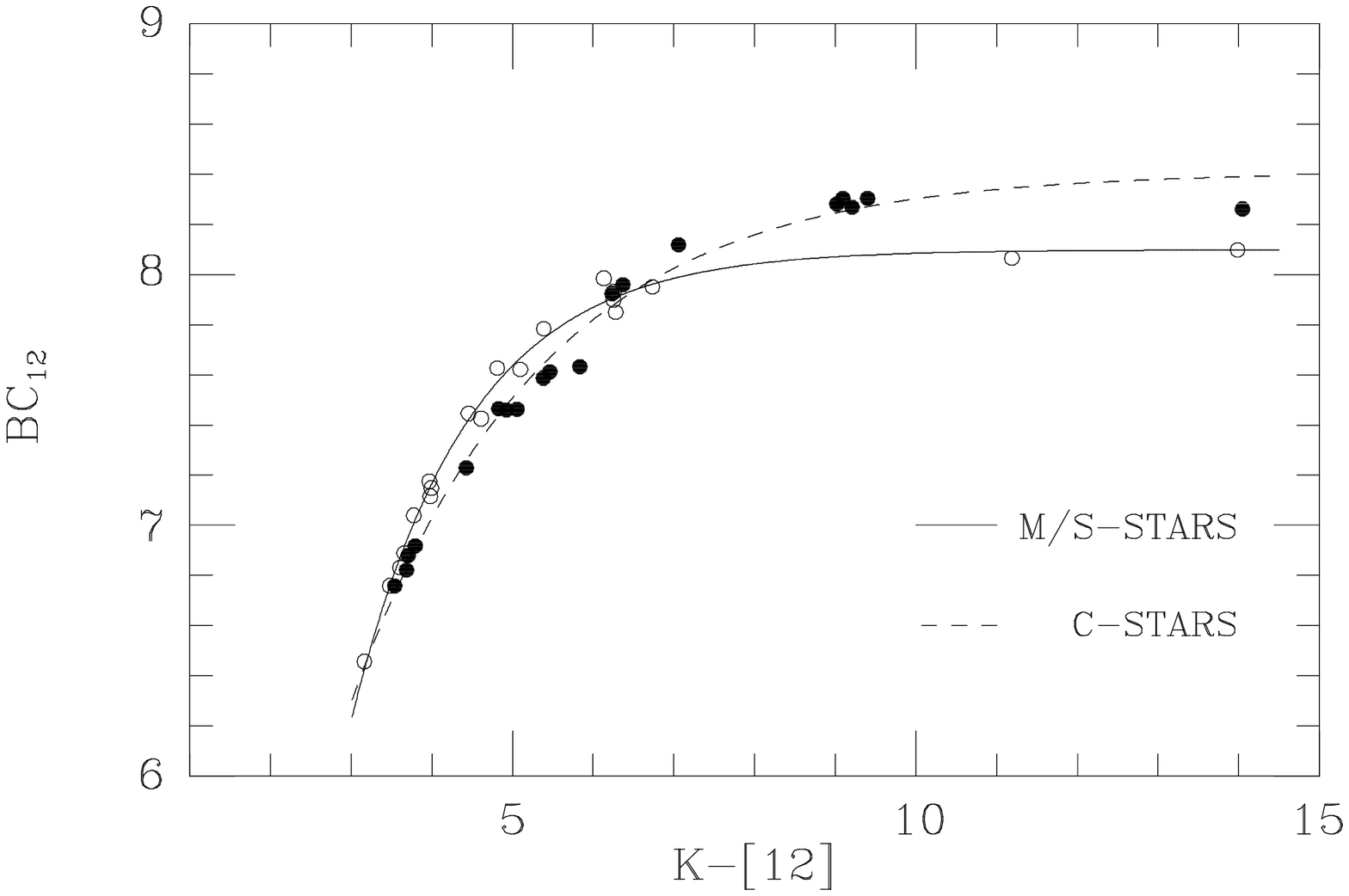}
\caption{$BC_{12}$ versus $K-[12]$. Symbols as in Fig.~5. Curves are those
given in equations~4 and 4.}
\end{figure}

The bolometric corrections at 12$\,\mu{m}$, $BC_{12}$, is 
plotted against the $K-[12]$ colour in Fig.~9, together with curves
given by:
\begin{equation}
BC_{12} = -15.4{e^{-0.70(K-[12])}} + 8.10 ,
\end{equation}
and 
\begin{equation}
BC_{12} = -7.59{e^{-0.43(K-[12])}} + 8.41 ,
\end{equation}
 which are least-square fits of this form to the oxygen- and carbon-rich
star data respectively. The rms difference between the data and these curves
is 0.04 and 0.06 mag respectively. For the remaining sources $m_{bol}$ was
calculated from [12] and  $K-[12]$ using equations~4 and 5.

The apparent luminosities derived here are typically rather smaller than
those estimated by JK89, because of the colour corrections applied to the
IRAS photometry.

\subsection{Absolute Bolometric Magnitudes and Luminosities}

For the 47 stars with periods the absolute bolometric magnitudes, $M_{bol}$,
was calculated from the period-luminosity (PL) relation, equation 6,
derived for LMC variables (Feast et al. 1989) assuming an LMC distance
modulus of 18.60 mag (Whitelock et al. 1997b) and the results given in
Table~7.
 \begin{equation}
M_{bol} = -2.34 \log P + 1.26
\end{equation}

The average luminosity, for all stars with periods less than 1000
days is $(0.98 \pm 0.17) \times 10^{4}\,L_{\odot}$, or $M_{bol}=-5.23\pm
0.19$.  For stars with unknown periods this mean value is assumed; which
seems reasonable since the carbon stars seem to have periods less than 1000
days and the remaining oxygen-rich stars have $[12]-[25]$ colours suggesting
that they too have periods less than a 1000 days.  This value of the mean
luminosity suggests that the $10^{4}L_{\odot}$ often assumed for the
luminosity of the tip of the AGB is consistent with the application of the
PL relation to this type of AGB star. Note this value may not be applicable to 
the Red Rectangle and the Egg Nebula, since they may have left the AGB already.

It is not really clear at this stage how well the various PL relations fit
long period variables or if different relations should be used for the
oxygen- and carbon-rich stars. Provided this caveat is born in mind when the
results are interpreted it seems reasonable to use the luminosities derived
as described above.

Note, in particular, that the PL relation used above was derived for stars
with periods less than 420 days and it may not be appropriate to use it for
the present sample.  There certainly are long period AGB ($P>420$ days)
stars in the LMC that lie above the PL relation (e.g. Feast et al. 1989).
However, recent work on dust-enshrouded stars in the LMC suggest that these
AGB stars are over-luminous because they are undergoing hot bottom burning
and that more normal long period ($P \geq 400$ day) carbon- and oxygen-rich
AGB stars fall very close to the PL derived for shorter period oxygen-rich
stars (Whitelock \& Feast 2000).

\subsection{Distances}
 Distances to all the stars in the sample (excluding the Red Rectangle and the
 Egg Nebula) were calculated from $m_{bol}$ and
$M_{bol}$ and are listed in Table~7. These distances are typically larger than
those estimated by JK89. As discussed above the apparent luminosities used
here are smaller than JK89's, while the absolute luminosities are on average
the same, except for the 3 stars with $P>1000$ where they are brighter. Thus
we find sources with distances up to 2\,kpc.

 The other plausible way of estimating distances to OH/IR stars is the phase
lag method, although it is notoriously difficult and time consuming. Van
Langevelde, van der Heiden \& Schooneveld (1990) measured phase lag
distances for two of the stars under discussion. For $01037+1219$ (WX~Psc)
they get $0.74\pm 0.15$ kpc, identical to the PL distance given in Table~7.
For $18348-0526$ (OH26.5+0.6) they give two values, $1.44\pm0.27$ and
$1.30\pm0.35$ kpc, both somewhat closer than the 1.9 kpc listed in Table~7.
Given the uncertainties in phase lag and the PL method this difference is not
problematic. West (1998) has measured a phase lag distance for $17411-3154$ 
(OH357.3--1.3) of $1.2\pm 0.4$ kpc, in agreement with the distance of 
0.99 kpc listed in Table~7.

\begin{figure}
\centering  
\epsfxsize=7.5cm
\epsffile{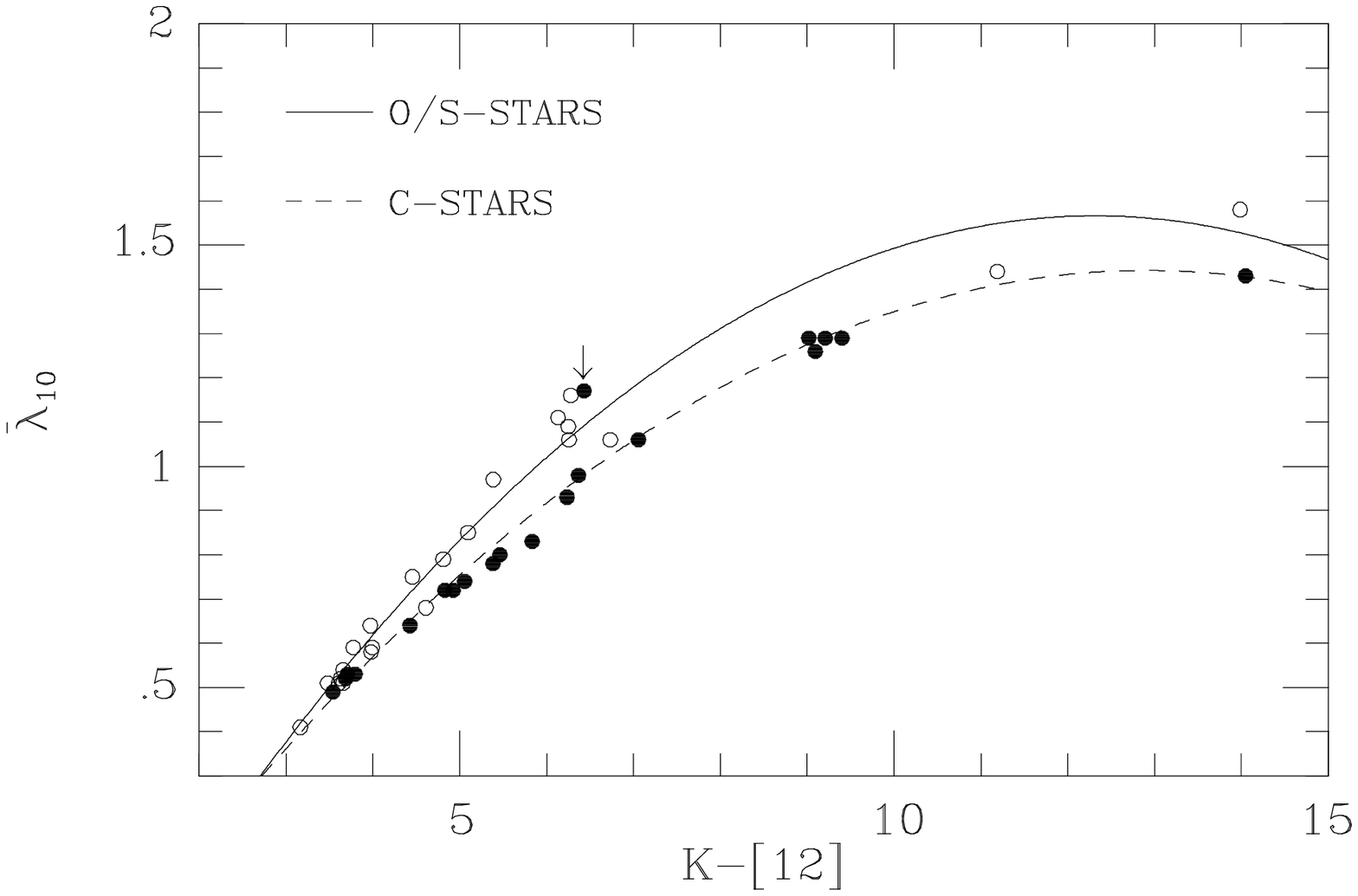}
\caption{$\bar{\lambda_{10}}$ versus $K-[12]$. Filled and open circles
represent carbon- and oxygen-rich stars, respectively. The arrowed point,
the Red Rectangle, is not used in the solution.}
\end{figure}

\section{Mass-loss rates}
 Mass-loss rates, $\dot{M}$, were $estimated$ using the expression given by
Jura (1987):
 \begin{equation}
\dot{M} = 1.7\times10^{-7}v_{15}r^{2}_{kpc}L^{-1/2}_{4}F_{\nu,60}
{\bar{\lambda}}_{10},
\end{equation}
 where $v_{15}$ is the outflow velocity of the gas in units of 15 $\rm
km\,s^{-1}$, $r_{kpc}$ the distance to the star in kpc, $L_{4}$ the
luminosity in units of $10^{4}L_{\odot}$, $F_{\nu,60}$ the flux measured at
$60\mu$m in Jansky and ${\bar{\lambda}}_{10}$ the mean wavelength of light
emerging from the star and its circumstellar dust-shell in units of
$10\mu$m.

Equation~7 assumes all the grains have an emissivity of $\rm 150\,cm^{2}g^{-1}$ 
at $60\mu$m and a dust to gas ratio of $4.5\times10^{-3}$ (Jura 1986). 
The mean wavelength of light emerging from the star and its circumstellar 
dust-shell, $\bar{\lambda}_{10}$, is given by:
\begin{equation}
\bar{\lambda}_{10} = \frac{\int_{0}^{\infty}{\lambda}{F_{\lambda}}d{\lambda}}
{\int_{0}^{\infty}{F_{\lambda}}d{\lambda}}   
\end{equation}
where $F_{\lambda}$ the monochromatic flux density in units of energy
received per unit area per unit time per unit wavelength interval. 

For the stars for which the function $F_{\nu}$ could be estimated (section
6.1), $\bar{\lambda}_{10}$ was calculated from equation 8 and is listed in
Table~7. The relationship between ${\bar{\lambda}}_{10}$ and $K-[12]$ is
illustrated in Fig.~10 and can be expressed as:
\begin{equation}
{\bar{\lambda}}_{10} = c + b(K-[12]) + a(K-[12])^{2},
\end{equation}        
where a, b and c are constants.

Least square fits of equation 9 to the data results in the following
values (a,b,c) for oxygen-rich (--0.0110,0.284,--0.392) and carbon-rich
(--0.0137,0.338,--0.513) stars, respectively, with an rms error of
$0.05\,\mu$m for both chemical types. The discrepant carbon star is the
Red Rectangle which was omitted from the solution.

Equation 9 with the appropriate parameters is used to estimate
${\bar{\lambda}}_{10}$ for the remaining stars in the sample; the results
are listed in Table~7

The outflow velocities, $v_{15}$, were taken to be the expansion velocities
of the shell obtained from CO($1\rightarrow0$) line measurements.
Appropriate values from the literature are listed in Table~7 with
references. For the three stars without measurements an average outflow
velocity, for stars with $v_{out}<25\, \rm km\,s^{-1}$, of $17\pm4\, \rm
km\,s^{-1}$ was used. Jones et al. (1983, henceforth JHG) have shown there
to be a correlation between outflow velocity and luminosity for a sample of
OH/IR stars (see fig.~1 in JHG). The outflow velocity versus luminosity for
stars with periods in our sample is illustrated in Fig.~11. It shows no
obvious correlation, which should not be entirely
unexpected, considering the large spread in outflow velocity at a given
luminosity (up to $10\,km\,s^{-1}$) in fig.~1 of JHG and the small range in
$M_{bol}$ in Fig.~11 (compared to that in fig.~1 of JHG).
\onecolumn
\begin{table}
\centering
\caption{Physical and Kinematic data.}
\footnotesize{
\begin{tabular}{lcccccccr}
\multicolumn{1}{c}{$\bf {NAME}$} & 
\multicolumn{1}{c}{$\bf {m_{bol}}$}&
\multicolumn{1}{c}{$\bf {M_{bol}}$}&
\multicolumn{1}{c}{$\bf {D}$}&
\multicolumn{1}{c}{$\bf {{\bar{\lambda}}_{10}}$}&
\multicolumn{1}{c}{$\bf {v_{out}}$}&
\multicolumn{1}{c}{$\bf {Log(\dot{M})}$}&
\multicolumn{1}{c}{$\bf {Log({\dot{M}}_{model})^{*}}$}&
\multicolumn{1}{c}{$\bf {v_{lsr}}$}\\
         & 
         &
         &
\multicolumn{1}{c}{$\bf {(kpc)}$}&
         &
\multicolumn{1}{|c|}{$\bf {(km\,s^{-1})}$}&   
         &
         &
\multicolumn{1}{c}{$\bf {(km\,s^{-1})}$}\\ 
\hline           
IRAS $00042+4248$&$(4.89)$ & $  -5.51$ & $ 1.20$ & $(0.91)$ & $21.7$ &               $ -4.87$ &  $-4.62$ &$-19.0$\\
IRAS $01037+1219$& $ 4.00$ & $  -5.35$ & $ 0.74$ & $  1.06$ & $21.6$ &               $ -4.64$ &  $-4.92$ &$-50.1$\\
IRAS $01159+7220$&$(5.16)$ & $  -5.30$ & $ 1.24$ & $(0.76)$ & $21.4$ &               $ -5.19$ &  $-5.46$ &$  9.9$\\
IRAS $02270-2619$& $ 4.64$ & $  -4.84$ & $ 0.79$ & $  0.49$ & $20.0$ &               $ -5.82$ &  $-5.82$ &$-68.1$\\
IRAS $02316+6455$&$(5.08)$ & $  -5.16$ & $ 1.12$ & $(1.00)$ & $15.4$ &               $ -5.10$ &  $-4.96$ &$  4.5$\\
IRAS $02351-2711$& $ 4.21$ & $  -5.05$ & $ 0.71$ & $  0.51$ & $16.2$ &               $ -5.77$ &  $-5.59$ &$-66.5$\\
IRAS $03229+4721$&$(4.39)$ & $  -5.17$ & $ 0.82$ & $(0.68)$ & $15.3$ &               $ -5.51$ &  $-5.26$ &$ -7.6$\\
IRAS $03507+1115$& $ 2.35$ & $  -5.03$ & $ 0.30$ & $  0.79$ & $22.0$ &               $ -5.31$ &  $-5.35$ &$-31.4$\\
IRAS $04307+6210$&$(4.87)$ & $  -5.23$ & $ 1.05$ & $(0.55)$ & $18.8$ &               $ -5.62$ &  $-5.10$ &$  9.9$\\
IRAS $04566+5606$&$(2.75)$ & $  -5.21$ & $ 0.39$ & $(0.55)$ & $18.7$ &               $ -5.59$ &  $-5.52$ &$  8.6$\\
IRAS $05073+5248$&$(5.44)$ & $  -5.33$ & $ 1.43$ & $(0.88)$ & $18.0$ &               $ -4.66$ &  $-5.24$ &$  7.8$\\
IRAS $05411+6957$&$(4.09)$ & $  -5.23$ & $ 0.73$ & $(0.69)$ & $21.2$ &               $ -5.37$ &  $-5.00$ &$ 20.1$\\
IRAS $05559+7430$&$(4.77)$ & $  -5.14$ & $ 0.96$ & $(0.47)$ & $15.1$ &               $ -5.82$ &  $-5.85$ &$ 22.9$\\
IRAS $06176-1036$& $ 4.99$ & $       $ & $     $ & $  1.17$ & $01.6$ &               $      $ &  $     $ &$-11.7$\\
IRAS $06300+6058$&$(4.33)$ & $  -5.23$ & $ 0.82$ & $(0.50)$ & $16.8$ &               $ -5.48$ &  $-5.26$ &$ 21.5$\\
IRAS $06500+0829$&$(4.09)$ & $  -5.15$ & $ 0.70$ & $(0.70)$ & $20.0$ &               $ -5.08$ &  $-5.21$ &$  4.1$\\
IRAS $08088-3243$& $ 5.45$ & $  -5.23$ & $ 1.37$ & $  0.98$ & $20.7$ &               $ -4.97$ &  $-5.20$ &$  0.4$\\
IRAS $09116-2439$& $ 4.94$ & $  -5.39$ & $ 1.17$ & $  1.29$ & $13.4$ &               $ -4.84$ &  $-4.92$ &$ 16.2$\\
IRAS $09429-2148$& $ 4.75$ & $  -5.35$ & $ 1.05$ & $  0.97$ & $14.0$ &               $ -5.05$ &  $-5.43$ &$ 23.3$\\
IRAS $09452+1330$& $ 0.38$ & $  -5.37$ & $ 0.14$ & $  1.26$ & $15.0$ &               $ -4.79$ &  $-4.32$ &$ 45.1$\\
IRAS $10131+3049$& $ 2.96$ & $  -5.35$ & $ 0.46$ & $  0.93$ & $16.9$ &               $ -5.10$ &  $-5.12$ &$ 56.0$\\
IRAS $10491-2059$& $ 3.04$ & $  -5.16$ & $ 0.44$ & $  0.52$ & $14.2$ &  $ -5.74$ &  $   $ &$ 33.6$\\
IRAS $12447+0425$& $ 4.93$ & $  -4.97$ & $ 0.95$ & $  0.53$ & $16.9$ &               $ -5.82$ &  $-5.08$ &$ 67.0$\\
IRAS $17049-2440$& $ 4.77$ & $  -5.54$ & $ 1.15$ & $  1.29$ & $20.7$ &               $ -4.54$ &  $-4.54$ &$  9.3$\\
IRAS $17119+0859$& $ 4.97$ & $  -5.23$ & $ 1.10$ & $  0.85$ & $14.0$ &               $ -5.26$ &  $     $ &$ -0.8$\\
IRAS $17297+1747$& $ 4.91$ & $  -5.14$ & $ 1.02$ & $  1.06$ & $17.0$ &               $ -4.90$ &  $-4.80$ &$ 25.3$\\
IRAS $17360-3012$& $ 5.65$ & $  -5.94$ & $ 2.08$ & $  1.16$ & $20.0^{\dagger}$ &     $ -4.34$ &  $     $ &$  0.5$\\
IRAS $17411-3154$& $ 3.80$ & $  -6.17$ & $ 0.99$ & $  1.58$ & $20.5$           &     $ -3.67$ &  $-5.06$ &$ -1.3$\\
IRAS $18009-2019$& $ 4.38$ & $  -5.23$ & $ 0.84$ & $  0.59$ & $17.0^{\ddagger}$&     $ -5.42$ &  $     $ &$  0.8$\\
IRAS $18040-0941$& $ 5.21$ & $  -5.23$ & $ 1.22$ & $  0.64$ & $22.0$           &     $ -5.31$ &  $-4.92$ &$  5.3$\\
IRAS $18194-2708$& $ 4.80$ & $  -5.42$ & $ 1.11$ & $  1.06$ & $23.0$           &     $ -4.77$ &  $-5.03$ &$ -6.2$\\
IRAS $18240+2326$& $ 4.92$ & $  -5.23$ & $ 1.07$ & $  1.29$ & $15.1$           &     $ -4.80$ &  $-4.82$ &$ 15.8$\\
IRAS $18333+0533$& $ 5.38$ & $  -5.57$ & $ 1.55$ & $  1.09$ & $20.0$           &     $ -4.52$ &  $-4.80$ &$  6.0$\\
IRAS $18348-0526$& $ 5.13$ & $  -6.26$ & $ 1.90$ & $  1.44$ & $10.6$           &     $ -3.89$ &  $     $ &$  0.6$\\
IRAS $18349+1023$& $ 3.81$ & $  -5.10$ & $ 0.60$ & $  0.58$ & $17.0$           &     $ -5.52$ &  $-5.26$ &$  7.8$\\
IRAS $18397+1738$& $ 4.41$ & $  -5.12$ & $ 0.81$ & $  0.72$ & $15.6$           &     $ -5.31$ &  $-5.04$ &$ 10.0$\\
IRAS $18398-0220$& $ 4.42$ & $  -5.28$ & $ 0.87$ & $  0.72$ & $34.5$           &     $ -5.00$ &  $-4.77$ &$  1.0$\\
IRAS $18413+1354$& $ 5.25$ & $  -5.26$ & $ 1.27$ & $  0.64$ & $18.5$           &     $ -5.36$ &  $-5.19$ &$  8.0$\\
IRAS $18560-2954$& $ 3.98$ & $  -5.24$ & $ 0.70$ & $  0.59$ & $13.2$           &     $ -5.55$ &  $-5.09$ &$-14.7$\\
IRAS $19008+0726$& $ 4.80$ & $  -5.24$ & $ 1.02$ & $  0.78$ & $17.4$           &     $ -5.28$ &  $-5.21$ &$  0.8$\\
IRAS $19059-2219$& $ 5.16$ & $  -5.12$ & $ 1.14$ & $  0.75$ & $22.2$           &     $ -5.10$ &  $-4.89$ &$-13.6$\\
IRAS $19093-3256$& $ 4.44$ & $  -4.82$ & $ 0.71$ & $  0.51$ & $15.0$           &     $ -5.72$ &  $     $ &$-18.4$\\
IRAS $19126-0708$& $ 3.36$ & $  -5.08$ & $ 0.49$ & $  0.68$ & $19.9$           &     $ -5.39$ &  $-5.07$ &$ -8.5$\\
IRAS $19175-0807$& $ 4.91$ & $  -5.40$ & $ 1.16$ & $  0.80$ & $35.5$           &     $ -4.77$ &  $-5.24$ &$-10.0$\\
IRAS $19321+2757$& $ 4.96$ & $  -5.32$ & $ 1.14$ & $  0.74$ & $24.4$           &     $ -5.04$ &  $-4.66$ &$  4.0$\\
IRAS $20077-0625$& $ 3.99$ & $  -5.40$ & $ 0.76$ & $  1.11$ & $16.0$           &     $ -4.76$ &  $-4.89$ &$-20.3$\\
IRAS $20396+4757$&$(3.58)$ & $  -4.92$ & $ 0.50$ & $(0.49)$ & $13.7$           &     $ -5.92$ &  $-5.52$ &$  3.8$\\
IRAS $20440-0105$& $ 4.58$ & $  -5.23$ & $ 0.92$ & $  0.41$ & $10.8$           &     $ -6.05$ &  $-5.40$ &$-25.8$\\
IRAS $20570+2714$& $ 5.31$ & $  -5.51$ & $ 1.46$ & $  0.83$ & $23.5$           &     $ -4.94$ &  $-5.35$ &$-12.0$\\
IRAS $21032-0024$& $ 4.55$ & $  -5.00$ & $ 0.81$ & $  0.53$ & $16.1$           &     $ -5.77$ &  $-5.42$ &$-29.6$\\
IRAS $21286+1055$& $ 5.10$ & $  -5.00$ & $ 1.05$ & $  0.51$ & $13.5$           &     $ -5.72$ &  $-5.55$ &$-28.2$\\
IRAS $21320+3850$&$(4.69)$ & $  -5.03$ & $ 0.88$ & $(0.52)$ & $14.0$           &     $ -5.80$ &  $-5.64$ &$ -9.4$\\
IRAS $21456+6422$&$(4.75)$ & $  -5.32$ & $ 1.03$ & $(0.42)$ & $17.0$           &     $ -5.77$ &  $     $ &$ 21.0$\\
IRAS $23166+1655$& $ 4.72$ & $  -5.44$ & $ 1.08$ & $  1.43$ & $14.5$           &     $ -4.37$ &  $-4.68$ &$-40.4$\\
IRAS $23320+4316$&$(4.38)$ & $  -5.31$ & $ 0.87$ & $(1.09)$ & $14.6$           &     $ -4.95$ &  $-4.68$ &$-17.1$\\
IRAS $23496+6131$&$(5.05)$ & $  -4.92$ & $ 0.98$ & $(0.83)$ & $19.0$           &     $ -5.11$ &  $     $ &$ -0.3$\\
RAFGL $1406$     &$(4.45)$ & $  -5.23$ & $ 0.86$ & $(0.75)$ & $10.9$           &    $(-5.14)$ &  $     $ &$ 33.9$\\
RAFGL $2688$     &$      $ & $       $ & $     $ & $      $ & $19.0$         &    $  $ &  $-3.85$ &$ -6.5$\\
\end{tabular}
}
\begin{flushleft}                   
\small{${*}$ - model mass-loss rates taken from Loup et al. (1993).\\
Almost all stars have outflow velocities taken from Loup et al.
(1993), with the following exceptions:\\
${\dagger}$ - outflow velocities taken from Jura \& Kleinmann (1989);\\
${\ddagger}$ - average outflow velocity of stars with $v_{out}<25\rm 
\,km\,s^{-1}$.}\\
\end{flushleft}                     
\end{table}
\twocolumn
\begin{figure}
\centering  
\epsfxsize=6.5cm
\epsffile{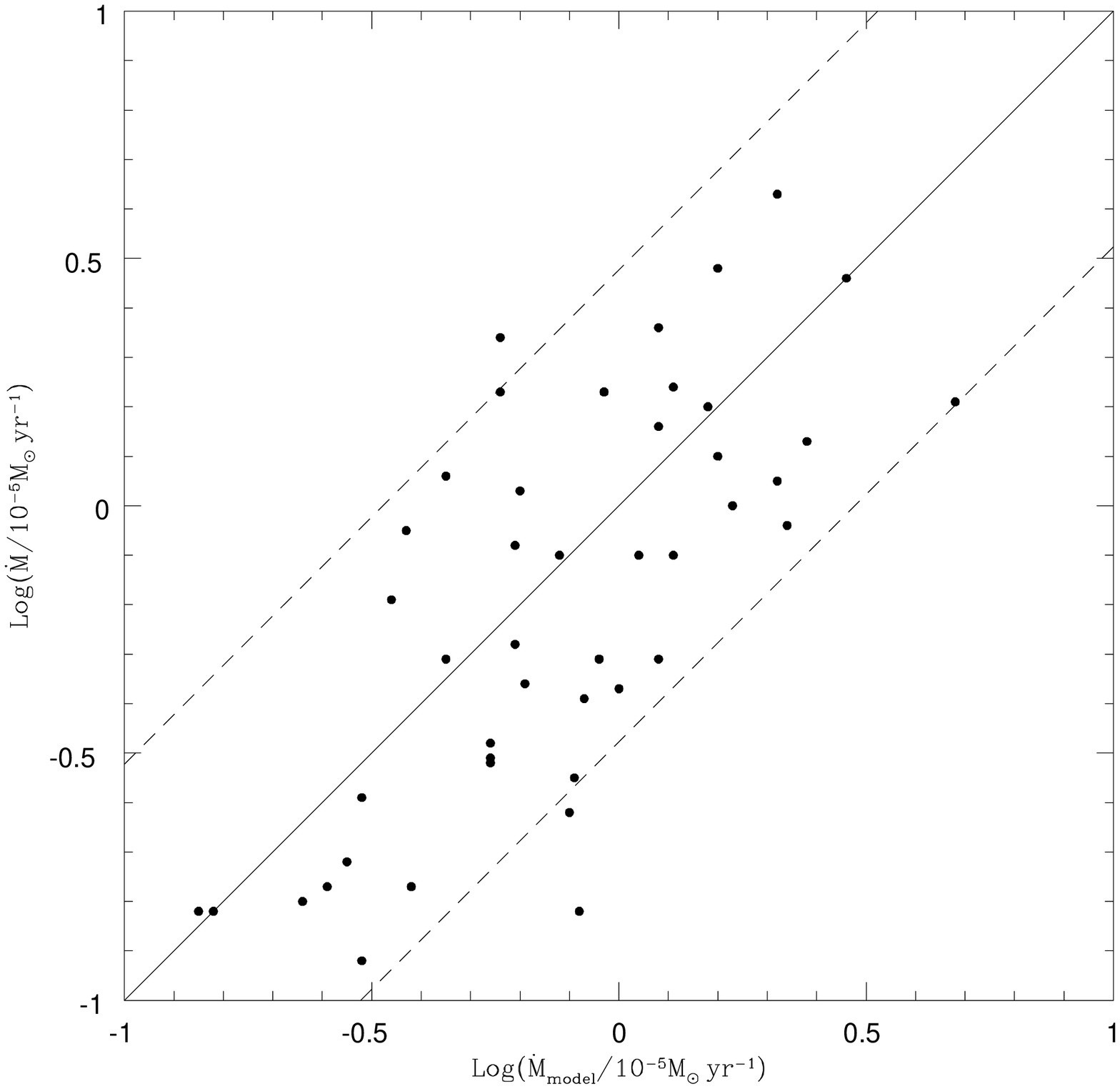}  
\epsfxsize=6.5cm
\epsffile{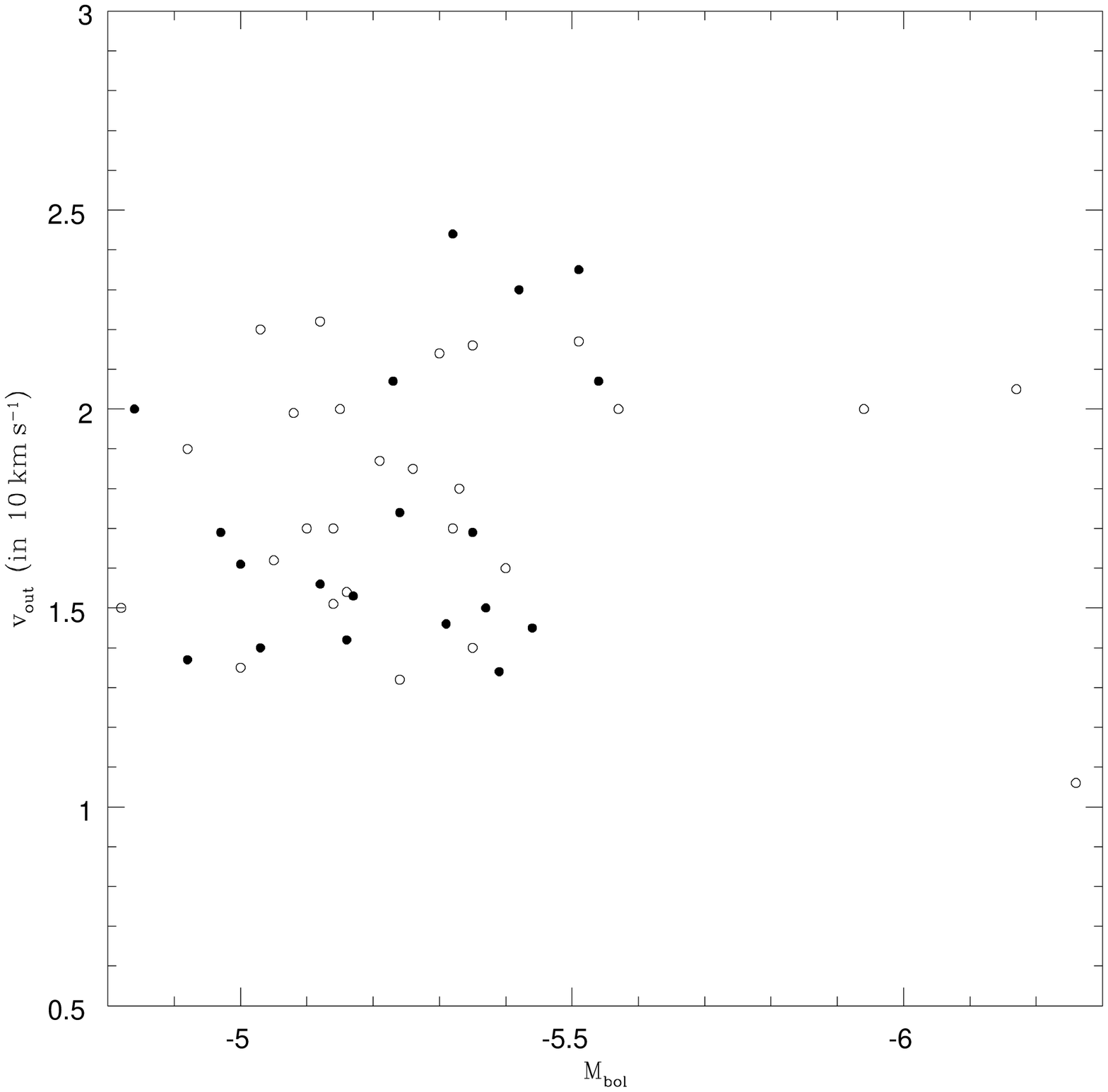}
\caption{A comparison of the mass-loss rates determined from equation 7
and listed in Table 7 with those given by Loup at al. ($\dot{M}_{model}$) in
the top panel and a plot of outflow velocity versus luminosity in the bottom
panel (symbols as in Fig.~10).}
\end{figure}
\begin{figure}
\centering  
\epsfxsize=7.5cm
\vspace{2cm}
\epsffile{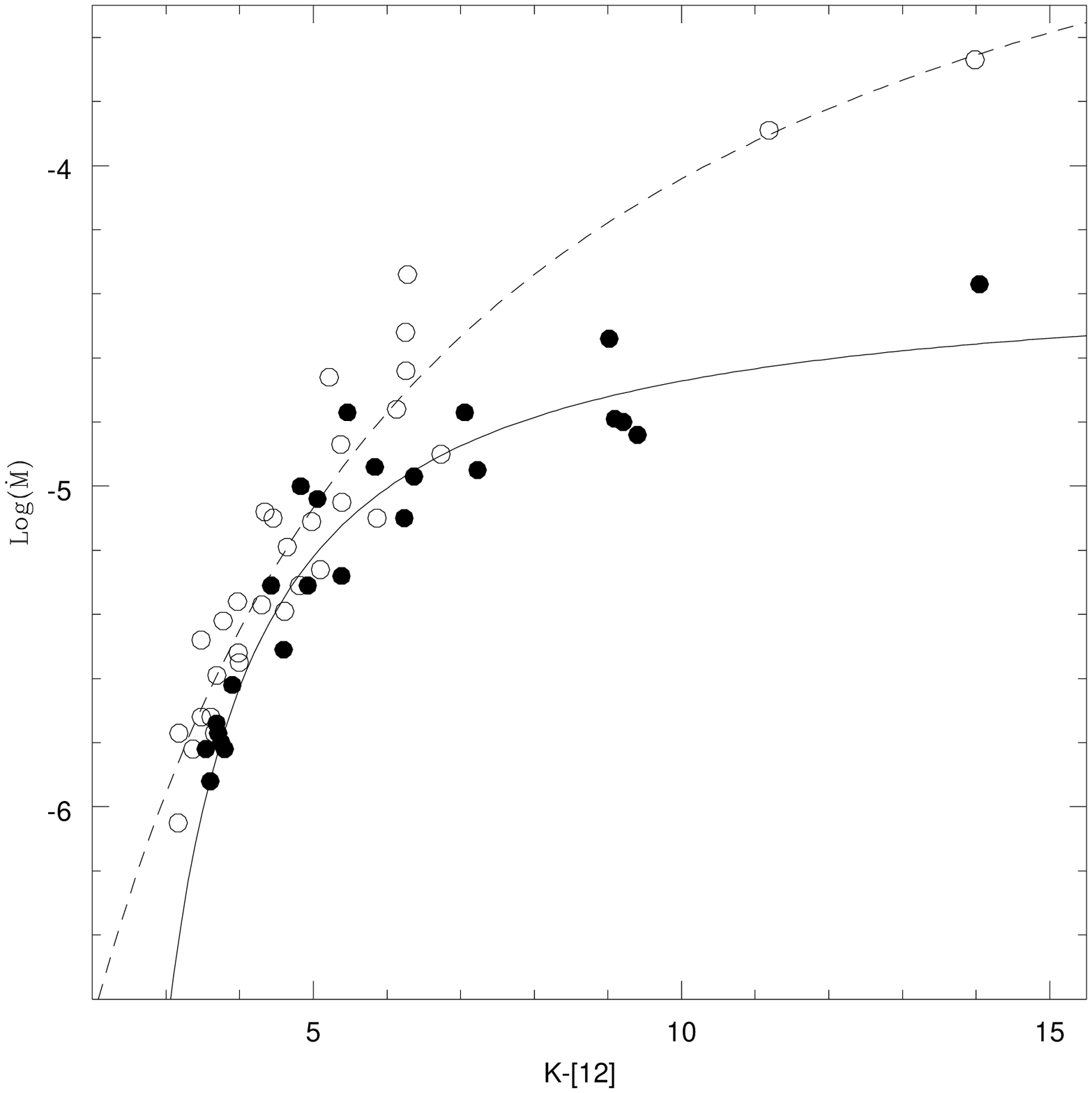}
\caption{Log($\dot{M}$) versus $K-[12]$. Filled and open circles represent 
carbon- and oxygen-rich stars respectively. The curves are described by equation~10.}
\end{figure}
The mass-loss rates for 58 of the stars in the sample were calculated from
equation~7 and are listed in Table~7 together with mass-loss rates,
${\dot{M}}_{model}$, taken from the compilation by Loup et al. (1993).
Where ${\dot{M}}_{model}$ was determined from detailed models of the CO
emission. A comparison of these values is illustrated in Fig.~11 which shows
that they agree within a factor of 3. This is reasonable given the
uncertainty in the various assumptions used to derive equation 7. 
$17411-3154$ does not appear in Fig.~11, it has an extremely high mass-loss
rate, $\dot{M}>10^{-4}M_{\odot}yr^{-1}$, from equation~7, much higher than
the value, $0.87\times10^{-5}M_{\odot}yr^{-1}$, given in Loup et al.
(1993).  Kastner (1992) applies a revised model of circumstellar CO emission,
and suggests that it is losing mass at a rate greater than
$\dot{M}>10^{-5}M_{\odot}yr^{-1}$, consistent with the value derived here.

Fig.~12 shows the relationship between mass-loss rate and $K-[12]$ ($\dot{M}$
correlates with other colours, but none as strongly as this). This relation
for Miras has been discussed by Whitelock et al. (1994) and Le Bertre \&
Winters (1998). A least squares fit can be made to the data in Fig.~12
assuming the form:
 \begin{equation}
\log (\dot{M}) = \frac{a}{(K-[12]+b)} + c,
\end{equation}  
where a, b and c are constants.

This results in values (a,b,c) for carbon- (--2.89, --1.79, --4.32) and
oxygen-rich (--21.34, 3.00, --2.40) stars respectively with rms uncertainties
of 0.14 and 0.16. This expression can then be used to estimate mass-loss rate
for the RAFGL 1406, which appear in brackets in Table~7.

\begin{figure}
\centering  
\epsfxsize=7.5cm
\epsffile{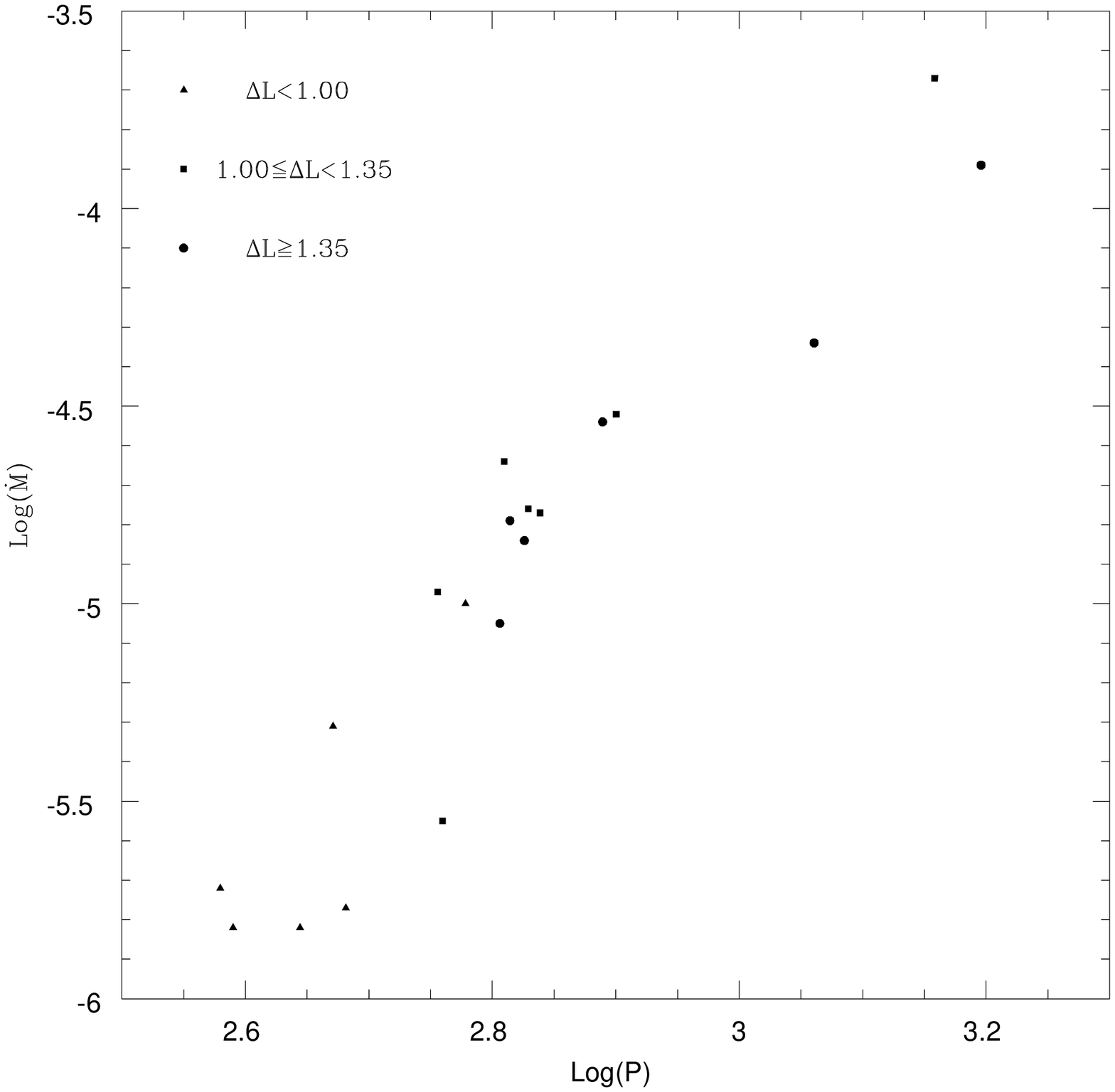}
\caption{Mass-loss rate ($\dot{M}$) as a function of pulsation period ($P$).}
\end{figure}
Fig.~13  shows that there is a correlation between mass-loss rate and the
period of pulsation. Some dependence of mass-loss rate on period might be
introduced by using the PL relation to calculate the quantities $L_4$ and
$r_{kpc}$ for equation 7. If we assume, for the sake of argument, that these
sources are all the same luminosity, then use of the PL will result in a
mass-loss/period dependence of the form $\dot{M} \propto P^{0.47}$, i.e. a
much weaker effect than is actually seen. A correlation between period and
mass-loss rate was also found by Whitelock et al. (1994).


The average mass-loss rate for stars in this sample is $ \dot{M} \sim
1.5\times10^{-5}M_{\odot}yr^{-1}$. Stars with $P>1000$ day (i.e.
$[12]-[25]>1.7$) have $\dot{M}\sim1.3\times10^{-4}M_{\odot}yr^{-1}$ and
stars with $P<1000$\,day (i.e. $[12]-[25]<1.7$) have $\dot{M}\sim
0.87\times10^{-5}M_{\odot}yr^{-1}$.  It will be seen later that
stars with $P>1000$\,day probably have main-sequence masses greater that
$4\,M_{\odot}$, while those with $P<1000$\,day have main-sequence masses
between 1 and $2\,M_{\odot}$. Thus intermediate mass stars lose mass 
an order of magnitude faster than do low mass-stars. The average
mass-loss rates of carbon- and oxygen-rich (including S-type) stars are
$\dot{M}\simeq(1.0\pm1.0)\times10^{-5}M_{\odot}yr^{-1}$ and
$\dot{M}\simeq(2.0\pm1.2)\times10^{-5}M_{\odot}yr^{-1}$ respectively. This 
slight difference may be a consequence of assuming the same grain emissivity 
and gas to dust ratio, in equation~7, for the two chemical types. Within the
limits of the assumptions used here, the mass-loss rates of the carbon and
oxygen-rich stars are plausibly identical.

One of the stars in this sample, RAFGL 1406 (IY Hya), is a known
super-lithium-rich carbon star (i.e. it has $\log\frac{n(Li)}{n(H)}>-8$) and
it is estimated that about 2 percent of carbon-stars are of this type (Abia
et al. 1993a). Furthermore there is evidence that these stars are currently
important contributors to lithium in the Galaxy (Abia et al. 1995 and
1998).

Abia et al. (1993b) estimate a lithium production rate of 
$<{\dot{m}}_{7}>=4.5\times{10^{-14}}\,M_{\odot}yr^{-1}$ per carbon star with a
return rate into the interstellar medium 
${P}_{7}=1.9\times{10^{-9}}\,M_{\odot}pc^{-2}\,Gyr^{-1}$. 
The sample of stars they studied have a typical mass-loss rate of
$2\times{10^{-7}}\,M_{\odot}yr^{-1}$, i.e. more than a order of magnitude less
than the average mass-loss rate of carbon-rich DE-AGB stars. Note that
Abia et al. (1993b) derive a average progenitor mass for the carbon stars in
their sample of 1.2-1.6$\,M_{\odot}$, which is similar to the average
progenitor mass estimated for this sample (see later).
Assuming a super-lithium-rich carbon star to have a typical lithium
abundance of $\log\frac{n(Li)}{n(H)}=-7$ (see fig.~3 in Abia et al. 1993a)
we estimate a lithium production rate from carbon-rich DE-AGB stars of 
$1.3\times{10^{-13}}\,M_{\odot}yr^{-1}$ per carbon
star, with the fraction of super-lithium-rich carbon stars in this sample
taken to be 0.02. The total return rate of lithium by these stars into the 
interstellar medium is then about
$1.0{\times}10^{-9}\,M_{\odot}pc^{-2}\,Gyr^{-1}$. This indicates that 
carbon-rich DE-AGB stars may contribute up to a third of  the lithium in the 
Galaxy produced from carbon-rich AGB stars. However, granting the underlying 
assumptions and uncertainty in the numbers above, further investigation into 
the problem is necessary (see e.g. Abia et al. 1993b).

\section{Scale-height, Progenitor Mass and Lifetime}
\subsection{Scale-height}
 Assuming the usually exponential distribution of stellar densities 
the number of stars per height interval, ${\rho}$, varies with 
distance from the galactic plane, $z$, as:
\begin{equation}
{\rho} = {\rho}_{0}e^{-|z|/H},
\end{equation}
 where ${\rho}_{0}$ is the density of stars at $z=0$ and $H$ is the
scale-height. For such a population the average height, $\bar{|z|}$, is
equal to the scale-height, $H$, provided the sample is reasonably complete
(as JK89 suggest it is).

The sample of stars discussed here were necessarily selected from a
limited region of the sky by JK98. Taking this into account we use only
stars with $14^{\circ}<l<230^{\circ}$ and find that $H=235$ pc.

\begin{figure}
\centering  
\epsfxsize=7.5cm
\epsffile{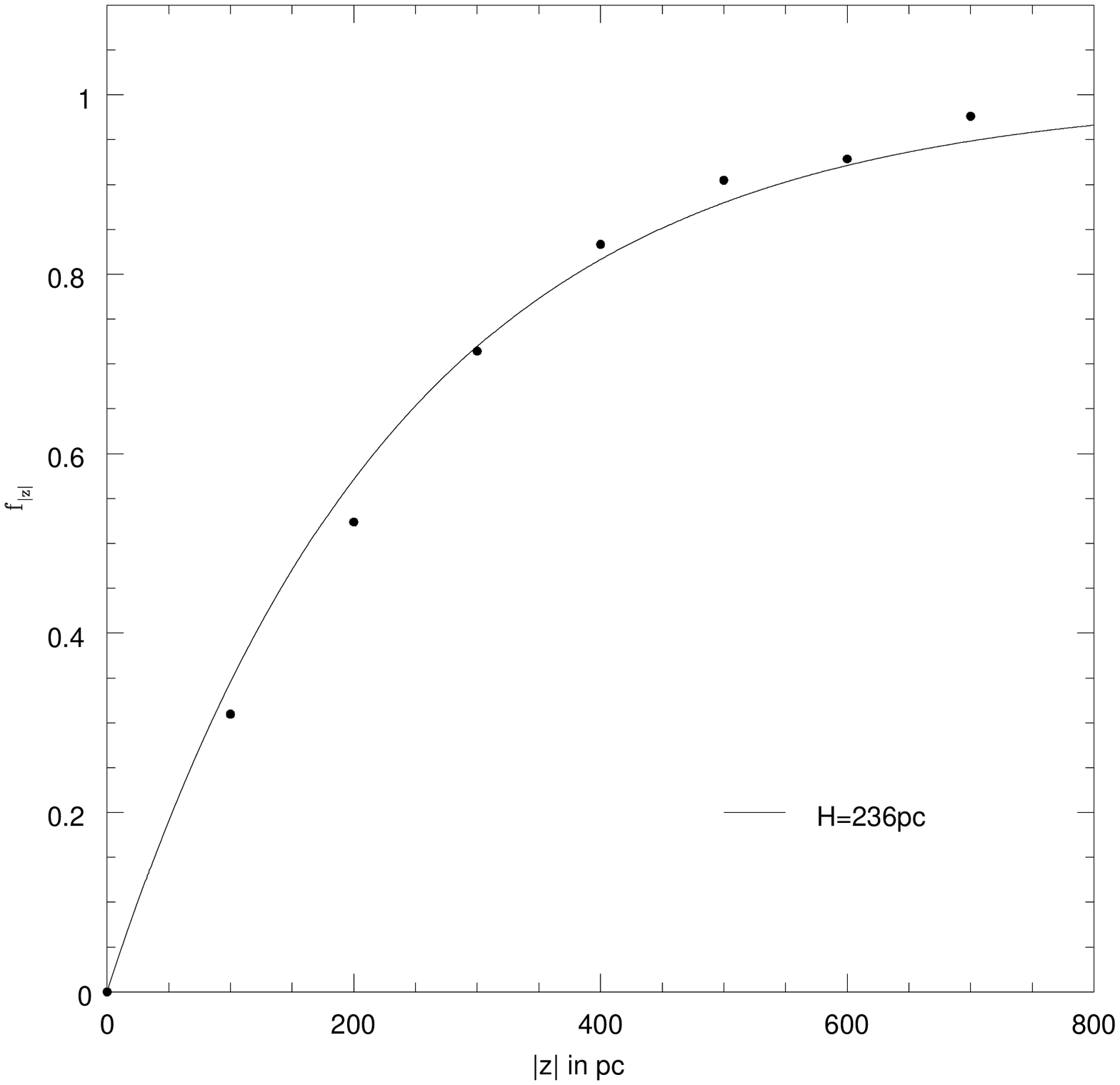}
\caption{The function $f_{|z|}$ gives the fraction of stars with distances
from the plane less than $|z|$. The curve is described by equation 12.}
\end{figure}
Alternatively, the fraction of stars, $f_{|z|}$, within a given $|z|$ of the
galactic plane is given by:
\begin{equation}
f_{|z|}= 1-e^{-|z|/H}.
\end{equation} 
 $f_{|z|}$ is plotted in Fig.~14 for each 100\,pc interval for the sample of
stars with $14^{\circ}<l<230^{\circ}$.  A curve of the form given by
equation~12, fitted by least squares, is also shown. This
fits the data very well (rms difference 0.03) and provides a scale-height of
$H=236\pm10$\,pc, comparable to the estimate made above. This value will be
used below as representative of the scale height of the sample.

\subsection{Progenitor Mass}
 The scale height derived above can be used to estimate a typical progenitor
mass for the AGB population. A comparison with Miller \& Scalo (1979 - their
table 1) indicates an average main-sequence mass $M_i\simeq1.2M_{\odot}$ for
the group under discussion, although it is difficult to estimate the
uncertainty on this value.

An alternative approach uses the data in table 3 of Miller \& Scalo (1979)
which gives the number of main-sequence stars pc$^{-2}(\log (M))^{-1}$,
$\phi_i$, as a function of $\log (M)$. A maximum likelihood 
straight line fit to these data over the range $-0.14 < \log (M) <0.92$ has a
slope $\alpha=-4.20\pm0.32$, a zero-point $\beta=1.20\pm0.14$ and fits the
data fairly well. Let $M_{lower}$ and $M_{upper}$ be the lower and upper
progenitor mass limits respectively of the stars in this sample. Then it is
straight forward to show that the average mass, $\bar{M}$, in $M_{\odot}$ is
given by:
\begin{equation}
\bar{M}={\frac{\alpha}{\alpha+1}}{\frac{M_{upper}^{\alpha+1}-M_{lower}^
{\alpha+1}}{M_{upper}^{\alpha}-M_{lower}^{\alpha}}}
\end{equation}

To get an idea of  the values of $M_{lower}$ and $M_{upper}$ the relation
between $M_{lower}$, $M_{upper}$ and the $average$ scale-height $\bar{H}$
was investigated. $\bar{H}$, for a given mass range
$[M_{lower},M_{upper}]$ is simply given by:

\begin{equation}
\bar{H}=\frac{\int_{M_{lower}}^{M_{upper}}H(M)dN(M)}{\int_{M_{lower}}^
{M_{upper}}dN(M)}
\end{equation} 
 where $H(M)$ is the scale-height as a function of mass $M$ (table 1 from
Miller \& Scalo), $dN(M)=\frac{\phi_i}{Mln(10)}dM$ the number of
stars$\,\rm pc^{-2}$ in the infinitesimally small interval $dM$ centred around
$M$.

\begin{figure}
\centering  
\epsfxsize=7.5cm
\epsffile{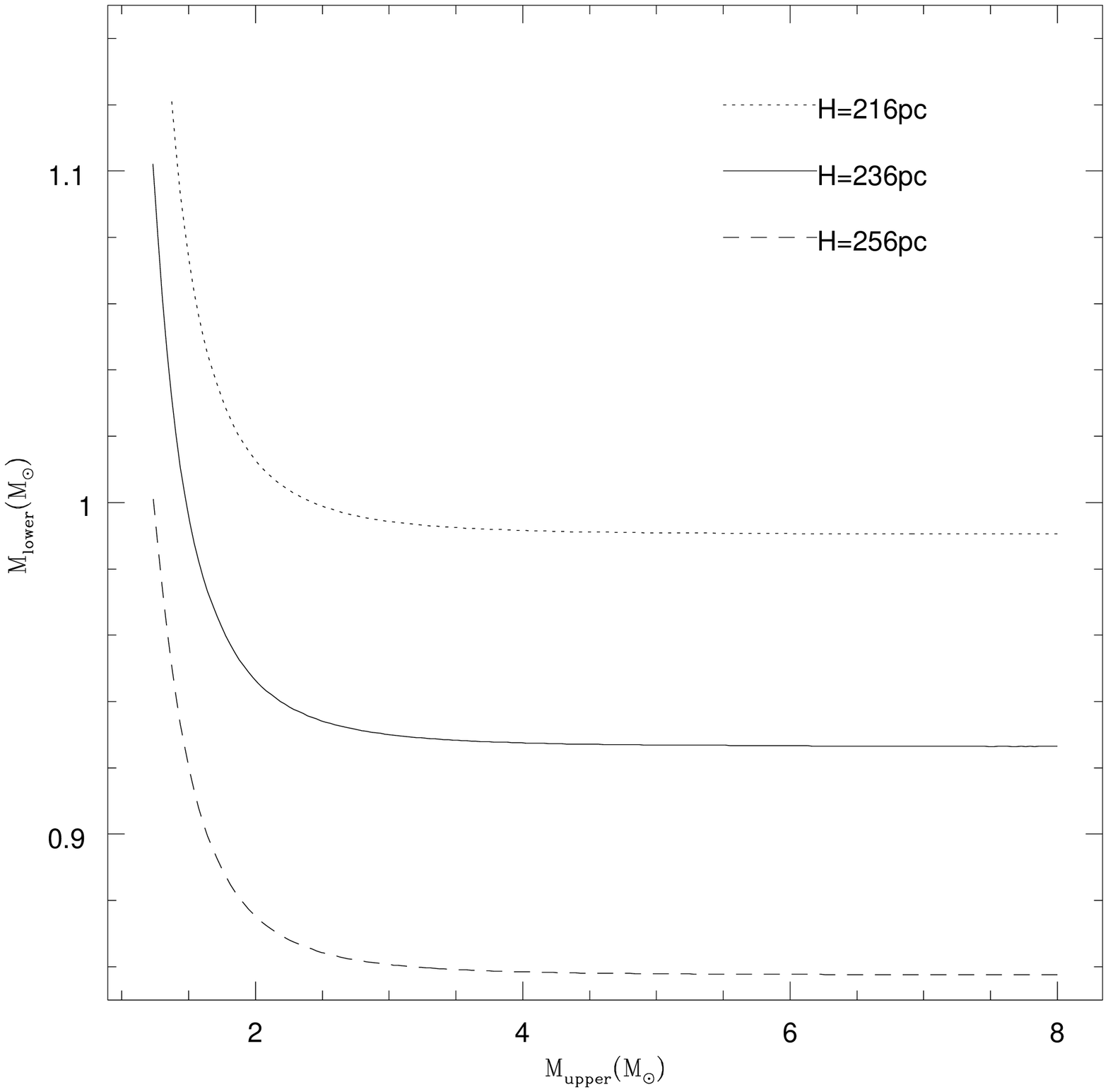}
\caption{The interdependence of lower and upper mass limits for different
scale heights.}
\end{figure}
Assuming an average scale-height, ${\bar{H}}_{0}$, then for a given upper mass 
limit $M_{upper}$ equation~14 can be solved numerically for 
$M_{lower}$, since the functions $H(M)$ and $\phi_i$ are known. This was
done for ${\bar{H}}_{0}=H=236$\,pc. The resulting relation between $M_{lower}$ 
and $M_{upper}$ is plotted in Fig.~15, from which it is clear
that for $M_{upper}>3M_{\odot}$, $M_{lower}$ has a constant value of
$0.96M_{\odot}$. 

Now since we expect to find dust envelope (DE) AGB main-sequence progenitors
up to about $8M_{\odot}$ and since the value of $\bar{M}$, from equation~13, is
relatively insensitive to the value of $M_{upper}$ if
$\frac{M_{upper}}{M_{lower}}>3$, values of $M_{lower}=0.93M_{\odot}$ and
$M_{upper}=8M_{\odot}$ will be assumed. It is important to note that the
value of $\bar{M}$ is essentially proportional to the value of
$M_{lower}$ and almost independent of $M_{upper}$ if $\frac{M_{upper}}{M_
{lower}}>3$. Also when calculating the surface density of progenitors, it
will be seen that this is very sensitive to the value of $M_{lower}$.
Changing the assumed average scale-height to 216\,pc or 256\,pc (i.e.
effectively changing the assumed scale-height by about $2\sigma$) gives
values of $M_{lower}$, assuming $M_{upper}=8M_{\odot}$, of $0.99M_{\odot}$ and 
$0.86M_{\odot}$ respectively.
Thus an error in $M_{lower}$ of $0.07M_{\odot}$ will be assumed.  This then
gives an average mass from equation~13 of $(1.22\pm0.18)M_{\odot}$.

From the discussion above, an average scale-height of 236\,pc
implies a progenitor mass range starting from about $0.96\,M_{\odot}$ up to
possibly $8\,M_{\odot}$. Hence using equation~18 (below), one can estimate 
the ratio of the number of stars in the mass range $2.0M_{\odot}<M_i<8.0\,
M_{\odot}$ to the number of stars in the mass range $1.0M_{\odot}<M_i<2.0\,
M_{\odot}$. This ratio is only 0.06. The three stars with periods 
greater than 1000 days have an average height
above the galactic plane of $20\pm2$\,pc. This indicates that the progenitors
of these stars were probably more massive stars, with $M_i>4.0M_{\odot}$, 
as can be seen from Miller \& Scalo's table 1. Interestingly the ratio 
of stars with $P>1000$\,day to the rest of the sample (which are expected to
have $P<1000$\,day from their colours) is 0.05, comparable 
to the ratio calculated above. The ratio, from equation~18, of the number of 
stars in the mass range $0.93M_{\odot}<M_i<1.0\,M_{\odot}$ to the number stars 
in the mass range $1.0M_{\odot}<M_i<2.0\,M_{\odot}$ is about 0.4. Thus the 
majority of the stars in this sample have main-sequence masses
in the range $1.0M_{\odot}<M_i<2.0\,M_{\odot}$. 

Limited discussion in the literature also supports this conclusion. Thus
Kahane et al. (2000) estimate the main-sequence progenitor of 
$09452+1330$, one of the stars in our sample, to be of low mass
($M_i\leq2M_{\odot}$) by comparing the predictions from AGB stellar
models of solar composition and various initial mass with the observed
chlorine isotopic ratio in its circumstellar shell. Also Nishida et al.
(2000) estimate the progenitor mass of three carbon-rich Miras, with
$P>450$\,day, in Magellanic Cloud clusters to be $1.5$-$1.6\,M_{\odot}$ by
main-sequence isochrone fitting.

The velocity dispersion of the stars with $P<1000$ provides another
indication of initial mass. Fifty three sources have velocities, relative to
the local standard of rest, tabulated by Loup et al. (1993), they show a
dispersion $\sigma=24\pm2\,\rm km\,s^{-1}$. After correcting for galactic
rotation, using the constants from Feast \& Whitelock (1997), this reduces
to $\sigma=19\pm2\,\rm km\,s^{-1}$. Table 8 lists the velocity dispersions
(calculated from table 2 of Delhaye 1965) and initial masses (from Allen
1973) for main-sequence stars between A0 and G5. The velocity dispersion
thus indicates that the mean sample has $M_i<2\,M_{\odot}$, consistent
with numbers given above.

\begin{table}
\centering
\caption{Initial Masses and Velocity Dispersions for Main-Sequence Stars}
\scriptsize{
\begin{tabular}{lccccc}
Sp (V) & A0 & A5 & F5 & G0 & G5 \\
$\sigma$ ($\rm km\,s^{-1}$) &11 & 14 & 21 & 22 & 23 \\ 
M ($M_\odot$) & 3.2 & 2.1 & 1.3 & 1.1 & 0.9\\
\end{tabular}}
\end{table} 

\begin{figure}
\centering  
\epsfxsize=7.5cm
\epsffile{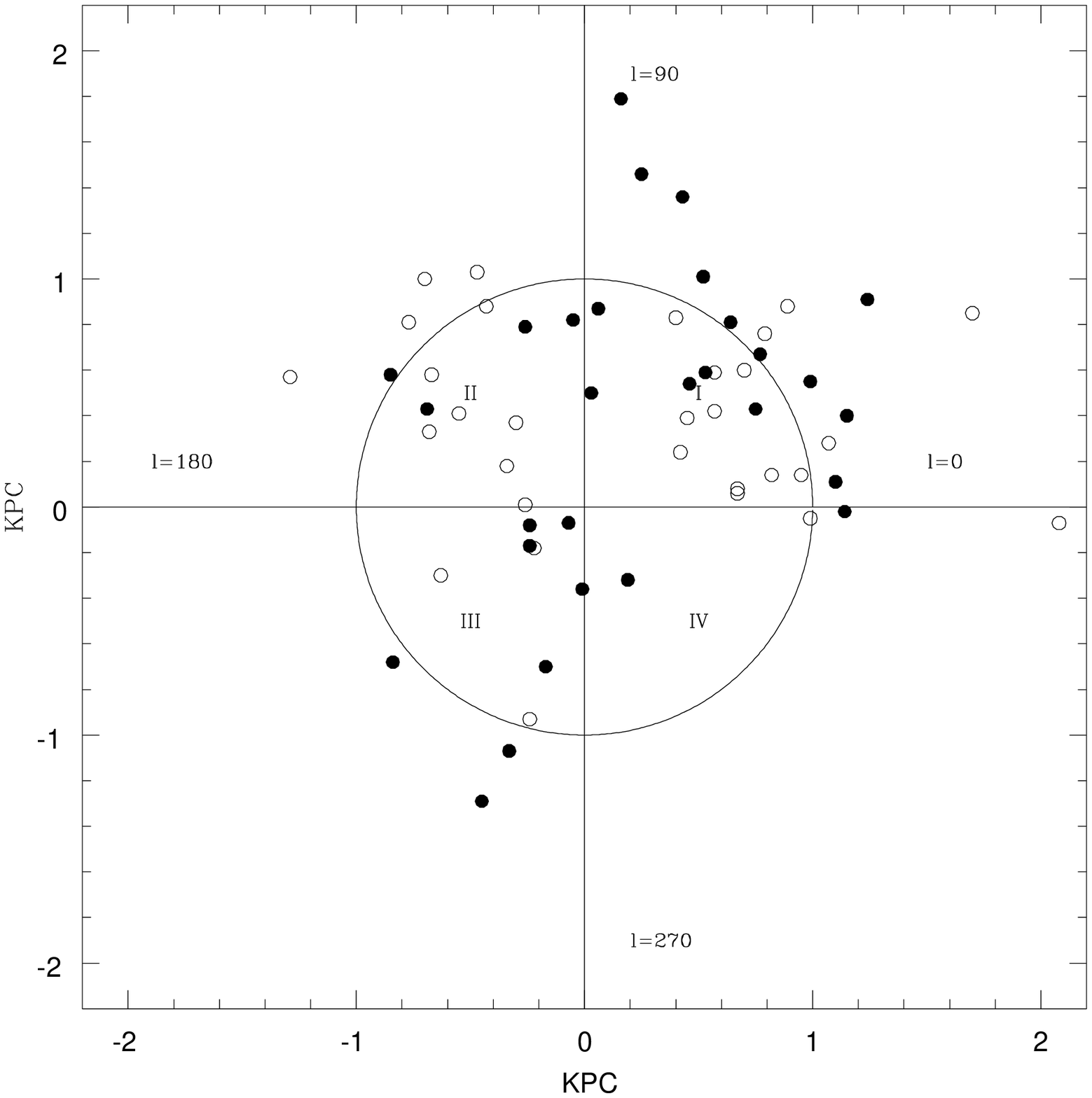}
\caption{The distribution of stars around the sun in the Galactic Plane.
Open and closed symbols represent oxygen- and carbon- rich stars
respectively. }
\end{figure}
\subsection{Surface Density}
 Fig.~16 shows the surface distribution of the stars in the galactic plane,
calculated from the distances derived previously. As can be seen, most of
the stars (90 percent) are within $1.4$\,kpc of the sun. There are very few 
in quadrant $IV$ due to positional selection effects in the original sample.
The surface density, $\sigma_{i}$ in 
region $i$, $i=I,II,III$, was estimated from:
\begin{equation}
\sigma_{i}=\frac{N_{i}}{\frac{1}{4}\pi(1.0)^{2}}\hspace{0.5cm}
\rm stars\,kpc^{-2}   
\end{equation}
where $N_{i}$ is the number of stars in a particular region $i$.

The average surface density, $\sigma_{DE-AGB}$ of DE-AGB stars in 
the solar neighbourhood was then estimated from:
\begin{equation}
\sigma_{DE-AGB}=\frac{2\sigma_{I}+\sigma_{II}+\sigma_{III}}{4}
\end{equation}
where $\sigma_{I}\simeq\sigma_{IV}$, assuming symmetry about $l=0$.\\ 
Using equations~15 and 16  we find
$\sigma_{DE-AGB}=15\pm4$ $ \rm stars\,kpc^{-2}$. 

\subsection{Lifetime}
 Assuming the average lifetime of stars in any given phase of evolution is
proportional to the number of stars in that phase, it follows that:
\begin{equation}
{\bar{\tau}}_{DE-AGB}=\frac{\sigma_{DE-AGB}}{\sigma_{MS}}{\bar{\tau}}_{MS}
\end{equation}
 where ${\bar{\tau}}_{DE-AGB}$ is the average lifetime of DE-AGB stars,
${\bar{\tau}}_{MS}$ and $\sigma_{MS}$ the average lifetime and surface
density, respectively, of the progenitors of DE-AGB stars on the
main-sequence.

From the linear relation between $\log(\phi_i)$ and $\log (M)$ it is 
straight forward to show that $\sigma_{MS}$ in the mass range
$[M_{lower},M_{upper}]$, is simply given by:
\begin{equation}
\sigma_{MS}=\frac{10^{\beta}}{\alpha{ln(10)}}[M_{upper}^{\alpha}-M_{lower}^
{\alpha}]
\end{equation}
 where $\sigma_{MS}$ is in units of $\rm stars\,pc^{-2}$.\\
It is clear from equation~18 and the value of $\alpha$ that $\sigma_{MS}$ 
is very sensitive to the value of $M_{lower}$ and $\propto{M^{\alpha}}_{lower}$ 
if $\frac{M_{upper}}{M_{lower}}>3.0$. Using the upper and lower mass limits
assumed above, gives $\sigma_{MS}=2.22\pm0.76$ $\rm stars\,pc^{-2}$.

Miller \& Scalo's table 1 also gives $\log(\tau_{MS})$ as a function of
$\log(M)$. ${{\bar{\tau}}_{MS}}$ is simply given by:
\begin{equation}
{\bar{\tau}}_{MS}=\frac{\int_{M_{lower}}^{M_{upper}}\tau_{MS}(M)dN(M)}
{\int_{M_{lower}}^{M_{upper}}dN(M)}
\end{equation}
Equation 19 was integrated numerically with the above mass limits,
giving ${\bar{\tau}}_{MS}=(5.5\pm1.5)\times10^{9}$\, years. The errors above were
estimated by changing the average scale-height (and hence $M_{lower}$) to 216 
and 256 respectively.

Using equation~17 an average lifetime for the stars in this sample was
estimated, giving ${\bar{\tau}}_{DE-AGB}=(3.7\pm1.9)\times10^{4}$\, years.
This value is in excellent agreement with lifetimes of similar stars at
the tip of the AGB estimated by various other investigators (see Habing 1996
p. 183).

\section{Evolutionary Status and Conclusions}
 The stars under discussion were selected to isolate high mass-loss, thick
dust-shell AGB stars within the Solar Neighbourhood.  This study indicates
that the selection criteria employed were fairly successful in this respect.
The $JHKL$ photometry, the IRAS two-colour diagrams and other published data
prove that the majority of these sources are cool variable stars with 
moderate to high mass-loss rates.

As shown above most of these stars will have had a main-sequence
progenitor with $ 1 \leq M \leq 2$ $M_{\odot}$. While the average progenitor
mass is $\sim1.3M_{\odot}$. Using this average value and an empirical
``initial/final mass''-relation (Weidemann 1987) an average final white
dwarf mass can be estimated at $\sim0.56M_{\odot}$.

Assuming the core masses of these stars are close to the final white dwarf
mass, it is found that the thermal interflash period (${t}_{if}$) will be,
${t}_{if}\simeq10^{5}$ years, using the core-mass interflash period relation
given by Boothroyd \& Sackmann (1988) (their equation 6 which is valid for
solar composition). A more recent core-mass interflash period relation given
by equation 11a of Wagenhuber \& Groenewegen (1998) also suggests a thermal
interflash period of ${t}_{if}\simeq10^{5}$ years; this is provided there is
no hot bottom burning and that the star has already undergone several helium
shell flashes (see Wagenhuber \& Groenewegen). The first assumptions is
reasonable as stars under 2 $M_{\odot}$ do not experience hot bottom
burning. The second is also an essential assumption for using the final
white dwarf mass for the current core mass and obviously the carbon stars,
at least, have undergone several flashes.

This interflash period, ${t}_{if}\simeq10^{5}$, is more than a factor of
two larger than the average lifetime of these DE-AGB stars,
$\bar{{\tau}}_{DE-AGB}= (3.7 \pm 1.9)\times 10^4$ years, deduced above.
Hence we can conclude that these DE-AGB stars will experience at most one
thermal pulse before leaving the TP-AGB phase.

The average mass-loss rate, ${\dot{M}}_{ave}$ for stars in the sample was 
found to be $\sim1.5\times10^{-5}M_{\odot}yr^{-1}$. Thus the
total mass lost, $M_{lost}$, by an average star in this phase is 
$M_{lost}={\dot{M}}_{ave}\bar{{\tau}}_{DE-AGB}\sim0.5M_{\odot}$. Which is
very close the average initial-final mass difference for these stars. 
However, this value for $M_{lost}$ is very uncertain due to the uncertainty
in the mass-loss rates.  Nevertheless, this indicates that a significant
fraction of the envelope mass is lost during this high mass-loss
dust-enshrouded TP-AGB phase.

The total rate of mass returned into the interstellar medium, 
${\dot{M}}_{ISM}$, by stars in this phase is ${\dot{M}}_{ISM}={\dot{M}}_{ave}
{\sigma}_{DE-AGB}\sim2.3\times10^{-4}M_{\odot}\,kpc^{-2}\,yr^{-1}$. Jura \&
Kleinmann (1990) estimate that M supergiants in the solar neighbourhood 
return mass to the interstellar medium at only a rate between     
$1\times10^{-5}$ and $3\times10^{-5}M_{\odot}\,kpc^{-2}\,yr^{-1}$. Thus
AGB stars must be important contributors of material to the interstellar
medium in the solar neighbourhood.

\subsection*{Acknowledgements} 
 We would like to thank Chris Koen, Michael Feast and Peter Wood for helpful
discussions and advice, Luis Balona for his Fourier analysis program and Tom
Lloyd Evans for allowing us to use observations in advance of publications.
We are also grateful to, Robin Catchpole and Dave Laney for making
near-infrared observations on our behalf. This paper made use of the SIMBAD
database and observations made from the South African Astronomical
Observatory.

\appendix
\section*{Appendix\\
Average Magnitudes and Colours}
\subsection*{Average Magnitudes}
 For variable stars, such as those discussed here, it is often useful to
know the average magnitude at a particular wavelength. The average magnitude
as discussed in section 3, was taken as the mean of the minimum and
maximum magnitude measured for the particular star. This average,
$A_{M}$, is compared in Table A1 with the average magnitude, $\bar{M}$,
derived from the fit given by equation 4.1  for the 19 stars with sufficient
data. The average absolute difference, ${\Delta}_{M}$, between these two
averages, for each near-infrared band, with the sample standard deviation,
${\sigma}_{\Delta}$, on this average is also listed in Table A1.  From these
values for ${\Delta}_{M}$ it can be seen that $A_{M}$ represents the average
magnitude fairly well. Obviously in instances where there are insufficient
observations to cover a complete cycle, $A_{M}$ will be a poor
representation of the mean.

\subsection*{Average Colours}
The stars in this sample not only vary in light output, but also in colour. 
In section 5 the average colour of a star is taken as the mean of the value
calculated at each Julian Date, $A$. This average is then compared
to the average colour, $A_{f}$, derived from the Fourier fits given by 
equation 4.1. $A_{f}$ is simply given by:
\begin{equation}
A_{f} = <M_{1}-M_{2}> = <M_{1}>-<M_{2}> = {M}_{1,av}-{M}_{2,av}
\end{equation}
 where $M_{1}$ and $M_{2}$ are the magnitudes in the bands making up the
colour. ${M}_{1,av}$ and ${M}_{2,av}$ are the average magnitudes in these
bands obtained from the Fourier fits given by equation 4.1. 

Both the averages, for each near-infrared colour, are listed in Table A2 for
the 18 stars with sufficient data. Also listed in the Table are the average
absolute difference, ${\Delta}_{A}$ and sample standard deviation,
${\sigma}_{\Delta}$, on this average. The two values are quite comparable.

\clearpage
\onecolumn
\begin{table}
\centering
\caption{Average magnitudes for variable sources with with good phase 
coverage}
\normalsize{
\begin{tabular}{lrrrrrrrr}\hline
\multicolumn{1}{c}{$\bf {NAME}$}&
\multicolumn{2}{c}{$\bf {J}$}&
\multicolumn{2}{c}{$\bf {H}$}&
\multicolumn{2}{c}{$\bf {K}$}&
\multicolumn{2}{c}{$\bf {L}$}\\ 
\multicolumn{1}{c}{$\bf {(IRAS)}$}& 
\multicolumn{1}{c}{$\bf {\bar{M}}$}&
\multicolumn{1}{c}{$\bf {A_{M}}$}&
\multicolumn{1}{c}{$\bf {\bar{M}}$}&
\multicolumn{1}{c}{$\bf {A_{M}}$}&
\multicolumn{1}{c}{$\bf {\bar{M}}$}&
\multicolumn{1}{c}{$\bf {A_{M}}$}&
\multicolumn{1}{c}{$\bf {\bar{M}}$}&
\multicolumn{1}{c}{$\bf {A_{M}}$}\\ \hline           
$01037+1219$& $ 7.84$ & $ 8.06$ & $ 4.57$ & $ 4.71$ & $ 2.34$ & $ 2.32$ & $-0.08$ & $-0.09$\\
$02270-2619$& $ 4.54$ & $ 4.57$ & $ 2.77$ & $ 2.74$ & $ 1.43$ & $ 1.42$ & $-0.04$ & $-0.06$\\
$02351-2711$& $ 2.75$ & $ 2.92$ & $ 1.57$ & $ 1.67$ & $ 0.93$ & $ 0.98$ & $ 0.17$ & $ 0.19$\\
$03507+1115$& $ 2.23$ & $ 2.57$ & $ 0.47$ & $ 0.69$ & $-0.59$ & $-0.47$ & $-1.83$ & $-1.79$\\
$08088-3243$& $ 8.45$ & $ 8.57$ & $ 5.87$ & $ 5.97$ & $ 3.77$ & $ 3.86$ & $ 1.19$ & $ 1.24$\\
$09116-2439$& $     $ & $     $ & $ 9.40$ & $ 9.68$ & $ 6.03$ & $ 6.04$ & $ 2.15$ & $ 2.22$\\
$09429-2148$& $ 5.50$ & $ 5.65$ & $ 3.75$ & $ 3.69$ & $ 2.40$ & $ 2.36$ & $ 0.72$ & $ 0.73$\\
$09452+1330$& $ 7.26$ & $ 7.28$ & $ 3.98$ & $ 3.99$ & $ 1.15$ & $ 1.17$ & $-2.54$ & $-2.55$\\
$12447+0425$& $ 4.89$ & $ 4.88$ & $ 3.13$ & $ 3.08$ & $ 1.83$ & $ 1.81$ & $ 0.28$ & $ 0.26$\\
$17049-2440$& $     $ & $     $ & $ 8.50$ & $ 8.57$ & $ 5.47$ & $ 5.51$ & $ 1.88$ & $ 1.93$\\
$17360-3012$& $ 8.37$ & $ 6.23$ & $ 5.98$ & $ 6.15$ & $ 3.99$ & $ 4.09$ & $ 1.66$ & $ 1.64$\\
$17411-3154$& $     $ & $     $ & $     $ & $     $ & $     $ & $     $ & $ 3.86$ & $ 3.94$\\
$18194-2708$& $ 9.22$ & $ 9.33$ & $ 6.06$ & $ 6.13$ & $ 3.67$ & $ 3.74$ & $ 0.89$ & $ 0.88$\\
$18333+0533$& $ 9.93$ & $10.38$ & $ 5.97$ & $ 6.20$ & $ 3.57$ & $ 3.73$ & $ 1.22$ & $ 1.29$\\
$18348-0526$& $     $ & $     $ & $     $ & $     $ & $ 8.20$ & $ 8.26$ & $ 2.01$ & $ 1.98$\\
$18398-0220$& $ 5.99$ & $ 5.91$ & $ 3.58$ & $ 3.51$ & $ 1.87$ & $ 1.79$ & $-0.13$ & $-0.19$\\
$18560-2954$& $ 3.03$ & $ 3.14$ & $ 1.60$ & $ 1.67$ & $ 0.80$ & $ 0.82$ & $-0.17$ & $-0.19$\\
$19093-3256$& $ 3.11$ & $ 3.14$ & $ 1.88$ & $ 1.90$ & $ 1.21$ & $ 1.22$ & $ 0.40$ & $ 0.39$\\
$20077-0625$& $ 6.70$ & $ 6.93$ & $ 3.92$ & $ 4.04$ & $ 2.10$ & $ 2.14$ & $ 0.18$ & $ 0.16$\\
\hline
\multicolumn{1}{c}{$\bf {\Delta}_{M}$}&
\multicolumn{2}{c}{$0.28$}&
\multicolumn{2}{c}{$0.11$}&
\multicolumn{2}{c}{$0.05$}&
\multicolumn{2}{c}{$0.03$}\\ 
\multicolumn{1}{c}{$\bf {{\sigma}_{\Delta}}$}&
\multicolumn{2}{c}{$0.16$}&
\multicolumn{2}{c}{$0.03$}&
\multicolumn{2}{c}{$0.02$}&
\multicolumn{2}{c}{$0.01$}\\ 
\hline
\end{tabular}
}
\end{table}
\begin{table}
\begin{center}
\caption{Average colours for variable sources with with good phase coverage}
\normalsize{
\begin{tabular}{lcccccccc}\hline
\multicolumn{1}{c}{$\bf {NAME}$}&
\multicolumn{2}{c}{$\bf {J-H}$}&
\multicolumn{2}{c}{$\bf {H-K}$}&
\multicolumn{2}{c}{$\bf {J-K}$}&
\multicolumn{2}{c}{$\bf {K-L}$}\\ 
\multicolumn{1}{c}{$\bf {(IRAS)}$}&
\multicolumn{1}{c}{$\bf {A_{f}}$}&
\multicolumn{1}{c}{$\bf {A}$}&
\multicolumn{1}{c}{$\bf {A_{f}}$}&
\multicolumn{1}{c}{$\bf {A}$}&
\multicolumn{1}{c}{$\bf {A_{f}}$}&
\multicolumn{1}{c}{$\bf {A}$}&
\multicolumn{1}{c}{$\bf {A_{f}}$}&
\multicolumn{1}{c}{$\bf {A}$}\\ \hline           
$01037+1219$& $3.28$ & $3.25$ & $2.23$ & $2.21$ & $5.51$ & $5.46$ & $2.42$ & $2.41$\\  
$02270-2619$& $1.77$ & $1.76$ & $1.34$ & $1.33$ & $3.11$ & $3.09$ & $1.48$ & $1.47 $\\ 
$02351-2711$& $1.18$ & $1.18$ & $0.65$ & $0.65$ & $1.82$ & $1.83$ & $0.76$ & $0.74 $\\ 
$03507+1115$& $1.77$ & $1.79$ & $1.06$ & $1.08$ & $2.83$ & $2.88$ & $1.24$ & $1.24 $\\ 
$08088-3243$& $2.58$ & $2.58$ & $2.10$ & $2.10$ & $4.68$ & $4.69$ & $2.58$ & $2.59 $\\ 
$09116-2439$& $    $ & $3.72$ & $3.37$ & $3.37$ & $    $ & $7.46$ & $3.88$ & $3.88 $\\ 
$09429-2148$& $1.74$ & $1.74$ & $1.35$ & $1.35$ & $3.10$ & $3.10$ & $1.68$ & $1.67 $\\ 
$09452+1330$& $3.27$ & $3.28$ & $2.83$ & $2.83$ & $6.10$ & $6.11$ & $3.69$ & $3.70 $\\ 
$12447+0425$& $1.76$ & $1.75$ & $1.30$ & $1.29$ & $3.06$ & $3.04$ & $1.55$ & $1.55 $\\ 
$17049-2440$& $    $ & $3.24$ & $3.03$ & $3.04$ & $    $ & $6.20$ & $3.60$ & $3.62 $\\ 
$17360-3012$& $2.39$ & $2.99$ & $1.99$ & $1.99$ & $4.38$ & $4.98$ & $2.33$ & $2.34 $\\ 
$18194-2708$& $3.16$ & $3.16$ & $2.40$ & $2.39$ & $5.55$ & $5.54$ & $2.78$ & $2.77 $\\ 
$18333+0533$& $3.96$ & $3.89$ & $2.40$ & $2.34$ & $6.35$ & $6.24$ & $2.36$ & $2.32 $\\ 
$18348-0526$& $    $ & $    $ & $    $ & $    $ & $    $ & $    $ & $6.20$ & $6.27 $\\ 
$18398-0220$& $2.40$ & $2.37$ & $1.71$ & $1.68$ & $4.12$ & $4.05$ & $2.00$ & $1.98 $\\ 
$18560-2954$& $1.44$ & $1.46$ & $0.79$ & $0.82$ & $2.23$ & $2.28$ & $0.98$ & $1.00 $\\ 
$19093-3256$& $1.23$ & $1.22$ & $0.67$ & $0.65$ & $1.90$ & $1.87$ & $0.81$ & $0.77 $\\ 
$20077-0625$& $2.79$ & $2.76$ & $1.82$ & $1.79$ & $4.60$ & $4.55$ & $1.92$ & $1.91 $\\ 
\hline
\multicolumn{1}{c}{$\bf {{\Delta}_{A}}$}&
\multicolumn{2}{c}{$0.06$}&
\multicolumn{2}{c}{$0.01$}&
\multicolumn{2}{c}{$0.07$}&
\multicolumn{2}{c}{$0.02$}\\ 
\multicolumn{1}{c}{$\bf {{\sigma}_{\Delta}}$}&
\multicolumn{2}{c}{$0.04$}&
\multicolumn{2}{c}{$0.01$}&
\multicolumn{2}{c}{$0.04$}&
\multicolumn{2}{c}{$0.01$}\\ 
\hline
\end{tabular}
}
\end{center}
\end{table}
\twocolumn

\end{document}